\documentclass[useAMS,usenatbib]{mn2e}

\usepackage{graphicx}
\usepackage[version=3]{mhchem} 
\usepackage{subfig}
\usepackage{lscape}

\title[The effect of the regular solution model in the condensation of protoplanetary dust]{The effect of the regular solution model in the condensation of protoplanetary dust}
\author[Pignatale et al.]
{F. C. Pignatale$^{1}$\thanks{E-mail: fpignatale@astro.swin.edu.au},
S. T. Maddison$^{1,2}$,
V. Taquet$^{1,2,3}$, 
G. Brooks$^{4}$,
K. Liffman$^{5}$\\
$^{1}$Centre for Astrophysics \& Supercomputing, Swinburne University, H39, PO Box 218, Hawthorn, VIC 3122, Australia\\
$^{2}$Laboratoire d'Astrophysique de Grenoble, UMR 5571 Universit\'e Joseph Fourier/CNRS, BP 53, 38041 Grenoble cedex 9, France \\
$^{3}$Magistere de Physique Fondamentale d'Orsay, Universite Paris-11, France\\
$^{4}$Mathematics Discipline, FEIS, Swinburne University, H38, PO Box 218, Hawthorn, VIC 3122, Australia\\
$^{5}$CSIRO/MSE, PO Box 56, Heightt, VIC 3190, Australia
}

\begin{document}

\date{Accepted 2011 February 16.  Received 2011 February 14; in original form 2010 July 22}

\pagerange{\pageref{firstpage}--\pageref{lastpage}} \pubyear{2010}

\maketitle

\label{firstpage}

\begin{abstract}

We utilize a chemical equilibrium code in order to study the condensation process which occurs in protoplanetary discs during the formation of the first solids. The model specifically focuses on the thermodynamic  behaviour on the solid species assuming the regular solution model. For each solution, we establish the relationship between the activity of the species, the composition and the temperature using experimental data from the literature. We then apply the Gibbs free energy minimization method and study the resulting condensation sequence for a range of temperatures and pressures within a protoplanetary disc.

Our results using the regular solution model show that grains condense over a large temperature range and therefore throughout a large portion of the disc. In the high temperature region ($T\ge1400$~K) hibonite and gehlenite dominate and we find that the formation of corundum is sensitive to the pressure. The mid-temperature region is dominated by Fe$_{(s)}$  and silicates such as  \ce{Mg2SiO4} and \ce{MgSiO3}.  The chemistry of forsterite and enstatite are strictly related, and our simulations show a sequence of forsterite--enstatite--forsterite with decreasing temperature and the abundance of the first high temperature peak of forsterite is also pressure sensitive. In the low temperature regions ($T\le600$~K) a range of iron compounds (FeS, \ce{Fe2SiO3}, \ce{FeAl2O3}) form.  We find that all the condensation sequences move towards lower temperature as the pressure decreases.

We also run simulations using the ideal solution model and see clear differences in the resulting condensation sequences with changing solution model. In particular, we find that the turning point in which forsterite replaces enstatite in the low temperature region is sensitive to the solution model. In this same temperature region, fayalite is the most stable compound for the regular solution, while magnetite replaces fayalite in the ideal solution model at the lowest values of temperature.

Our results show that the ideal solution model is often a poor approximation to experimental data at most temperatures important in protoplanetary discs. We find some important differences in the resulting condensation sequences when using the regular solution model, and suggest that this model should provide a more realistic condensation sequence.

\end{abstract}


\section{Introduction}

Studying protoplanetary discs, and more particularly their dust content, is crucial for understanding the creation of the solar system. Solar system bodies such as planets, meteorites and comets are all created from solids during the protoplanetary disc phase. 

The quantity and the quality of observations of protoplanetary discs around young stars has increased substantially in the last years: Spitzer Space Telescope has provided a large amount of useful data whose interpretation provides information on the composition, growth and thermal processing of dust grains in discs \citep[]{2009A&A...507..327O, 2009A&A...497..379M, 2008ApJ...683..479B, 2006ApJ...639..275K, 2006A&A...459..545G}. The processes of grain formation are intrinsically related to the chemistry of gas and dust. Indeed, the material in the discs is a mixture of condensates with different chemical composition and properties \citep[]{2009arXiv0911.1010H, 2009arXiv0908.3708B}. 

The study of objects in our solar system, including asteroids, meteorites and comets provides more direct and accessible evidence of the chemistry of the young solar system. Meteorites are characterized by heterogeneous compositions \citep[]{2005mcp..book...83K} and petrographic and mineral analysis show mixtures of compounds that are the result of condensation processes in different environments and times during the protoplanetary disc phase  \citep[]{2005mcp..book..143S}. Even comets, among the oldest and most distant objects of our solar system, show this heterogeneity \citep[]{2005mcp..book..663B}. The Stardust mission \citep{2003JGRE..108.8111B}  returned samples from comet Wild 2 and studies of their composition \citep[]{2008Sci...321.1664N, 2006Sci...314.1735Z} show complex thermodynamic and dynamic scenarios where particles probably formed in the inner regions of the solar system \citep{2006M&PS...41....3L} which experienced high temperature processes were found close to other material produced by cold thermodynamics. 

This paper is the first in a series in which we aim to determine the composition of both the gas and solid phases in various regions of protoplanetary discs in order to understand their chemistry and study the dust content. We assume thermodynamic equilibrium and solve for the condensation sequence using the regular solution model \citep{DeHoff1993} for the behaviour of solid solutions.  In this paper we present the regular solution model and compare the  condensation results with those obtained using an ideal solution for a range of temperatures at fixed pressures. In our next papers, we will apply this technique to specific 1D and 2D discs models.

The Gibbs energy minimization technique for equilibrium calculation is a powerful method for understanding the evolution of chemistry in a complex system \citep{DeHoff1993}. However, the predictions using this technique are sensitive to  providing rigorous mathematical descriptions of solution behaviour, in particular, providing allowances  for the effect of concentration and temperature on the activity of species dissolved into a solid state solution \citep{DeHoff1993}.  In this study, we include the effect of temperature of solution behaviour by using the regular solution model, which is superior to the assumption used in previous studies in which ideal mixing is assumed for many of the solid  phases and the effect of temperature on solution behaviour is ignored \citep[e.g.][]{Pasek2005,Gail1998}. It is well established that these effects are important in complex phase equilibria. 

The outline of this paper is as follows: in section 2 we describe the physical and chemical conditions assumed to be  present in protoplanetary discs. We also introduce the Gibbs free energy minimization method to solve for the condensation sequence and the steps followed to determine our final thermodynamic model. In section 3 we discuss the behavior of solid solutions and present a range of activity-composition relations derived from laboratory experiments.  In section 4, we present the results of our simulations and  we compare them with the ideal solution models (section 5)  and with previous modeling work and recent observations (section 6). Conclusions are presented in section 7.


\section{The Model}
\label{sec-model}

Protoplanetary discs results from conservation of the angular momentum of a collapsing, rotating cloud of gas and dust. Detailed studies have been made to define the structure of protoplanetary discs \citep[]{2008A&A...480..859B, 2004IAUS..221..403D, 1999ApJ...527..893D,Dalessio1998, 2004A&A...413..571G, 2001A&A...378..192G, Gail1998} and the physics involved in the evolution of the resulting gas--dust system \citep[]{2010A&A...513A..79B, 2008PhST..130a4015D}.  In this work, we calculate the condensation sequence for an initial chemistry over a range of temperatures and at fixed values of pressure which are appropriate for protoplanetary discs. 

In this section, we describe the conditions within our model, the chemical equilibrium code used to study the condensation sequence, along with assumptions and limitations of the code.

\subsection{Physical properties of the disc}
\label{modeltp}

Different disc models return different temperature and pressure distributions. These can be described as functions of the disc radius ($r$) in 1D models and both the radius and the distance from the mid-plane ($r$,$z$) for 2D models. In this model, we do not subscribe to any specific disc model, but instead chosen range of temperatures and specific pressure values that are relevant in protoplanetary discs. These values are not meant to represent specific locations in the disc but to constrain the values of temperature and pressure we use in the our simulations and define three main regions (high, middle, low temperatures) in the disc in which condensation occurs. Furthermore, condensation sequences calculated over a range of temperatures at fixed values of pressure allow us to compare our results with previous work and thermodynamical predictions.

For this work, we use the disc model of \citet{Dalessio1998} to guide us to find suitable values of $T$ and $P$. \citet{Dalessio1998}  studied the vertical structure, temperature, and surface density of a disc whose heating processes include viscous dissipation, radioactive decay, cosmic rays, and stellar irradiation. They chose a typical T Tauri star with $M_{\ast}=0.5 \, {\rm M}_{\odot}$, $R_{\ast}=2 \, {\rm R}_{\odot}$,  $T_{\ast}=4000$~K, and $\dot{M}=10^{-8} \, {\rm M}_{\odot} {\rm yr}^{-1}$. 

We assume an irradiated disc with a constant accretion rate and consider only the inner 5~AU. Most of the dust mass is located near the mid-plane of the disc, so the study of temperature and pressure was done at the mid-plane. 
\citet{Dalessio1998} calculated the radial temperature profile at the mid-plane, the surface density, and the different vertical heights above the mid-plane.
The temperature at the mid-plane, $T_{c}$, ranges from 60 to 1900~K for radii between 5 and 0.05~AU. The values of $P$ and $T_{c}$ are shown in Table~\ref{table1} for three typical disc radii: $r=0.05, 1.0$ and 5~AU.  In the condensation studies we present in this work, we use a temperature range from 50 to 1850~K for fixed pressures $10^{-3}, 10^{-4},...,10^{-8}$ bar.

\begin{table}
\centering
\caption{Temperature and pressure values for three typical radii calculated from the model of \citet{Dalessio1998}.}
\begin{tabular}{l l l l}
\hline
Parameters & $r=0.05$ AU & $r=1$ AU & $r=5$ AU \\
\hline
$T_{c}$ (K) & 1900 & 240 & 60 \\
$P$ (bars) & $2.1\times 10^{-3}$ & $2.2\times 10^{-7}$ & $9.4\times 10^{-9}$ \\
\hline
\end{tabular}
\label{table1}
\end{table}

\subsection{Gibbs free energy minimisation} 
\label{gibbsfreeenergy}
Our condensation model of protoplanetary discs is based on the determination of the equilibrium of the initial system given a set of temperatures, pressures and initial composition. For transformations occurring at constant temperature and pressure, the relative stability of the system is determined by its Gibbs free energy. The general definition of Gibbs free energy is
\begin{equation}
\label{gib}
G=H-TS \, ,
\end{equation}
where $H$ is the enthalpy, $T$ the absolute temperature and $S$ the entropy. The laws of classical thermodynamics determine that the system will be in its most stable state if  the Gibbs free energy is minimized \citep[][]{DeHoff1993,Porter1992}.
In a system with multiple species, the total Gibbs free energy, $G_T$, is given by 
\begin{equation}
G_T=\sum_{i=1}^{N} x_i G_{i} \, ,
\label{eqn-G}
\end{equation}
where $x_i$ is the number of moles of species $i$, and $G_{i}$ is the Gibbs free energy of species $i$, and $N$ is the total number of possible species defined for the system. $G_i$ is a function of the temperature, the mole fraction and solution parameters 
\begin{equation}
G_{i}= G_{i}^{0} + R T \ln \gamma_i + R T \ln X_i \, ,
\label{eqn-gibbs}
\end{equation}
where $G_{i}^{0}$ is the standard Gibbs free energy of formation of the species $i$, $R$ is the gas constant, $T$ the temperature, $X_i$ is the molar fraction of the species in the solution phase,
and $\gamma_i$ is the activity coefficient of species $i$. The variation of $\gamma_i$ with temperature and composition is determined experimentally. 

The determination of the equilibrium is equivalent to finding the set of values which minimizes the {\bf function} in equation~(\ref{eqn-G}) and satisfies the mass balance constraint given by 
\begin{equation}
\sum_{i=1}^{N} a_{ij}  x_i = b_j \,\,\, (j=1,2,...,m) \,,
\end{equation}
where there are $m$ different types of {\bf elements} and $a_{ij}$ is the number of atoms of element $j$ in species $i$, $x_i$ is the number of moles of species $i$ (such that the molar fraction $X_i=x_i/x_t$, where $x_t$ is the total number of moles), 
and $b_j$ is the total number of {\bf moles} of element $j$.  

Therefore, the parameters required to solve equation~\ref{eqn-gibbs} include the temperature, $T$, the number of moles $x_i$ of each of the $N$ species $i$ and the total number of moles $x_t$ in the initial composition of the system, the standard Gibbs free energy of formation of each species $i$, $G_{i}^{0}$, and the activity coefficients $\gamma_i$, which depend on the behaviour of the species $i$ in the solution, for a given temperature, $T$.  

The Gibbs energy minimisation method is a technique widely used for geology, chemistry, astrophysics, metallurgy, materials and chemical engineering and industrial purposes for understanding complex equilibrium calculation at high temperature \citep[]{Hack1996}. For our purpose, we use the HSC\footnote{HSC website: http://www.outotec.com} software package \citep{Roine2007},  which uses the Gibbs free energy minimization method of \citet{White1958}. 

The thermodynamic data for each compound in our list  are taken from the database provided by HSC. The list of references is very long and not reported here. For each compound the database reports the enthalpy, $H$, entropy, $S$, and heat capacity, $C$, along with a range of other important parameters. We refer the reader to \citet{Roine2007} and references within, plus the {\it HSC Database module} for full details.


\subsection{Method and limitations}

In the previous section, we introduced the concept of equilibrium which assumes that the system has reached the most stable state for a given $T, P$ and composition. This is not completely true for a protoplanetary disc. For example, the existence of non-crystalline material in protoplanetary discs is a sign that the disc is not in equilibrium and at low temperatures reaction rates and transport phenomena can indeed be so slow as to make equilibrium unachievable. However it is assumed that the lifetime of a disc ($\sim$6--30 Myrs) \citep[]{2005sptz.prop20069C,2001ApJ...553L.153H} is long enough to overcome the kinetic barriers such as mass transfer of species, reaction rates and surface phenomena, thus allowing the disc to reach an equilibrium state. \citet {2006GeCoA..70.5035T} studied the reaction rates of many compounds in our list (e.g. forsterite, anorthite, spinel, corundum, pyroxene, melilite) also focusing on the influence of time, temperature and gas composition in the reaction rates. Their results show that a mean time of 1 hour is required to return stable phases.
Furthermore, computational techniques based on the assumption of equilibrium have been successfully used to study and understand reactions in metallurgical systems and in the Earth's geology processes \citep[and references therein]{Belov2002}.

We assume that the system is only composed of solid and gas states. While the condensation model developed by \citet{Yoneda1995} showed that liquids can be formed at atmospheric pressures, the typical pressure of a protoplanetary disc is much lower than this ($P \leq 10^{-2}$ bar -- see Table~\ref{table1}). The absence of liquid phase in this pressure regime can also be seen in the $T$--$P$ phase diagrams of compounds in our list, and hence we ignore the liquid phase.

The range of possible species produced in the model are derived from a limited number of elements. We choose the fifteen most abundant elements of the Sun taken from \citet{Pasek2005}, who use the solar photosphere abundance observations of \citet{Grevesse1998}, \citet{2002ApJ...573L.137A} and \citet{2001ApJ...556L..63A} (see Table~\ref{table2}) from a total amount equal to 100~kmol.   We assume that the Sun's photosphere has changed little over the time and so the current composition can safely be used for the initial conditions in our model. 

The chondrites, which formed early in the history of the solar system, have the same abundance for most elements as the photosphere \citep{Scott2006}, which supports our assumption of a relatively constant stellar composition. (While \citet{2009ARA&A..47..481A} have a revised photospheric composition, we use the values of \citet{Pasek2005} to compare our results to previous work.)

The potential number of species that can be formed from the combination of these 15 elements considered in this system is over 700. Our aim is to improve upon previous works in this field by using the regular solution model, and for the non-ideal case the HSC software is limited in the number of species it can handle. 
Using the 15 elements in Table~\ref{table2}, we follow \citet{Pasek2005} to prepare an initial list of species likely to form.  Using the tables in Appendix A1, we first remove all compounds that include the extra 8 solar elements which we exclude (i.e. Cr, P, Mn, Cl, K, Ti, Co, F).  We next use this list and assume ideal solution behaviour for just two phases, gas and solids, which are assumed to be pure phases. 
We run this ideal model for temperatures ranging from 50--1850~K for fixed pressures $10^{-3}, 10^{-4},...,10^{-8}$ bar.  Species which  did not appear in the system within the $T$ and $P$ ranges of our models were deleted, such as carbides, nitrides and some sulfide species.  This  procedure eliminated species that are very unlikely to be formed, while allowing the software to run efficiently.  However, we kept some species, such as solid Al and various oxides which are clearly important in the chemistry of the condensation sequence of the solar system and other protoplanetary discs, to verify if these compounds, for our range of temperatures and pressures, are involved in the formation of more complex species or not. 

Carbon is only present in our system in the main gaseous compounds together with solid graphite. Ideal calculations made during the definition of our initial system show that carbon solid compounds are very unlikely to form for our range of pressures and temperatures in equilibrium conditions. The majority of the initial C(g) is involved in the formation of gas compounds like \ce{CO} and \ce{CH4} (see section~\ref{sec-sims}), due to the presence of large amounts of oxygen and hydrogen in our system. Indeed the Gibbs free energies of these gas compounds makes  carbon gases the most stable carbon compounds. Furthermore, the formation of graphite is strictly related to the environmental condition in which a large amount of long chains of carbon gas is found \citep{Pasek2005}. Our equilibrium calculations show that lowering the temperature results in C(g) being replaced by CO(g) and there is not enough C(g) in the range of condensation temperature available to form graphite.

The complete list of species and phases used in our model is shown in Table~\ref{table3}.

\begin{table}
\centering
\caption{Abundance of elements in our system, taken from \citet{Pasek2005}, for a total amount equal to 100~kmol.}
\begin{tabular}{l r}
\hline
Element & Abundance (kmol) \\
\hline
H & 91 \\
He & 8.89 \\
O & $4.46 \times 10^{-2}$ \\
C & $2.23 \times 10^{-2}$ \\
Ne & $1.09 \times 10^{-2}$ \\
N & $7.57 \times 10^{-3}$ \\
Mg & $3.46 \times 10^{-3}$ \\
Si & $3.30 \times 10^{-3}$ \\
Fe & $2.88 \times 10^{-3}$ \\
S & $1.44 \times 10^{-3}$ \\
Al & $2.81 \times 10^{-4}$ \\
Ar & $2.29 \times 10^{-4}$ \\
Ca & $2.04 \times 10^{-4}$ \\
Na & $1.90 \times 10^{-4}$ \\
Ni & $1.62 \times 10^{-4}$ \\
\hline
\end{tabular}
\label{table2}
\end{table}

\begin{table*}
\centering
\caption{List of phases with their species used in our model.}
\begin{tabular}{l l l l l l l}
\hline
\multicolumn{2}{c}{\textbf{Gas phase}} & & \multicolumn{4}{c}{\textbf{Solid phases}} \\ 
\hline
Al	&	Mg	&	SiS	&	 \textbf{Olivine}	&	\textbf{Melilite}	&	\textbf{Magnesiowustite} 	&	\textbf{Pure phases}	 \\
C	&	N	&		&	\ce{Mg2SiO4}	&	 \ce{Ca2Al2SiO7}	&	FeO	&	\ce{Na2SiO3}	 \\
\ce{CH4}	&	\ce{N2}	&		&	\ce{Fe2SiO4}	&	\ce{Ca2MgSi2O7}	&	MgO	&	 \ce{CaAl12O19}	 \\
CO	&	\ce{NH3}	&		&		&		&		&	 \ce{H2O}	 \\
Ca	&	Na	&		&	 \textbf{Orthopyroxene}	&	\textbf{Fassaite}	&	\textbf{Sulfide}	&	\ce{Al2O3}	 \\
CaO	&	Ne	&		&	\ce{MgSiO3}	&	\ce{CaAl2SiO6}	&	FeS	&	CaO	 \\
Fe	&	Ni	&		&	\ce{FeSiO3}	&	\ce{CaMgSi2O6}	&	NiS	&	 \ce{Fe2O3} 	 \\
FeO	&	O	&		&		&		&	\ce{Ni3S2}	&	 \ce{Fe3O4}	\\
H	&	\ce{O2}	&		&		&		&		&	\ce{SiO2}	\\
\ce{H2}	&	OH	&		&	 \textbf{Plagioclase}	&	 \textbf{Spinel}	&	\textbf{Metal}	&	C	\\
\ce{H2O}	&	S	&		&	\ce{CaAl2Si2O8} 	&	\ce{MgAl2O4}	&	Fe	&		 \\
HS	&	Si	&		&	\ce{NaAlSi3O8}	&	\ce{FeAl2O4} 	&	Ni	&		 \\
\ce{H2S}	&	SiO	&		&		&		&	Si	&		 \\
He	&	\ce{SiO2}	&		&		&		&	Al	&		 \\
 \hline
\end{tabular} 
\label{table3}
\end{table*}

\section{Behaviour of Solid Solutions}
\label{sec-solidsoln}
We assume that solid species can mix together to form solid solutions, as these phases are observed on Earth as minerals and in meteorites. The choice of solid solutions is justified from previous thermodynamic studies.  The behaviour of each solid solution is studied, in order to determine their relation between composition, temperature and activity parameter. 

The choice of the solution behaviour which we use in this study is based on both laboratory and theoretical thermodynamic studies for each phase found in the literature. Most of these works focused on the relations between species composition and activity for one temperature, whereas observations of activities clearly show non-ideal solution behaviour evolving with temperature.
The solution behaviours can be quite complex with several parameters involved in the reactions.
There is no comprehensive scientific treatment of solution behaviour and no single approach to modelling multicomponent solutions, but there are several developed models available \citep{DeHoff1993}.
We choose the regular solution model to establish relations between the temperature in which the experiments were made, called reference temperature, the composition of the solution, a range of temperatures, the activity and the activity coefficient describing a particular compound in its phase. 

Whilst the thermodynamics of pure species is largely uncontroversial, solution behaviour is difficult to measure and model, particularly for complex oxide and sulfide phases. Therefore critically examining the solution behaviour of species important for understanding protoplanetary dust formation is vital in developing a rigorous understanding of the system.  In particular, we have sourced the best available solution data from geological and metallurgical studies to incorporate into studies of astrophysical dust, as the underlying physical chemistry of these different systems is fundamentally the same. In this work we have attempted to include the best available solution behaviour and where we have made assumptions, we state these clearly as a means by which to identify experimental conditions that need to be studied in the laboratory and stimulate the development of approximate solution models.

\subsection{The regular solution model}

The regular solution model is based on two fundamental assumptions \citep{DeHoff1993}: (i) the entropy of mixing is the same as that for an ideal solution, and
(ii) the enthalpy of solution is not zero, as in ideal solution, but is a function of
composition. As such, the model gives a simple relation between the activity coefficient and the temperature: if 
$\gamma_{T_{1}}$ is the activity coefficient of one species
at a temperature $T_1$ and $\gamma_{T_{2}}$ is the activity coefficient at another temperature
$T_2$ (for the same composition) then
\begin{equation} 
\frac{\ln \gamma_{T_{1}}}{\ln \gamma_{T_{2}}}=\frac{T_{2}}{T_{1}} \ ,
\end{equation}
or
\begin{equation}
\gamma_{T_{1}} = \gamma_{T_{2}}^{T_2/T_1} ,
\label{or}
\end{equation}
In the absence of activity data for a species ($k$), a simple relation between the temperature
and the activity coefficients was introduced, based on the observations
of the evolution of the behaviour of the species  with temperature given by
\begin{equation}
\gamma_{k}(T)=0.99^{T_0/T} \ ,
\label{eqn-regular}
\end{equation}
where $T_0$ is the reference temperature for the study of the phase.  Thus, the activity coefficient is near the ideal (i.e. $\gamma \simeq 1$) for high temperatures when $T \ge T_{0}$, while it moves away from the ideal as the temperature drops (i.e. $T \le T_{0}$).
This assumption provides a reasonable method to extrapolate existing solution data to conditions where there are no measurements of solution behaviour, and is tested in section~\ref{testing}.

Most thermodynamic literature report experimental values of activity at several values of composition and $T_0$.
The activity, $a$, and the activity coefficient, $\gamma$, are linked by the relation 
\begin{equation}
\label{act}
a_{k}(X_{k},T_0)=\gamma_{k}X_{k} \ ,
\end{equation} 
where $X$ is the composition of the studied species $k$.
The relations in equations ~(\ref{act}) and~(\ref{or}) allow us to derive a polynomial for the activity coefficient given by

\begin{equation}
\label{actcoef}
\gamma_{k}(X_k,T)=\frac{a_{k}(X,T_0)}{X_{k}}=\gamma_{k}(X_k,T_0)^{(T_0/T)}.
\end{equation} 

The meaning of activity and activity coefficient can be understood from equation~(\ref{act}).  If $\gamma=1$ (ideal) the activity of component {\it k} is equal to its mole fraction $X_k$ and the behavior of {\it k} is completely determined by its composition. If the value of $\gamma_{k} \ne 1$  (as in the regular solution model) the component acts as if the solution contains more ($\gamma_k >1$) or less ($\gamma_k <1$) than the mole fraction X suggests. Furthermore, from equation~(\ref{act}) the activity coefficient $\gamma$ provides information about the departure from the ideal behavior.

The regular solution model is a quantitative explanation of the non-ideal behaviour of solid solution, but there are some limitations. In a dilute solution of component {\it b} (solute) in the component {\it a} (solvent), $X_{b}$ approaches 0 or $X_{a}$ approaches 1. In this limit the activity coefficient for the solute component $\gamma_{b}$  becomes close to the Henry's law coefficient. 
A similar behaviour for this kind of solution can be seen in the opposite limit of the composition range when component {\it a} is dilute in component {\it b}. This symmetry in the extreme values of composition implies that the properties associated with an atom of component {\it b} surrounded by atoms of component {\it a} are the same as the opposite case in which an atom of component {\it a} is dilute in a component {\it b}  \citep{DeHoff1993}. Clearly this is not the case. Furthermore, the regular solution model returns a parabolic $\Delta H^{mix}$ (heat of mixing) function that is also symmetrical with respect to composition, whereas it is often unsymmetrical, especially in liquid solutions  \citep{Porter1992}.

\subsection{Solid solutions and their behaviour}
\label{ssb}
For each solution and each species, using thermodynamic laboratory experiments and theoretical models from the literature, and the use of the regular solution model as described above, we can now establish a relationship between the composition, temperature and activity coefficient of the nine solid phases listed in Table~\ref{table3}.

As a general principle when choosing the data for the activity, we have tried to find direct experimental data reporting activity values of compounds. These data are preferred over indirect methods derived via mathematical techniques, such as the Margules and Gibbs Duhem equations, since these techniques inherently approximate solutions 
\citep[see e.g.][]{Bodsworth1965,Anderson2005,Darken1953} which could affect the results. Thus where possible we preferred to choose data derived from direct laboratory experiments.

\subsubsection{Magnesio-wustite phase}

\ce{MgO} and \ce{FeO} can mix together to form the magnesio-wustite phase. \citet{Nafziger1973} studied the behaviour of these compounds in the \ce{MgO}--\ce{FeO}--\ce{F2O3} system at $T_0=1573.15$~K, providing activities of both monoxides according their composition in the system.

\begin{figure}
\centering
\includegraphics[scale=0.7]{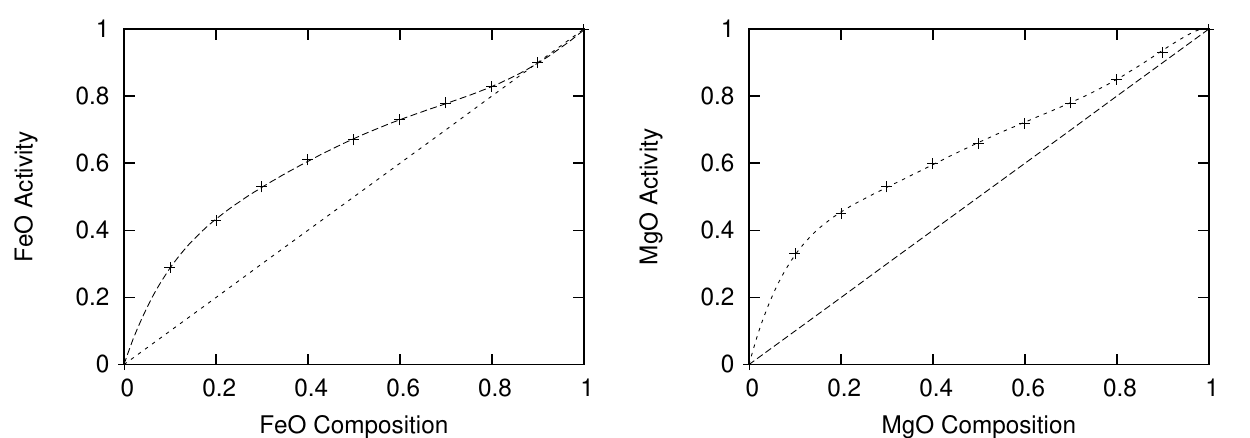}
\includegraphics[scale=0.7]{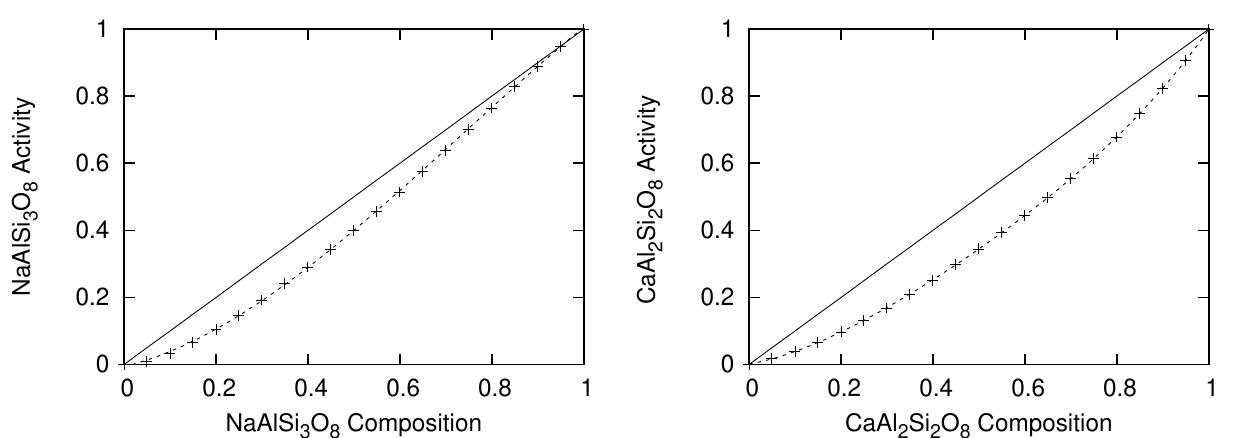}
\includegraphics[scale=0.7]{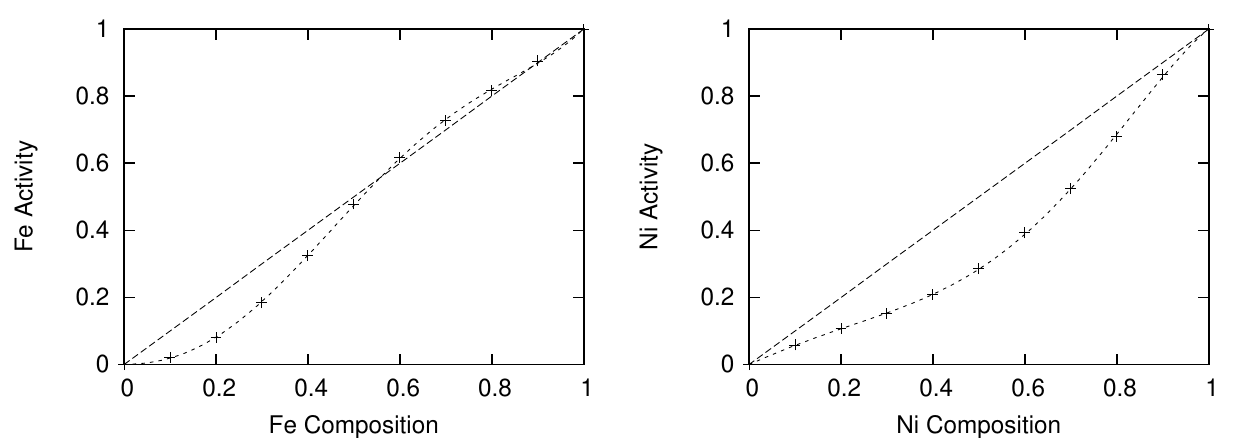}
\caption{Activity-composition functions for various phases. The data points are from experimental data and the curves are our polynomial fit to the data.  The straight line is the ideal solution. {\bf Top (a):} Magnesio-wustite solution: activity-composition functions for FeO (left) and MgO (right) at $T_0=1573.15$~K. Data from \citet{Nafziger1973}. {\bf Middle (b):} Plagioclase solution: activity-composition functions for albite (\ce{NaAlSi3O8} -- left)  and anorthite (\ce{CaAl2Si2O8} -- right). Data from  \citet{Newton1980}.  {\bf Bottom (c):} Metal solution: activity-composition functions for Fe and Ni at $T_0=1273.15$~K. Data from \citet{Conard1979}. } 
\label{fig-ActCompRelations}
\end{figure}

Using these experimental data, an explicit relation between the activity and the composition for FeO and MgO can be deduced. We fit a polynomial function to the data using a least squares fit algorithm with weight and standard deviation to determine the best polynomial order.
Thus an activity coefficient--composition relation could be derived for $T_0=1573.15$~K and then for the entire temperature range using the regular solution model (equation~\ref{actcoef}) given by

\begin{eqnarray}
\label{gammafeo1} 
\lefteqn{ \gamma_{FeO}(T,X_{FeO}) = (-10.62X_{{FeO}}^5+38.59X_{{FeO}}^4-54.48X_{{FeO}}^3 } \nonumber \\
                               & &+38.86X_{{FeO}}^2 -15.43X_{{FeO}}+4.07)^{T_0/T} \, ,
\end{eqnarray}
\begin{eqnarray}
\label{gammamgo1} 
\lefteqn{  \gamma_{MgO}(T,X_{MgO}) = (-23.36X_{{MgO}}^5 + 77.79X_{{MgO}}^4  } \nonumber \\
            & & - 102.20X_{{MgO}}^3 + 67.88X_{{MgO}}^2 -24.20X_{{MgO}}  \nonumber \\
            & & + 5.09)^{T_0/T} \, .
\end{eqnarray}
These relations are shown in Fig.~\ref{fig-ActCompRelations}a.

%
\subsubsection{Sulfide phase}

The sulfide phase is composed of FeS (troilite), NiS, and \ce{Ni3S2}. The activity data of FeS and \ce{Ni3S2} was studied by \citet{Byerley1972}, who measured  the sulfur activity for the Fe-Ni-S  solid ternary system in equilibrium with gaseous mixtures of \ce{H2S} and \ce{H2} at $T_0=1523.15$~K, and using the Gibbs-Duhem equation.  
The activity data for these two species were presented in an isoactivity Fe-Ni-S ternary diagram in \citet{Byerley1972}. An analytical function of composition for the activity coefficient could not be deduced, and hence the determination of the activity had to be made using the composition  values of sulfide compounds found in our first test simulations using the ideal solution model. 

For most temperatures in our calculations where FeS and \ce{Ni3S2} appear,
 $N_{\ce{FeS}}=1.43\times 10^{-3}$~kmol, $N_{\ce{Ni3S2}}=6.13\times10^{-5}$~kmol, and $N_{\ce{NiS}}=1.5\times 10^{-6}$~kmol, where $N$ is the number of mols. Thus the composition of Fe, S and Ni in the system are
$X_{S}=0.49, X_{Fe}=0.46$ and $X_{Ni}=0.05$.

Plotting these data on the isoactivity diagram of \citet{Byerley1972}, the activity was calculated to be 
$a_{\ce{Ni3S2}}=0.01$ and $a_{\ce{FeS}}=0.93$,  with a composition rate of
$X_{FeS}=0.96$ and $X_{\ce{Ni3S2}}=0.04$, and 
$\gamma_{FeS}=0.97$ and $\gamma_{\ce{Ni3S2}}=0.25$ at $T_0=1523.15$~K. Thus
\begin{eqnarray}
\gamma_{FeS}(T) &=& 0.97^{T_0/T} \, ,\\
\gamma_{\ce{Ni3S2}}(T) &=& 0.25^{T_0/T} \, .
\end{eqnarray}

As no activity data could be found in literature for NiS,  equation~(\ref{eqn-regular}) provides the following relationship
\begin{equation}
\label{nisgamma}
\gamma_{NiS}(T)=0.99^{T_0/T} \, ,
\end{equation}
where $T_0=1523.15$~K. 
\subsubsection{Olivine phase}

This phase is composed of forsterite (\ce{Mg2SiO4}) and fayalite (\ce{Fe2SiO4}). The study of the activity of this phase is based on data from \citet{Nafziger1967}, who experimentally determined the activity of olivine species at $T_0=1473.15$~K for the MgO-FeO-\ce{SiO2} system.

Depending on the initial composition of MgO, FeO, and \ce{SiO2}, five solid solutions can exist -- see Figure 1 of \citet{Nafziger1967}. Assuming ideal solution behaviour, our calculations show that olivine and pyroxene species coexist for the range of temperature in our system. So in this study
we chose to focus on the olivine $+$ pyroxene case. 

Pyroxene and olivine species are produced by the following reactions

\begin{eqnarray*}
\ce{Fe + SiO2 + 1/2O2} &\rightarrow& \ce{FeSiO3}   \, \, \rm{(pyroxene phase)} ,\\
\ce{2Fe + SiO2 + O2} &\rightarrow& \ce{Fe2SiO4}  \, \, \rm{(olivine phase)} ,
\end{eqnarray*}
and are in equilibrium following the reaction:
\begin{equation*}
\ce{2FeSiO3 + Mg2SiO4} \rightleftharpoons \ce{Fe2SiO4 + 2MgSiO3}\, .
\end{equation*}

Setting the oxygen partial pressure and enstatite composition, and using relations between the standard free energy of the reaction, activity, and composition \citep{Nafziger1967}, the activity-composition of olivine species could be deduced.

With these data we are able to extract activities for \ce{Fe2SiO4} and \ce{Mg2SiO4} using the relations
\begin{eqnarray}
a_{\ce{Fe2SiO4}} &=& a_{\rm{FeSi_{0.5}O_2}}^2 ,\\
\label{amezzo2}
a_{\ce{Mg2SiO4}} &=& a_{\rm{MgSi_{0.5}O_2}}^2 ,
\label{amezzo1}
\end{eqnarray}
presented by \citet{Anderson1993}.
Fitting a polynomial to the experimental data presented by \citet{Nafziger1967} for the olivine phase, an activity coefficient--composition relation for the two species could be deduced, firstly for $T_0= 1473.15$~K and then for all temperatures, given by
\begin{eqnarray}
\label{acoeffaya1}
\gamma_{Fay}(T,X_{Fay})&=& (3.80X_{Fay}^4 - 08.19X_{Fay}^3 + 5.60X_{Fay}^2 \nonumber \\
                        & &-0.45X_{Fay} + 0.24)^{T_0/T} \, ,\\
\label{acoefforst1}
\gamma_{Fo}(T,X_{Fo})&=& (7.58X_{Fo}^4 - 19.19X_{Fo}^3 + 16.91X_{Fo}^2 \nonumber \\
& &-5.19X_{Fo} + 0.88)^{T_0/T} \, ,
\end{eqnarray}
where $Fay$ is the fayalite (\ce{Fe2SiO4}) and $Fo$ the forsterite (\ce{Mg2SiO4}).

\subsubsection{Orthopyroxene phase}

The activities of enstatite (\ce{MgSiO3}) and ferrosilite (\ce{FeSiO3}) in orthopyroxene solution were determined experimentally by \citet{Saxena1971}.
They assume that the silicate framework does not change significantly and its contribution to the change in activity with temperature may be neglected, and therefore $a_{\rm Fe}^{\rm (Opx)}$ is similar to $a_{\rm FeSiO_{3}}^{\rm (Opx)}$. 

The most complete data from their study was for $T_{0} = 873.15$~K, and we used these data to establish our activity coefficient as a function of temperature and composition.

From the experimental data curves, we fitted a polynomial and using the regular solution model, the activity coefficients as a function of temperature and composition for the enstatite ($En$) and ferrosilite ($Fr$) were derived as follows:

\begin{eqnarray}
\label{genstatite1} \gamma_{En}(T,X_{En})&=&(12.90X_{En}^4-35.62X_{En}^3+35.61_{En}^2 \nonumber \\
& &-15.54_{En}+3.63)^{T_0/T} \, , \\
\label{gferrosilite1}\gamma_{Fr}(T,X_{Fr})&=&(5.72X_{Fr}^4-13.04X_{Fr}^3+10.44X_{Fr}^2 \nonumber \\
& &-4.12X_{Fr}+2.01)^{T_0/T} \, ,
\end{eqnarray}
where $T_0=873.15$~K.

\subsubsection{Plagioclase phase}

The activities of anorthite (\ce{CaAl2Si2O8}) and albite (\ce{NaAlSi3O8}) of the plagioclase solution must be treated with a non-ideal model, as shown by \citet{Newton1980}.
\citet{Kerrick1975} derived an ideal model for plagioclase solution from statistical mechanical considerations, based on the aluminium avoidance principle (Al-O-Al bonds in \ce{CaAl2Si2O8} and \ce{NaAlSi3O8} species are believed energetically unfavorable). They deduced activity--composition relationships
\begin{eqnarray}
\label{21}
a_{Ab}(X) &=& X_{Ab}^{2} (2-X_{Ab}) \, ,\\
\label{22}
a_{An}(X) &=& \frac{1}{4} X_{An} (1+X_{An})^{2}  \, ,
\end{eqnarray}
for albite ($Ab$) and anorthite ($An$) respectively.
\citet{Newton1980} used experimental studies of plagioclase to show that the solid solution of \citet{Kerrick1975} does not follow ideal behaviour. Specifically, they found that the measured enthalpy of solution is non-zero, contrary to ideal model assumption.  \citet{Newton1980} found activity coefficient--composition relationships from the measured excess enthalpy of mixing to be given by
\begin{eqnarray}
\gamma_{Ab} &=& \exp [X_{An}^{2} (W_{Ab}+2 (W_{An}-W_{Ab}) X_{Ab})/RT] \, ,\\
\gamma_{An} &=& \exp [X_{Ab}^{2} (W_{An}+2 (W_{Ab}-W_{An}) X_{An})/RT] \, ,
\end{eqnarray}
where $W_{An}$ and $W_{Ab}$ are the Margules parameters (which are constants at constant temperature and pressure), and $X_{An}$ and $X_{Ab}$ are respectively the mol fractions of anorthite and albite.  In their system, \citet{Newton1980} found  $W_{An}=2.025$~Kcal and $W_{Ab}=6.746$~Kcal.

In order to derive the activity and then the activity coefficient polynomials using equation~(\ref{actcoef}), these coefficient must be multiplied by the composition functions  for activity found by \citet{Kerrick1975} (equations (\ref{21}) and (\ref{22})), and thus the theoretical activity coefficients (from $\gamma = a/X$) which are required by the HSC software were found to be:
\begin{eqnarray}
\label{gdefalbite1}
\gamma_{Ab}(T,X_{Ab}) &=& (0.87X_{Ab}^3-1.57X_{Ab}^2+1.45X_{Ab} \nonumber \\ 
                      & &+0.24)^{T_0/T} \, ,\\
\label{gdefanort1}
\gamma_{An}(T,X_{An})&=&  (-0.55X_{An}^2+1.23X_{An}+0.33)^{T_0/T} \, ,
\end{eqnarray}
where $T_0 = 1000$~K. These relations are shown in Fig.~\ref{fig-ActCompRelations}b.

\subsubsection{Melilite phase}

In order to determine  the behaviour of the melilite phase and derive the activity--composition relations of gehlenite (\ce{Ca2Al2SiO7}) and akermanite (\ce{Ca2MgSi2O7}) in solid solution, we used the work of \citet{Charlu1981} and \citet{Waldbaum1973} and the analysis of the melilite system from \citet{1994Metic..29...41B}. We also made use of the MELTS Supplemental Calculator described by \citet{Ghiorso1995} and  \citet{Asimow1988} and  references therein in order to derive the activity--composition  data for gehlenite and akermanite.

While the small deviation from ideality shown by \citet{Charlu1981} for the $\Delta G$ of mixing seems to suggest that the ideal solution is a good approximation to describe the behaviour of this phase, several studies suggest that the melilite phase is in fact non-ideal. Models and results from \citet{1994Metic..29...41B} show how melilite deviates from the ideal, suggesting that this behaviour has to be taken in account when calculating phase equilibria involving this phase. More recent  studies by \citet{Merlini2004} and \citet{Gemmi2007} further support the non-ideality of this system.

The MELTS Supplemental Calculator was used to calculate thermodynamical properties of the melilite binary system using data from \citet{Charlu1981} and \citet{Waldbaum1973}, extrapolating activities for  both gehelenite and akermanite at fixed values of temperature, pressure and mole fractions.  We set  $T_0=1000$~K, $P=1$~atm and used values of mole fractions from $0\le X_{Ge} \le 1$ and  $1 \ge X_{Ak} \ge 0$.
Fitting polynomials to these data, we derived functions for the activity  $a$ and then, using the regular solution model, for the activity coefficients given by
\begin{eqnarray}
\label{gammaaker} \gamma_{Ak}(X_{Ak},T)&=& (-8.98X_{Ak}^5+28.71X_{Ak}^4-29.08X_{Ak}^3 \nonumber \\
& &+8.00X_{Ak}^2+2.11X_{Ak}+0.22)^{({T_o}/{T})} ,\\
\label{gammaghele} \gamma_{Ge}(X_{Ge},T)&=&(4.05X_{Ge}^5-15.05X_{Ge}^4+17.75X_{Ge}^3 \nonumber \\
& &-5.97X_{Ge}^2-1.24X_{Ge}+1.47)^{({T_o}/{T})} ,
\end{eqnarray}
for $T_{0}=1000$~K.

\subsubsection{Fassaite phase}

Activity and activity coefficient data for the fassaite phase were deduced by \citet{Benisek2007}. They studied the thermodynamic properties of  Ca-tshermak (\ce{CaAl2SiO6}) and diopside (\ce{CaMgSi2O6}), and found activity coefficient relations
\begin{eqnarray}
\label{actcfdiopside} 
RT\ln\gamma_{Di}&=& X_{CaT}^2[W_{Di}+2(W_{CaT}-W_{Di}) \nonumber \\
& &\times(1-X_{CaT})]   ,\\
\label{actcfcats} 
RT\ln\gamma_{CaT}&=&[W_{CaT}+2(W_{Di})-W_{CaT})X_{CaT}] \nonumber \\
& &\times(1-X_{CaT})^2 ,
\end{eqnarray}
for Ca-tshermak ($CaT$) and diopside ($Di$), 
where W is the Margules parameter reported by \citet{Benisek2007} for both Ca-tshermak and diopside. 

The relation that links the activity coefficients and activity reported by \citet{Benisek2007} is
\begin{equation}
a=a^{id}\gamma ,
\end{equation}
where
\begin{eqnarray}
\label{actdioid}
a_{Di}^{id} &=& (1-X_{CaT})[(2-X_{CaT})/2]^2  \, ,\\
\label{actcatsid}
a_{CaT}^{id} &=& (4X_{CaT})(X_{CaT}/2)(2-X_{CaT}/2) \, .
\end{eqnarray}

Using these relations, we can calculate activities of Ca-Ts and diopside using $a=\gamma X$
together with equations~(\ref{actdioid}) and (\ref{actcatsid}) for $X_{CaT}$ and $X_{Di}$, and equations~(\ref{actcfdiopside}) and (\ref{actcfcats}) for $\gamma$.

Fitting polynomials to the calculated data, we derived equations for $a_{CaT}$ as a function of $X_{CaT}$ and $a_{Di}$ as a function of $X_{Di}$ (where $X_{Di}=1-X_{CaT}$), and then for $\gamma$ according the regular solution model.  Finally we obtained solutions for the activity coefficients
\begin{eqnarray}
\label{gammacats1} \gamma_{CaT}(T,X_{CaT}) &=& (0.43X_{CaT}^{3}-1.79X_{CaT}^{2}+2.32X_{CaT} \nonumber \\
                                 & &+0.04)^{T_0/T} ,\\
\label{gammadio1} \gamma_{Di}(T,X_{Di}) &=&  (0.17X_{Di}^{3}-0.018X_{Di}^{2}+0.52X_{Di} \nonumber \\
& &+0.32)^{T_0/T},
\end{eqnarray}
where $T_0=1500$~K.

\subsubsection{Spinel phase}

The activity data for spinel species (\ce{MgAl2O4} and \ce{FeAl2O4}) was taken from activity-composition measurements made by \citet{Jacob1998} at $T_0=1300$~K. They measured  the electromotive force for different compositions of the spinel solution, which is linked with \ce{FeAl2O4} activity. They deduced activity--composition relationship for \ce{MgAl2O4} using the Gibbs-Duhem Equation. 

Using the experimental data of \citet{Jacob1998} and fitting a polynomial, a relation between activity coefficient and composition of spinel species at $T_0=1300$~K was determined via
\begin{eqnarray}
\label{fegs1} 
\gamma_{FeSp}(T,X_{FeSp})&=&(-0.175X_{FeSp}^2+0.373X_{FeSp} \nonumber \\
 & &+0.800)^{T_0/T} ,\\
\label{gmgs1} 
\gamma_{MgSp}(T,X_{MgSp})&=&(0.035X_{MgSp}+0.96)^{T_0/T} \, .
\end{eqnarray}

\subsubsection{Metal alloy phase}

Results from our ideal model simulations show that Fe is among the most abundant species in the solid solution. Indeed, it represents between 90 and 94 mole-percent across the entire temperature range where the metal phase is present, with Ni representing between 6 and 10 mole-percent, and Si and Al less than $10^{-7}$ mole-percent.

The activity coefficient for Fe and Ni can be studied in a Fe-Ni solution, since the amounts of Si and Al are negligible and hence they will have little effect on the solution behaviour.
\citet{Conard1979} studied the Fe and Ni behaviour for the liquid-solid iron-nikel alloys for three specific values of temperature in the range of 1273.15--1873.15~K.
The range of interest to us is 1273.15--1573.15~K where the Fe-Ni system is in solid phase.  Fitting polynomials to the experimental data, we derive the activity and then the activity coefficients given by 

\begin{eqnarray}
\label{gfe} \gamma_{Fe}(X_{Fe},T) &=& (5.012X_{Fe}^4-10.27X_{Fe}^3+4.854X_{Fe}^2 \nonumber \\
                           & &+1.401X_{Fe}+0.003)^{{T_0}/{T}} ,\\ 
\label{gni} \gamma_{Ni}(X_{Ni},T) &=& (-2.91X_{Ni}^4+4.92X_{Ni}^3-1.33X_{Ni}^2 \nonumber \\
& &-0.27X_{Ni}+0.60)^{{T_0}/{T}} ,
\end{eqnarray}
for $T_{0}$=1273.15~K.
These relations are shown in Fig.~\ref{fig-ActCompRelations}c.

The activity data for the silicon as solute in Fe-solution were taken from \citet{Sakao1975}. They studied the activity of Si in solid Fe-Si alloys in the region $0.028\leq X_{Si}\leq 0.084$ and $1373.15 \leq T(\rm{K}) \leq 1643.15$. 
The experimental results clearly show a relation between Si activity coefficient, $T$ and $X_{Si}$ given by
\begin{equation}
\log \gamma_{Si} = 1.19 - \frac{7070}{T} + \left(-6.30 + \frac{18.30}{T} \right) X_{Si} .
\end{equation}
Our ideal results show that Si has an amount  of $\sim10^{-7}$ mole-percent so the equation could be simplified for very low Si-composition as
\begin{equation} 
\gamma_{Si}(T)=10^{(1.19-7070/T)} \, .
\end{equation}

No data for aluminium in solid Fe-Al alloys was found at very low Al composition, so the Al activity coefficient relationship was chosen as ideal at high temperature following the equation (\ref{eqn-regular})
\begin{equation} 
\label{algamma}
\gamma_{Al}(T)=0.99^{T_0/T} \, .
\end{equation}
where $T_0=1873.15$~K.

\subsection{Testing the regular solution model}
\label{testing}
When we have experimental data for the same solution at different reference temperatures we can test the regular solution model.
According to the regular solution model
\begin{equation}
\label{testg}
\gamma_{T_i}=\gamma_{T_j}^{T_j/T_i} ,
\end{equation}
and for different experimental data at three different reference temperatures, $T_{0}=T_{i,j,k}$, we find
\begin{equation}
\label{testg1}
\gamma_{T_i}=\gamma_{T_j}^{T_j/T_i}=\gamma_{T_k}^{T_k/T_i} .
\end{equation}
Using the set for activity coefficients by \citet{Conard1979} for iron in the metal phase for three different values of $T_{0}$,  we derived activity coefficient polynomials at  $T_{j}$=1473~K and $T_{k}$=1573~K and tested the appropriateness of the our regular solution model given by equation (\ref{gfe}) where the reference temperature is $T_{i}$=1273~K.
\begin{figure}
\begin{center}
\includegraphics[width=8.5cm]{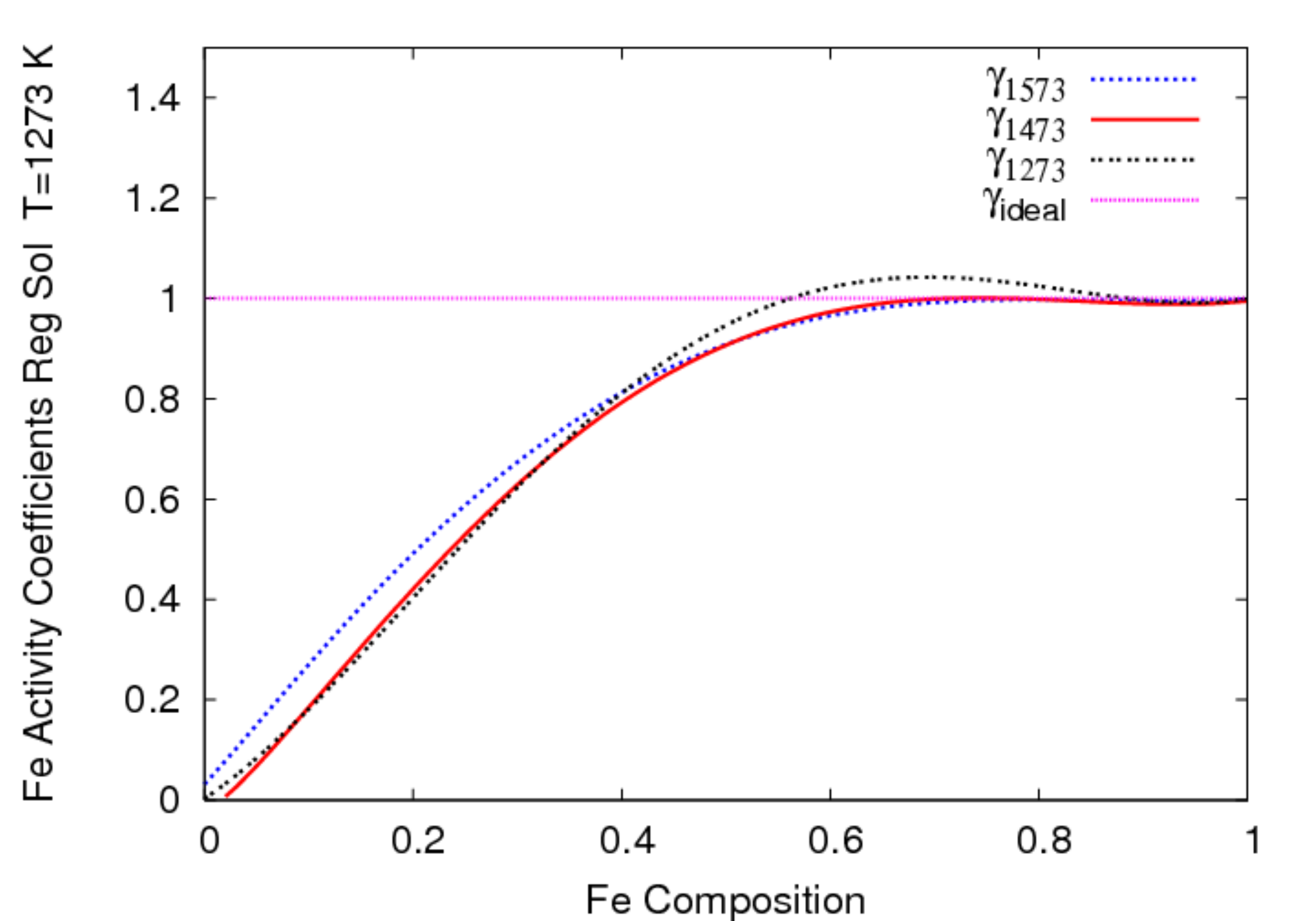}\\
\caption{Test of the regular solution model for iron given by equation~(\ref{gfe}) using equation~(\ref{testg1}).
Here $T_{i}$=1273~K, $T_{j}$=1473~K and $T_{k}$=1573~K. Polynomials of different activity coefficients are derived using data provided by \citet{Conard1979}. The ideal solution is shown as a straight line at $\gamma=1$.}
\label{1273}
\end{center}
\end{figure}
Fig.~\ref{1273} shows the application of equation~(\ref{testg1}). The test returns good agreement across the entire range of composition, with profiles given by the second and the third terms of equation~(\ref{testg1}) fitting the first function with a standard deviation of 0.03 ($\gamma_{1573}$) and 0.02 ($\gamma_{1473}$). Fig.~\ref{1273} also clearly show how the behaviour of Fe deviates from the ideal at lower concentrations ($X_{Fe}\le0.5$).

We can also use experimental data to test the assumption of equation~(\ref{eqn-regular}). Following the results reported by  \citet{1994Metic..29...41B} for gehlenite and akermanite, which show activities moving toward the  ideal with increasing temperature, we ran a test in which we compared results using equations~(\ref{gammaaker}) and~(\ref{gammaghele}) with equation~(\ref{eqn-regular}). 
Differences between the results are reported in Table~\ref{table4}. As can be seen, the variations in the temperature of  appearance, $T_a$, and disappearance, $T_d$, and the maximum amount of composition, $A_{max}$, between the results of the two models are small. Extrapolations made by MELTS Supplemental Calculator (equations~(\ref{gammaaker}) and~(\ref{gammaghele})) and results given by equation~(\ref{eqn-regular}) are in good agreement at the reference temperature $T_{0}=1000$~K for the melilite phase considered here, giving us confidence in equation ~(\ref{eqn-regular}) when applied to species for which we do not have experimental data, such as \ce{NiS} (equation~\ref{nisgamma}) and \ce{Al} (equation~\ref{algamma}).

\begin{table*}
\centering
\caption{Temperature of appearance, $T_a$, and disappearance, $T_d$, and the maximum amount, $A_{max}$, for gehlenite and akermanite calculated by HSC using polynomials for the activity coefficients derived by MELTS Supplemental Calculator (equation~(\ref{gammaaker}) and equation~(\ref{gammaghele})) and equation~(\ref{eqn-regular}) with $T_0=1000$~K.}
\begin{tabular}{|c|c|c|c|c|c|c|}
\hline
 & \multicolumn{3}{|c|}{$\gamma_{ge}=$ eq.~(\ref{gammaghele}) $\gamma_{ak}=$eq.~(\ref{gammaaker}) } & \multicolumn{3}{|c|}{$\gamma_{ge}=\gamma_{ak}=0.99^{T_0/T}$} \\
P & $T_a $ &  $T_d $  & $A_{max}$ & $T_a  $ & $T_d $ & $A_{max}$ \\
(bar) &  (K)  &   (K)   & (kmol) &  (K)   & (K)  & (kmol)  \\
\hline
\multicolumn{7}{|c|}{\bf Gehelenite} \\
$10^{-3}$ & 1556 & 1445 & $9.47\times10^{-5}$ & 1566 & 1445  & $9.43\times10^{-5}$  \\
$10^{-4}$ & 1506 & 1374 &  $1.00\times10^{-4}$ & 1496  & 1374  & $9.52\times10^{-5}$   \\
$10^{-6}$ & 1364  & 1253 &  $9.63\times10^{-5}$ & 1364 & 1253  &  $9.63\times10^{-5}$   \\
$10^{-8}$ & 1202 & 1152 &  $9.70\times10^{-5}$ & 1212  & 1152  & $9.70\times10^{-5}$   \\
\multicolumn{7}{|c|}{\bf Akermanite} \\
$10^{-3}$ & 1556 & 1445 & $1.80\times10^{-5}$ & 1556 & 1445 &  $2.30\times10^{-5}$  \\
$10^{-4}$ & 1496 & 1374 &  $1.58\times10^{-5}$ & 1496  & 1374  & $1.96\times10^{-5}$   \\
$10^{-6}$ & 1364 & 1273 &  $5.37\times10^{-6}$  & 1354  & 1253  & $1.29\times10^{-5}$   \\
$10^{-8}$ & 1202 & 1152 &  $7.58\times10^{-6}$  & 1212 & 1152  &  $5.97\times10^{-6}$   \\
\hline
\end{tabular}
\label{table4}
\end{table*}

\section{Simulation Results}
\label{sec-sims}

Now that we have all the thermal data, we can use our solar composition for a given pressure, $P$, and temperature, $T$, to solve for the equilibrium condensation sequence using the HSC software.
As shown by \citet{Dalessio1998}, the temperature and particularly the pressure in the disc span a very large range (see section~\ref{sec-model}.1). Runs were made for fixed values of pressure ($10^{-3},10^{-4},...,10^{-10}$ bar) over the temperature range $50\le T(\rm{K}) \le1850$. The condensation sequence was calculated as a function of temperature and results are reported using temperature--amount diagrams. 

In this section we describe the overall results of our equilibrium calculations for both gases and solids. Our simulations show that the condensation temperatures depends on pressure, with sequences moving towards lower temperatures as the pressure decreases. This behaviour is seen more clearly at higer temperatures than in the cooler regions (see for example Fig.~\ref{fig-majorgases} and Fig.~\ref{fig-hight}). 
The temperature of appearance (disappearance) is defined as the temperature at which the amount of a compound raises above (falls below) $10^{-7}$ kmol.  In our data tables, we indicate with {\it trace} those compounds which are present with an amount between  $10^{-9}$ and $10^{-7}$~kmol.
The following discussion is made using the results of the condensation sequence at $P=10^{-3}$ bar unless otherwise specified. 

\subsection{Gases}
\label{gassequence}

\begin{figure}
\centering
\includegraphics[width=8.5cm]{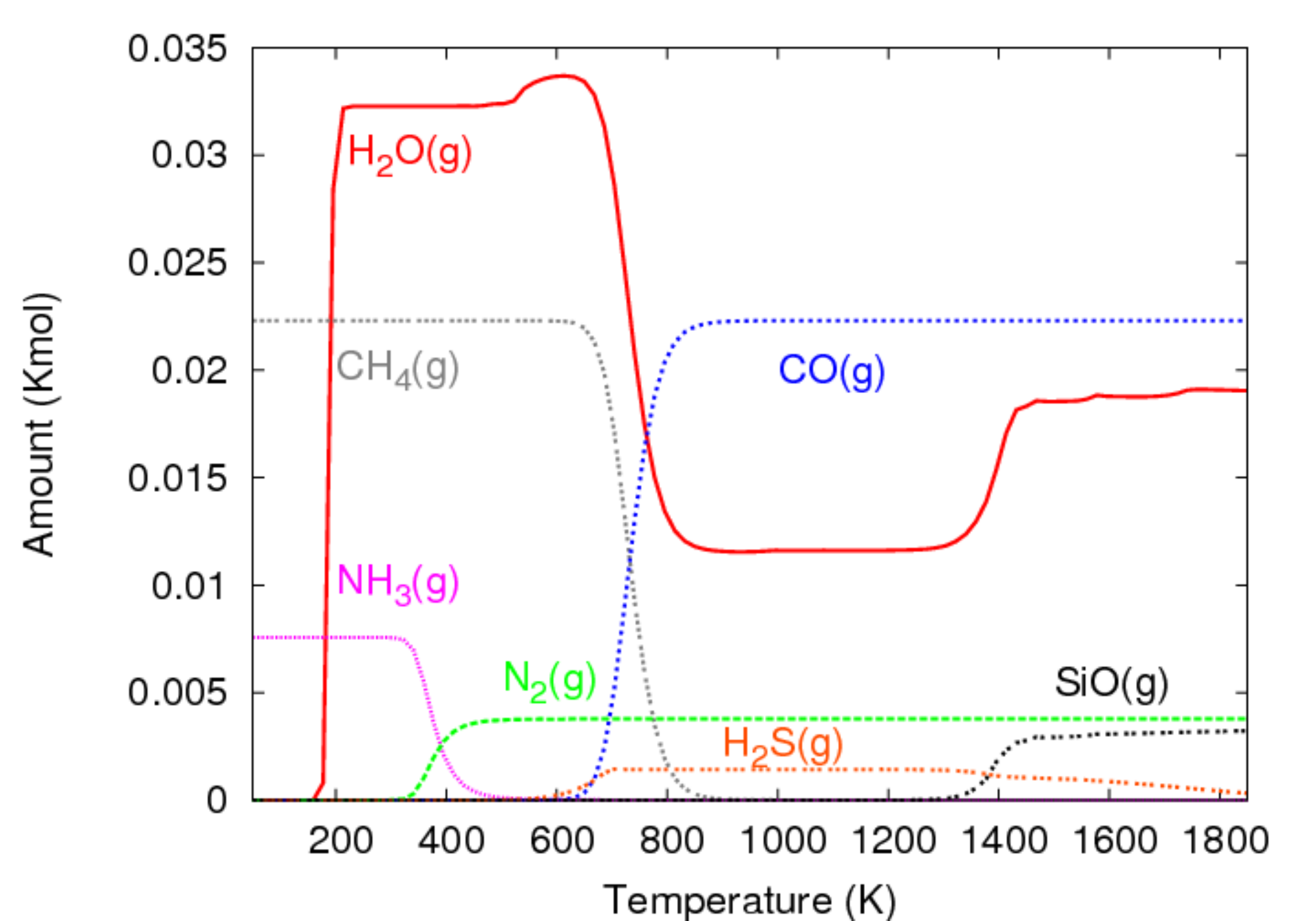}
\includegraphics[width=8.5cm]{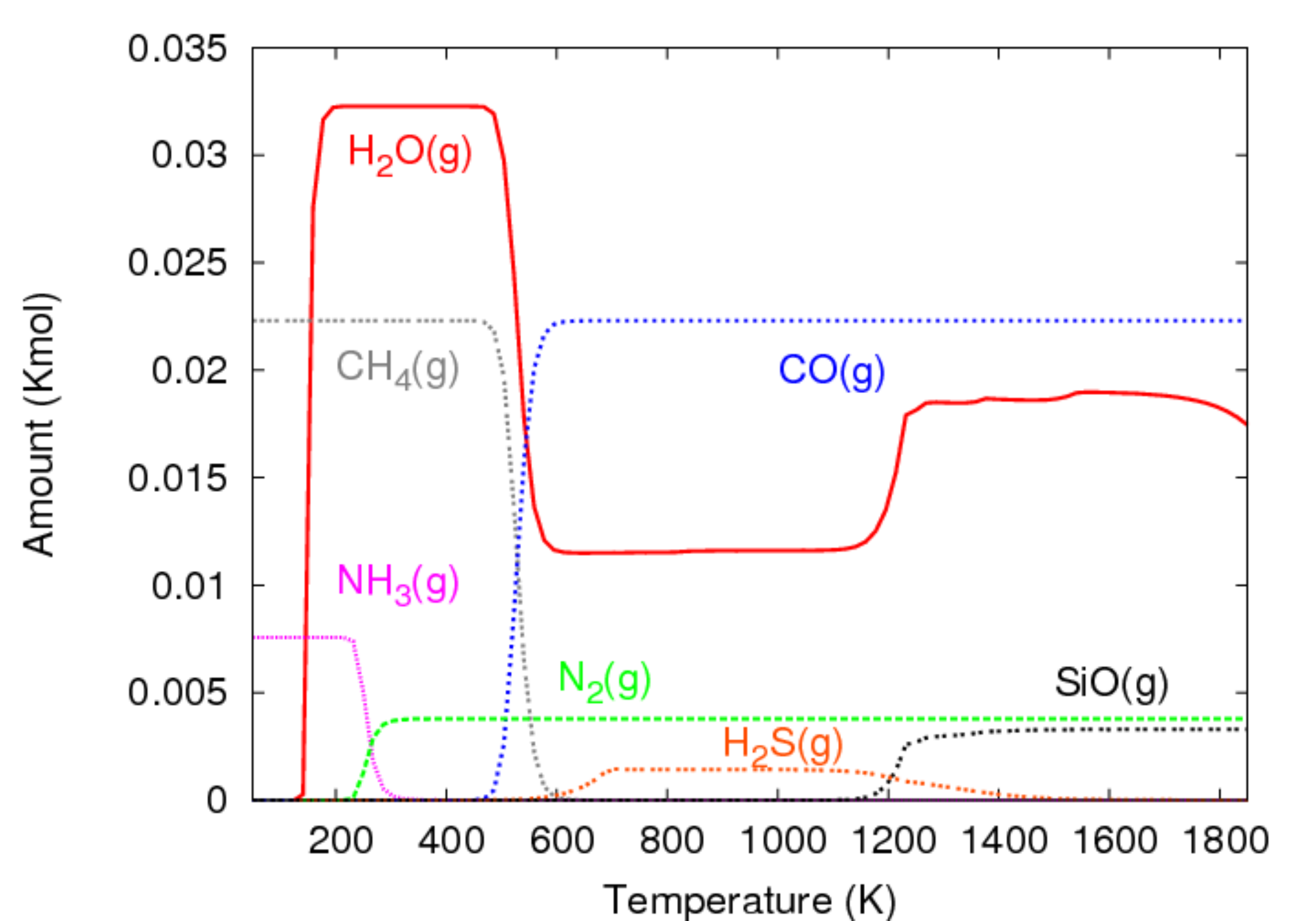}
\caption{Evolution of the major gases as a function of temperature at (top) $P=10^{-3}$~bar and (bottom) $P=10^{-6}$~bar.}
\label{fig-majorgases}
\end{figure}

H, \ce{H2} and  He are the most abundant gas species in the system and they reach a maximum amount of $8.56\times10^{-1}$ kmol, $45.48$ kmol and $8.89$ kmol respectively in different condensation regions.
Looking at the entire temperature range, \ce{H2O}, SiO, CO, \ce{N2}, \ce{H2S}, \ce{CH4} and \ce{NH3} are the major gas compounds in our results (Fig.~\ref{fig-majorgases}).
SiO, CO and \ce{H2O} characterize the higher temperature region with \ce{CO} being the most abundant gas. The region between $600\le T(\rm{K})\le850$ shows an important transition with \ce{CH4} replacing CO, and \ce{H2O} becoming the most important gas. \ce{N2} is present along almost the entire temperature range until $T\sim400$~K, where it is replaced by \ce{NH3}. Among the less abundant gases, sulfur gases (HS, SiS) are also formed. 

Moving to  lower pressure, we see the entire sequence move to lower temperatures, including the transition between \ce{CO} and \ce{CH4} -- see Fig.~\ref{fig-majorgases}. The maximum amount of each gas at equilibrium changes little with pressure. 

\subsection{Solids}
\label{solidsequence}

\begin{figure}
\centering
\includegraphics[width=8.5cm]{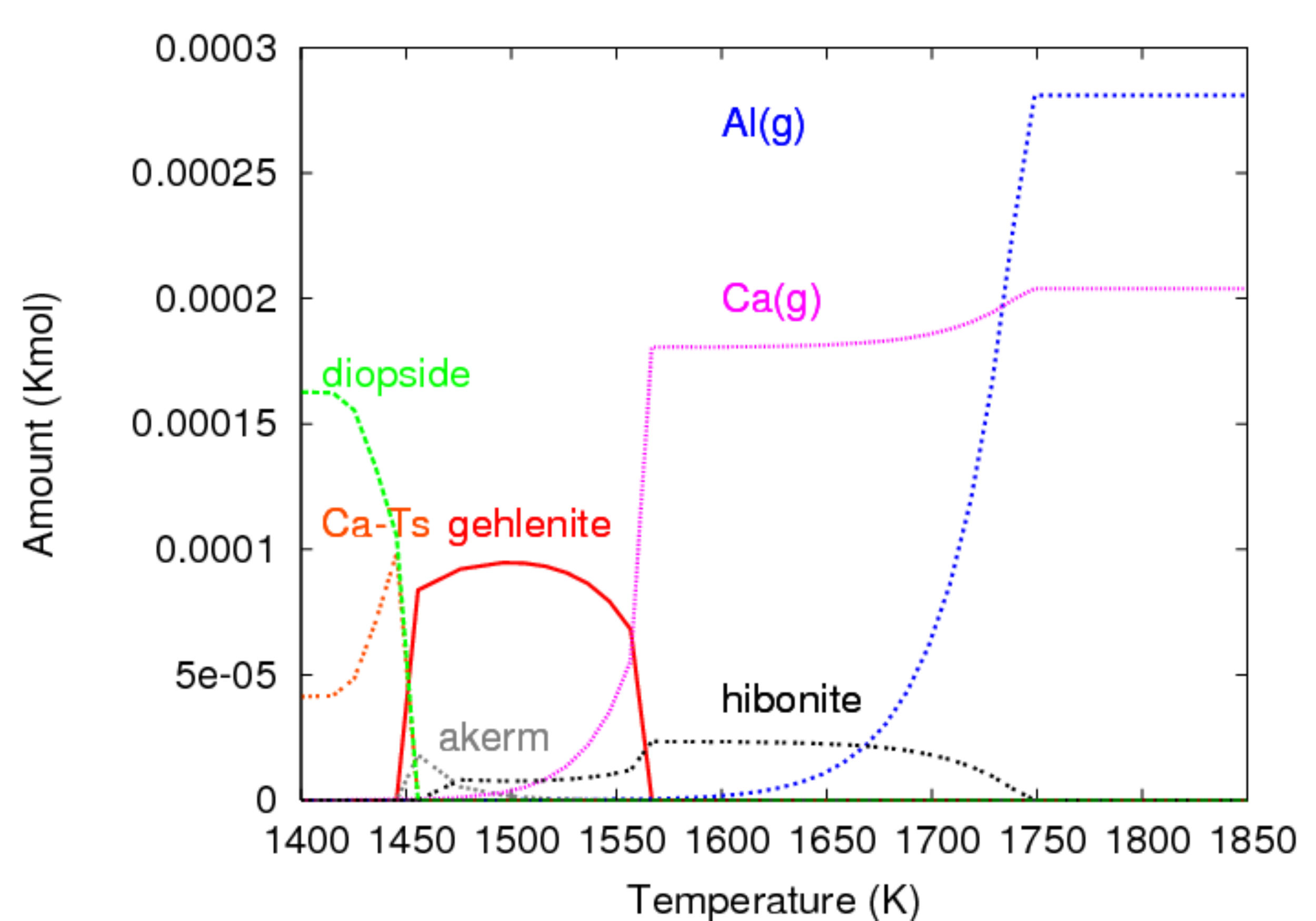}
\includegraphics[width=8.5cm]{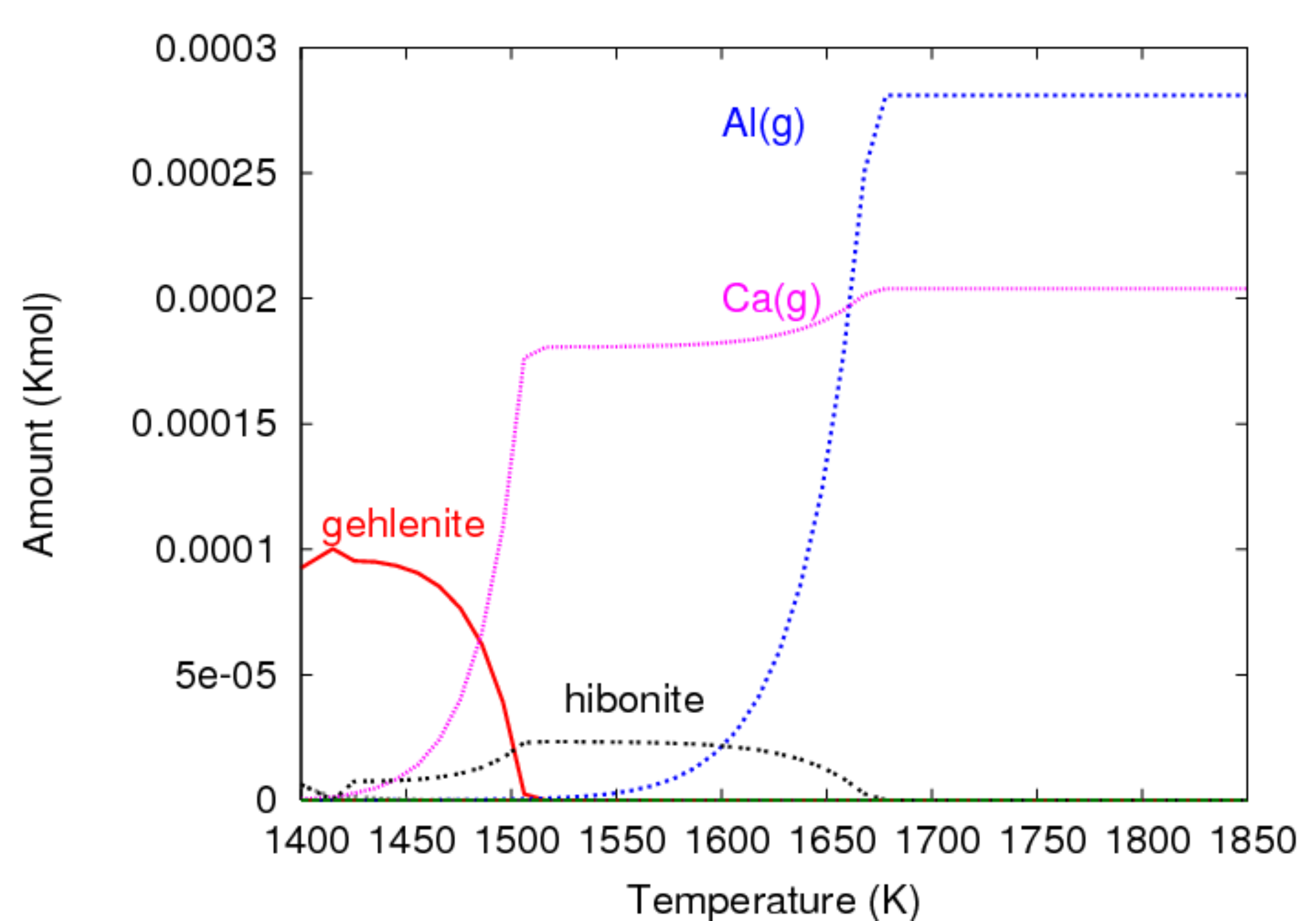}
\includegraphics[width=8.5cm]{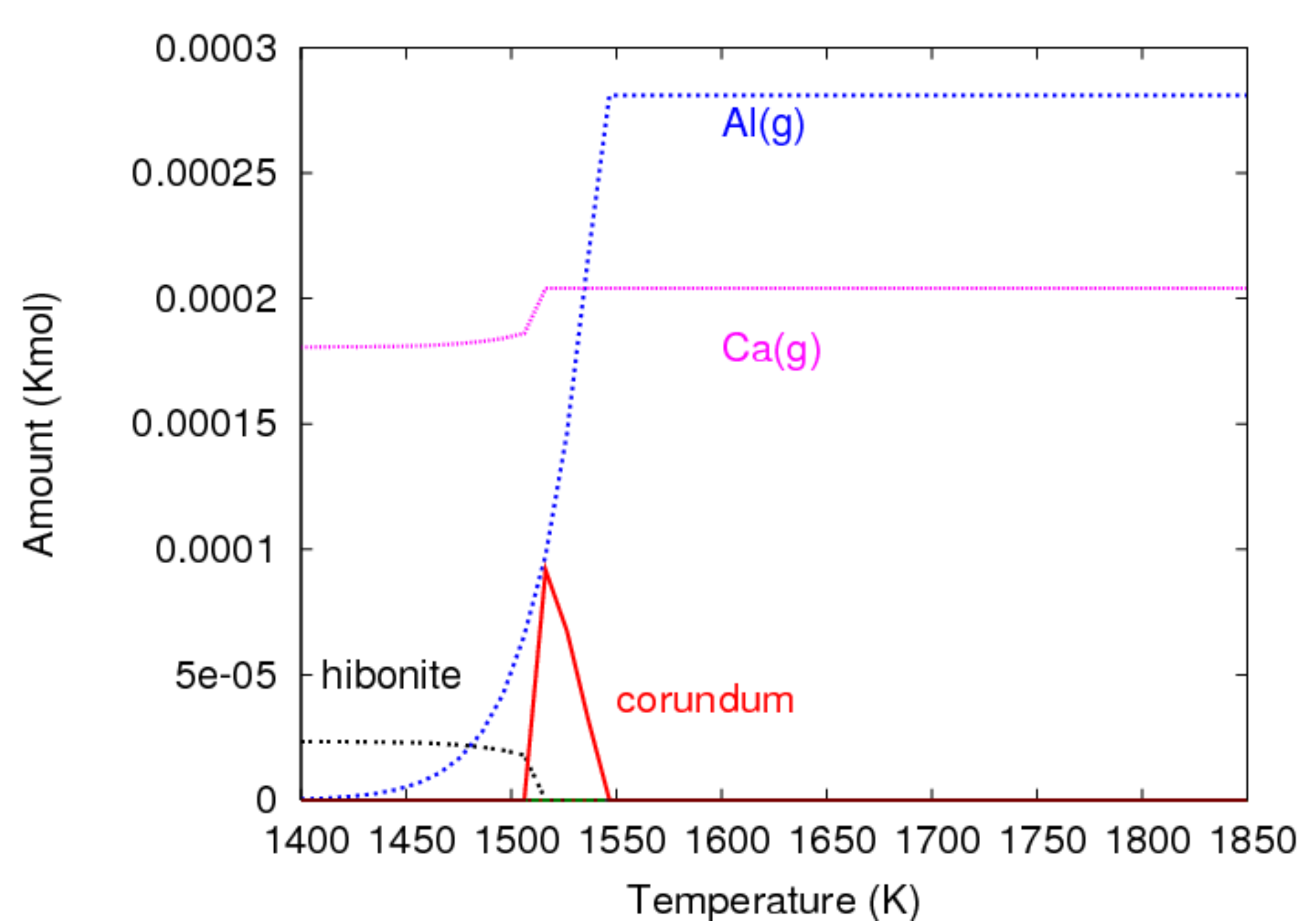}
\caption{Condensation sequence in the high temperature region as function of temperature at (top) $P=10^{-3}$~bar, (middle) $P=10^{-4}$~bar and (bottom) $P=10^{-6}$~bar.}
\label{fig-hight}
\end{figure}

The high temperature region ($1400\le T(\rm{K}) \le1850$) is  the place where the calcium--aluminium solids start to form, with several compounds in sequence from hibonite (\ce{CaAl12O19}) to gehlenite (\ce{Ca2Al2SiO7}), followed by the first calcium--magnesium silicates, akermanite (\ce{Ca2MgSi2O7}) and calcium-Tchermak (\ce{CaAl2SiO6}) -- see Fig.~\ref{fig-hight}, which shows the high temperature condensates for three pressures.  The condensation temperatures of Ca(g) and Al(g) are very close and sensitive to the pressure. At $P=10^{-3}$ bar, Ca(g) and Al(g) start to condense at $T =1738$~K and the first, most stable solid compound formed is hibonite (\ce{CaAl12O19}). Lowering the pressure to  $10^{-6}$ bar changes the sequence with the Al(g) ($T=1536$~K) condensing before Ca(g) ($T=1506$~K) and the first most stable solid compound formed is the corundum (\ce{Al2O3}). 

\begin{figure}
\centering
\includegraphics[width=8.5cm]{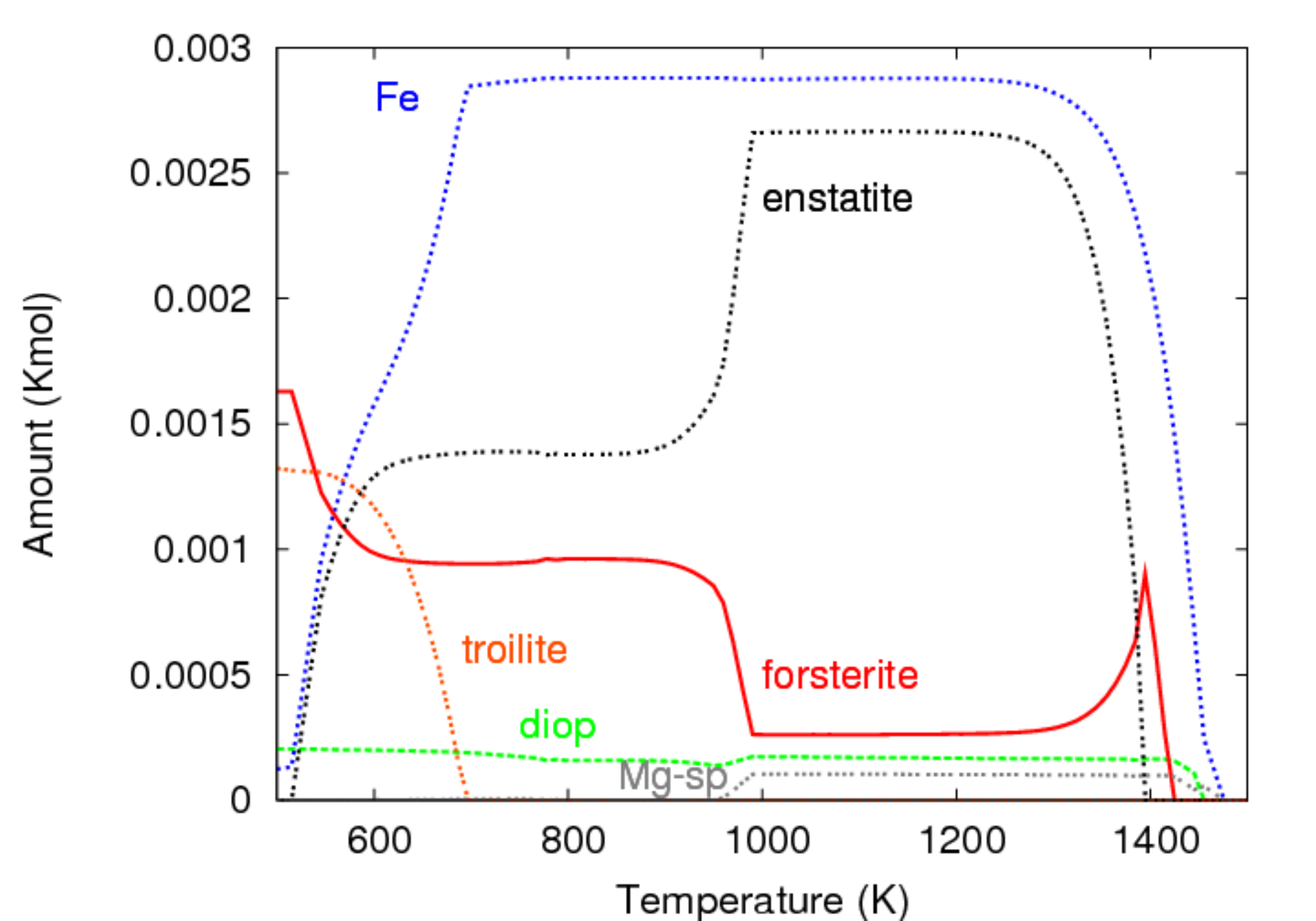}
\includegraphics[width=8.5cm]{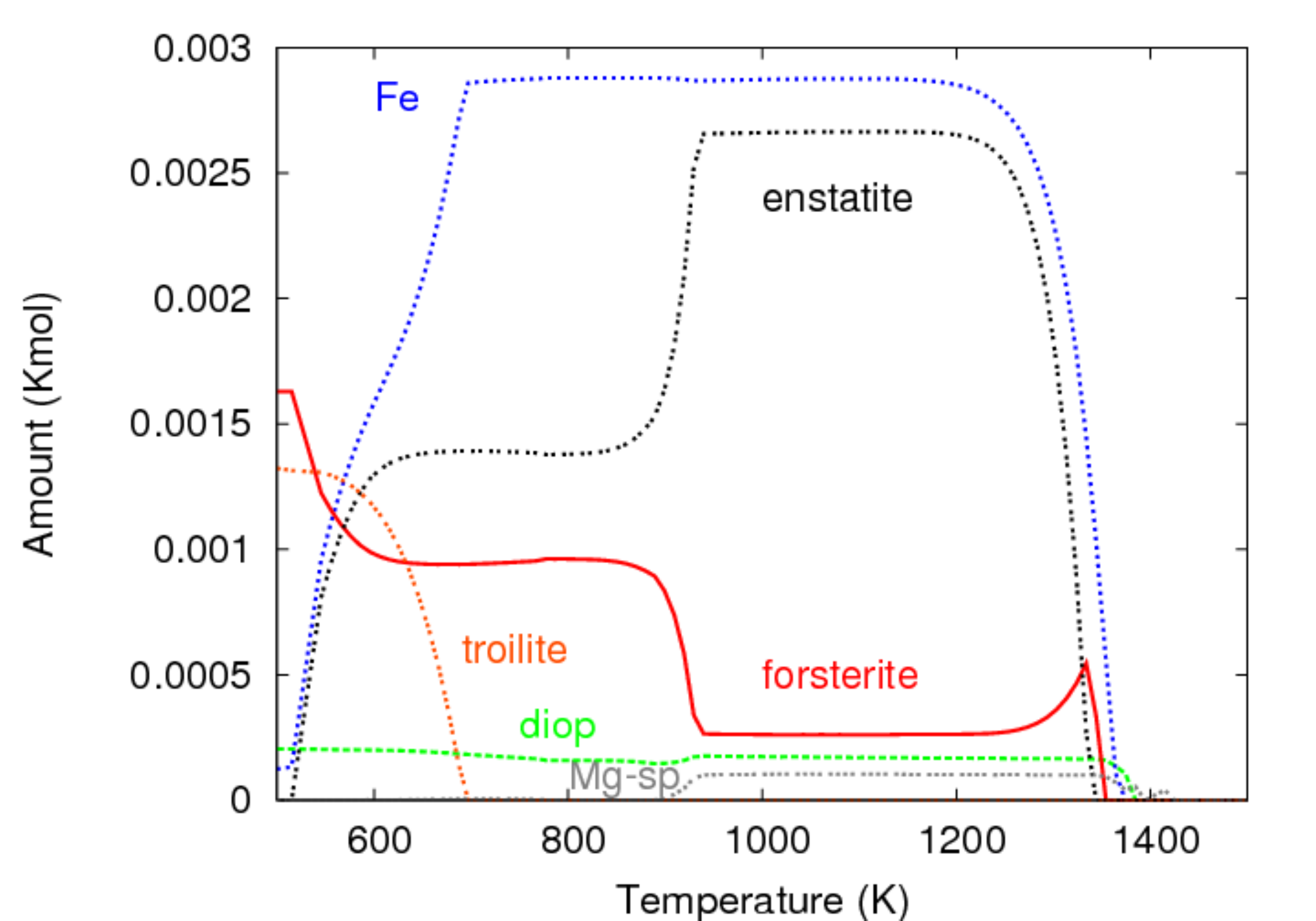}
\includegraphics[width=8.5cm]{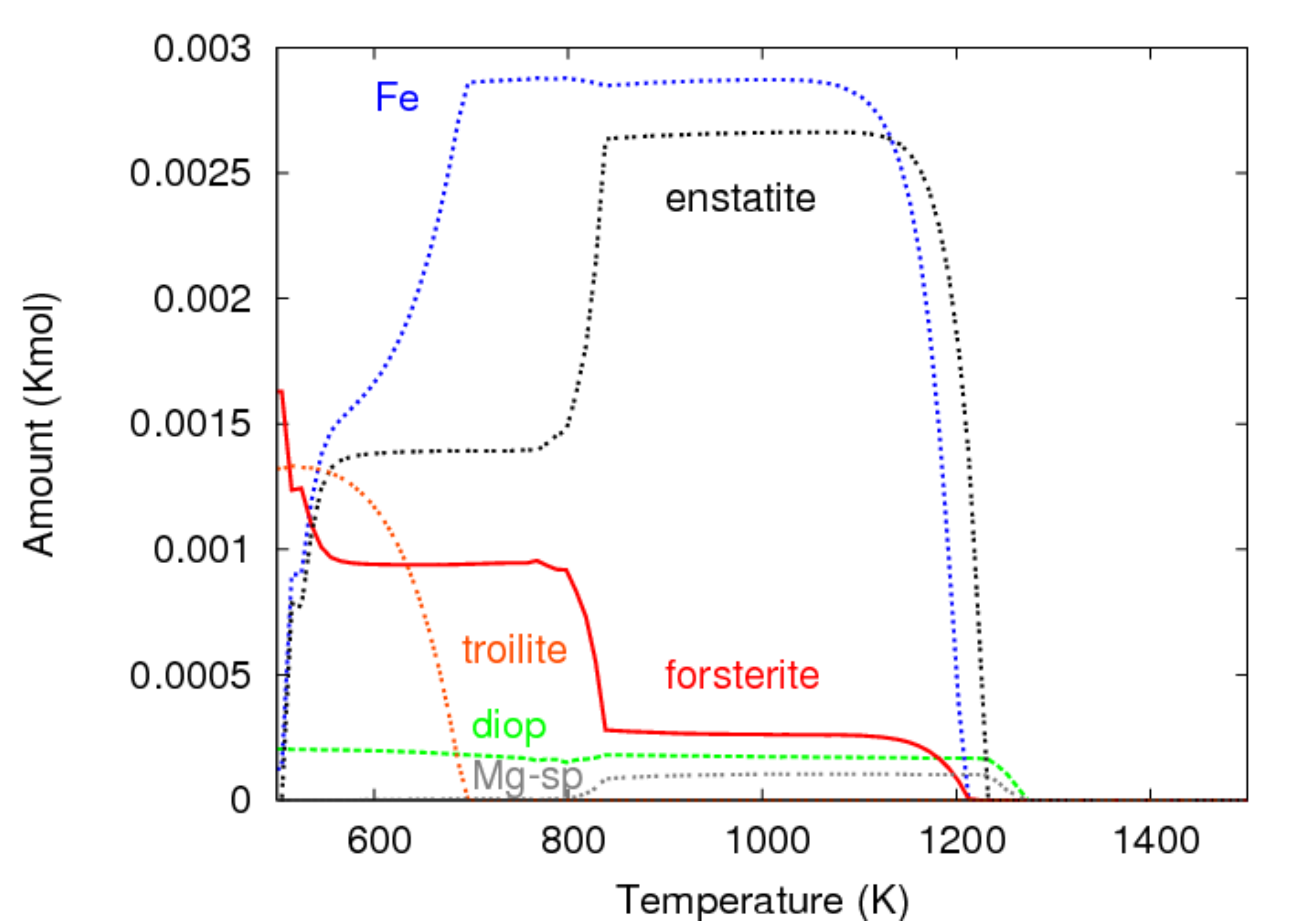}
\caption{Condensation sequence in the middle temperature region as function of temperature at (top) $P=10^{-3}$~bar, (middle) $P=10^{-4}$~bar and (bottom) $P=10^{-6}$~bar.}
\label{fig-middlet}
\end{figure}

The central zone of our temperature range ($600\le T(\rm{K}) \le 1400$) is dominated by Mg--silicates, like forsterite (\ce{Mg2SiO4}) and enstatite (\ce{MgSiO3}), plus metal  iron -- see Fig.~\ref{fig-middlet}. The formation of enstatite and forsterite is chemically related. The condensation sequence of these two compounds shows that the first peak of forsterite formed before enstatite is pressure sensitive: lowering the pressure decreases  the amount of forsterite formed in its first peak until enstatite become the most dominant of the two (Fig.~\ref{fig-middlet}). Mg--spinel (\ce{MgAl2O4}) and diopside (\ce{CaMgSi2O6}) also condense in this region.

\begin{figure}
\centering
\includegraphics[width=8.5cm]{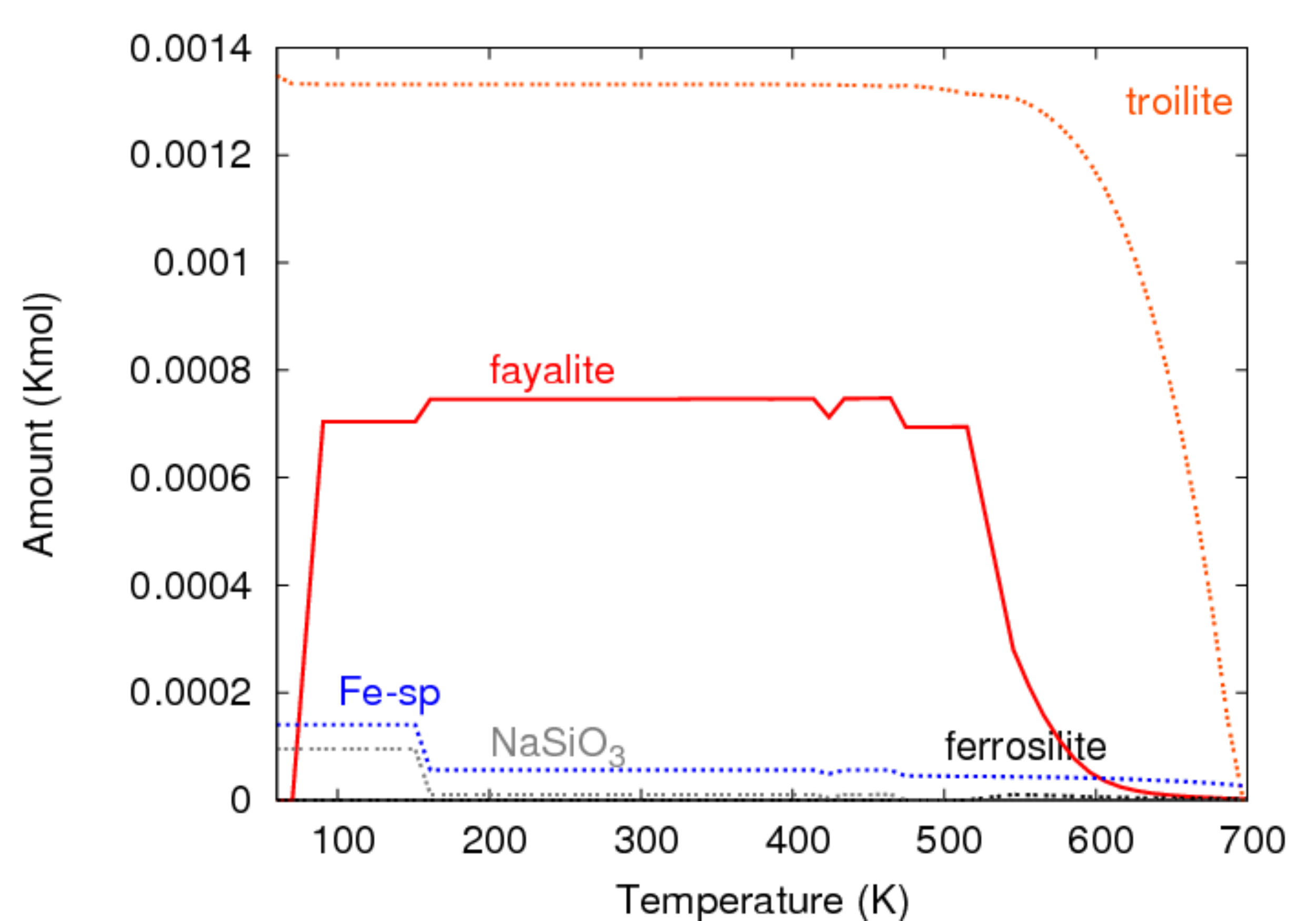}
\includegraphics[width=8.5cm]{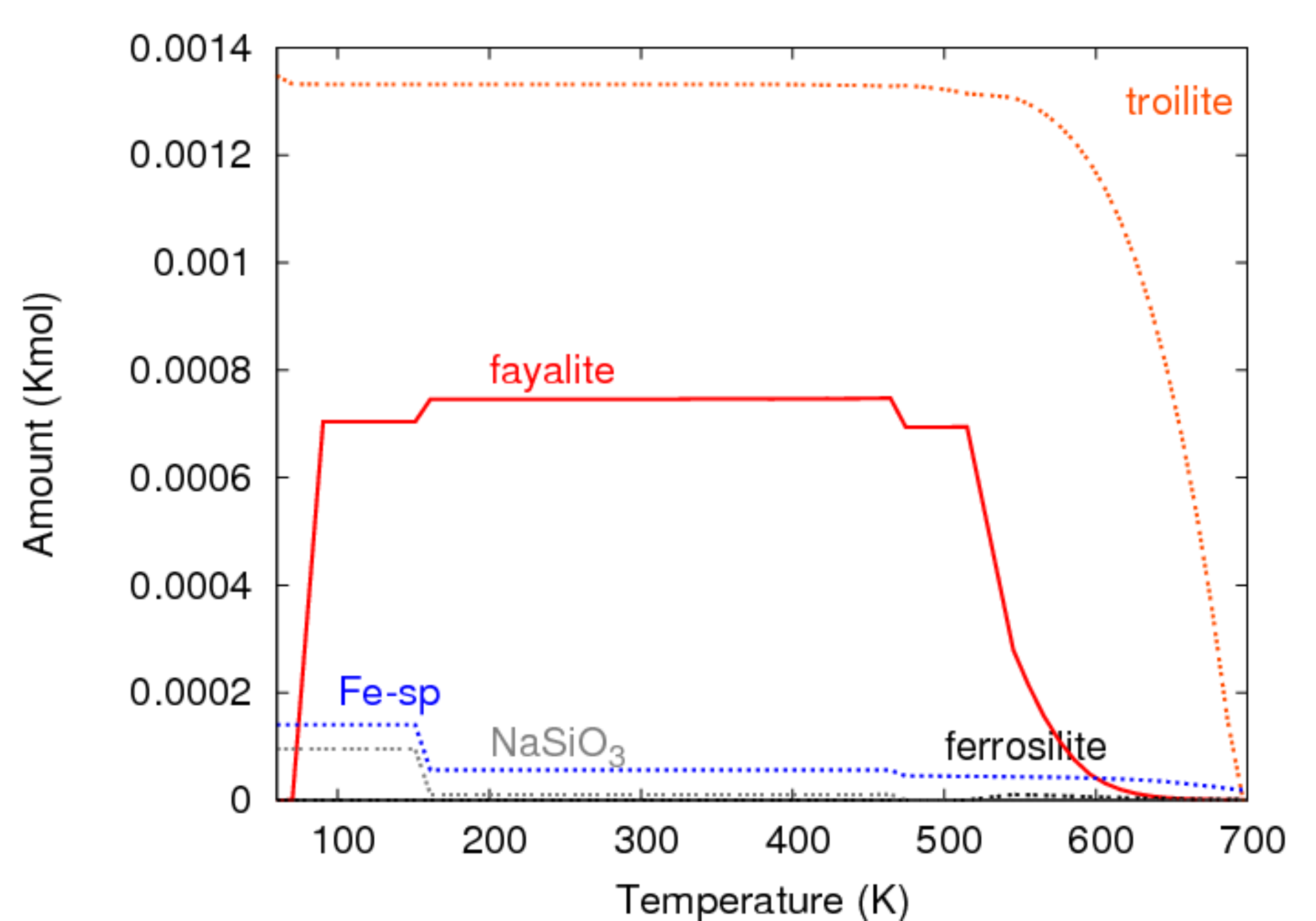}
\includegraphics[width=8.5cm]{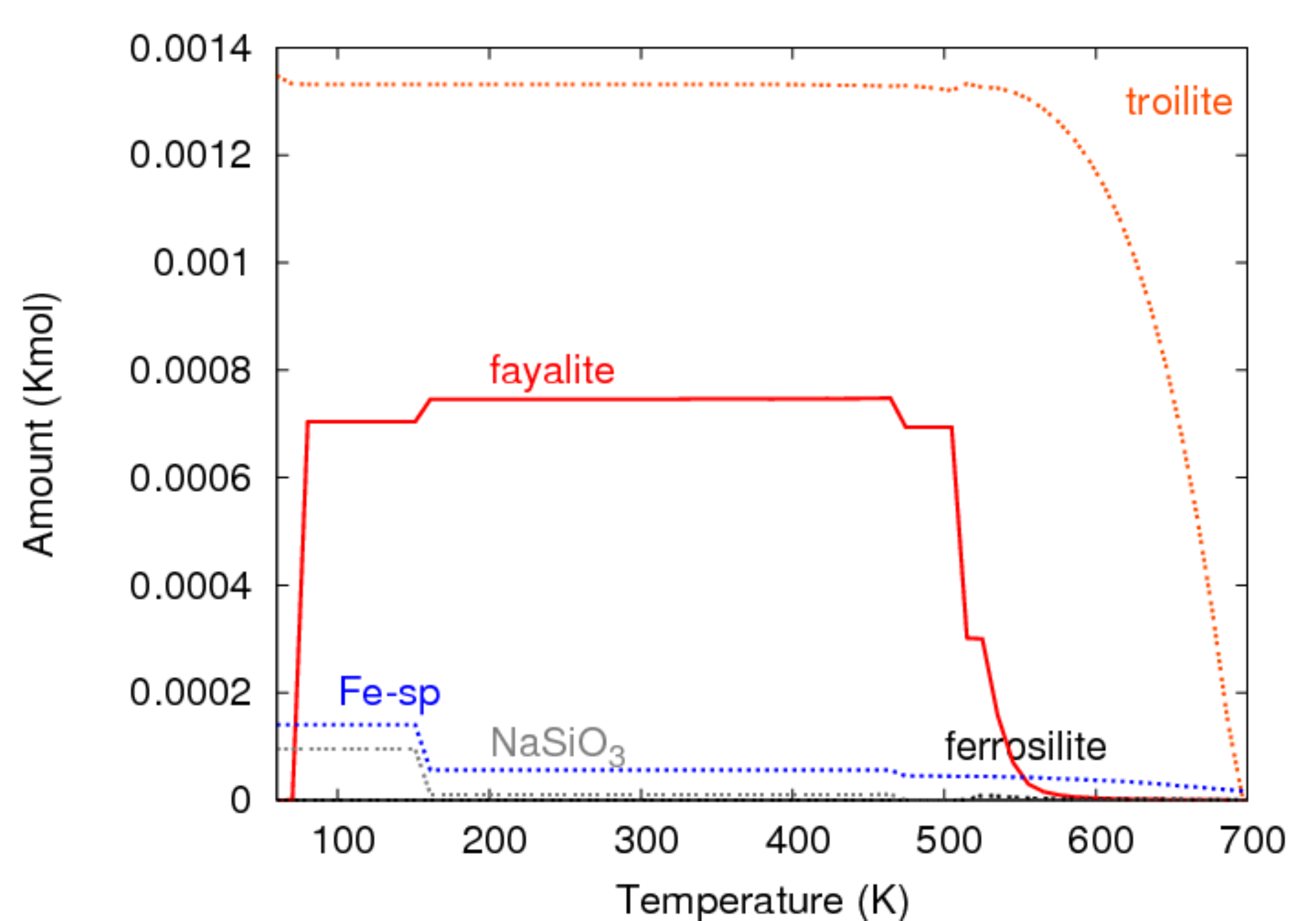}
\caption{Condensation sequence in the low temperature region as function of temperature at (top) $P=10^{-3}$~bar, (middle) $P=10^{-4}$~bar and (bottom) $P=10^{-6}$~bar.}
\label{fig-lowt}
\end{figure}

 The cooler region ($50\le T(\rm{K}) \le600$) shows the condensation of many Fe--compounds: troilite (\ce{FeS}), fayalite (\ce{Fe2SiO4}), ferrosilite (\ce{FeSiO3}) and Fe-spinel (\ce{FeAl2O4}) -- see Fig.~\ref{fig-lowt}. We also see that in this temperature zone the condensation sequence is not very sensitive to the pressure, unlike the high temperature regions. 
 From our definition of the temperature of appearance, which is when the amount of a compound reaches $10^{-7}$~kmol, it may seem that compounds have not yet condensed, when in fact they are in trace amounts. 
For example, with the olivine and orthopyroxene phases, the amount of fayalite (\ce{Fe2SiO4}) is non-zero in the temperature range from $T=1415$~K (the temperature of appearance of forsterite) to $T=717$~K (when fayalite finally reaches an abundance greater than $10^{-7}$ kmol).  Similarly, the amount of ferrosilite (\ce{FeSiO3}) ranges from about $1 \times 10^{-8}$ to $8 \times 10^{-8}$ ~kmol over the temperature range from $T=1384$~K (the condensation temperature of enstatite) and $T=757$~K  (when the amount of ferrosilite reaches $10^{-7}$~kmol, i.e. when it appears).

 The full list of compounds formed is given in Appendix~\ref{tab-tot}, together with their condensation temperatures for different values of pressure.

%


\subsection{Dust distribution}

\begin{figure}
\centering
\includegraphics[width=8.5cm]{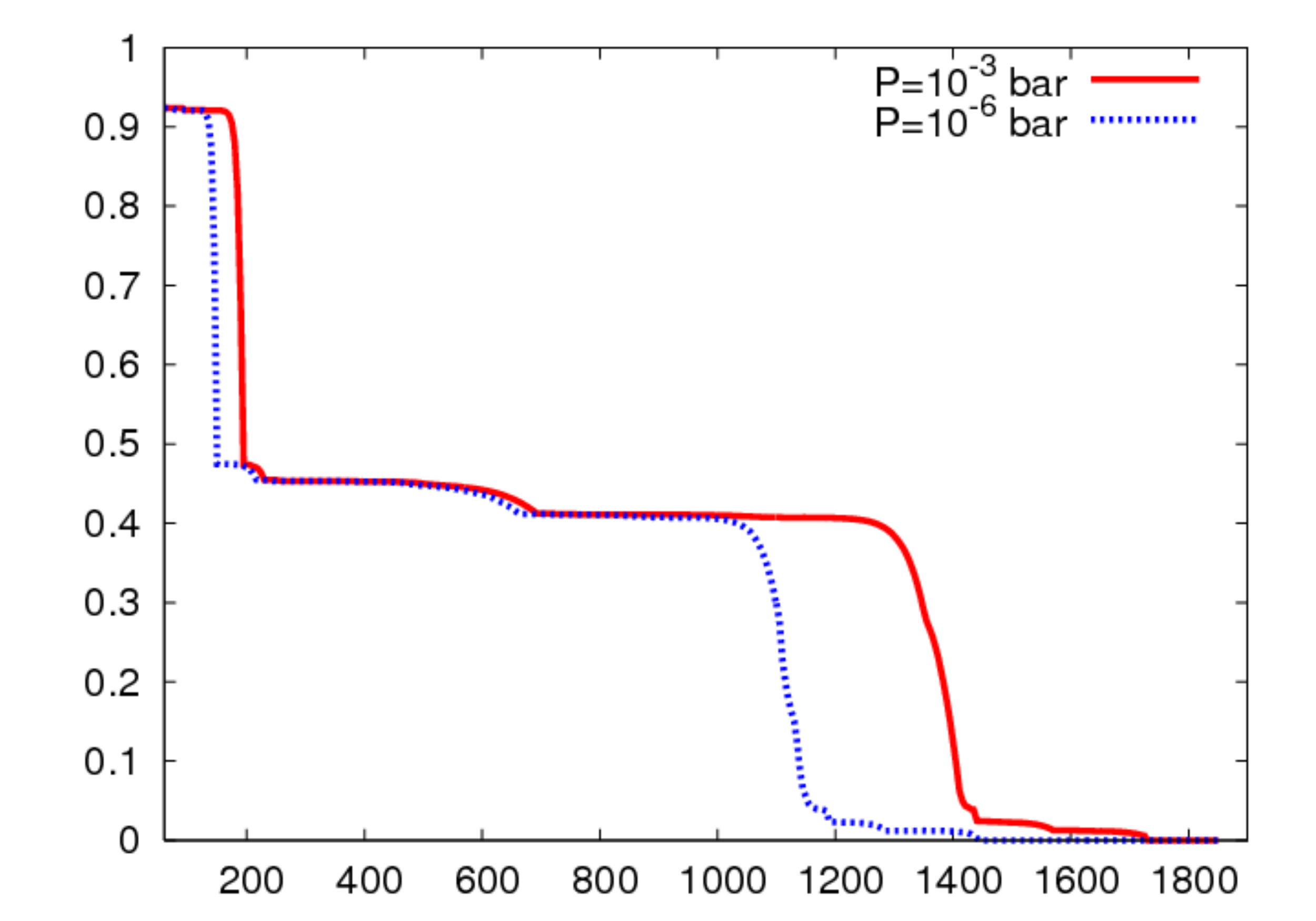}
\caption{Percentage by mass of dust in the system as a function of temperature for two pressures: $P=10^{-3}$ and 
$10^{-6}$~bar, solid and dotted line respectively.}
\label{fig-solidmass}
\end{figure}

Figure~\ref{fig-solidmass} shows the percentage by mass of dust in the system for  $P=10^{-3}$ and $10^{-6}$~bar. It shows that the dust mass distribution is a function of temperature, and thus a function of radius, assuming that the dust temperature decreases as radius increases.

Fig.~\ref{fig-solidmass} shows an abrupt change between 150 and 200~K, which is where  \ce{H2O} ice forms (see table~\ref{tab-tot}). In a protoplanetary disc, the inner edge of the region where the temperature falls below the condensation temperature of water is called the snow line \citep{2000ApJ...528..995S}.  Our simulation at $P=10^{-3}$ bar show that $T=191$~K is the point where the process of condensation of \ce{H2O}(g) starts, while $T=141$~K is the temperature at which the amount of \ce{H2O}(g) drops below $10^{-7}$~kmol.
At $P=10^{-6}$ bar, the temperature in which the process of vapor condensation starts is $T=151$~K  while $T=110$~K is the temperature at which the amount of \ce{H2O}(g) drops below $10^{-7}$~kmol.

\section{Regular versus Ideal solution models}
\label{rvsi}
\begin{figure}
\begin{center}
\includegraphics[width=8.5cm]{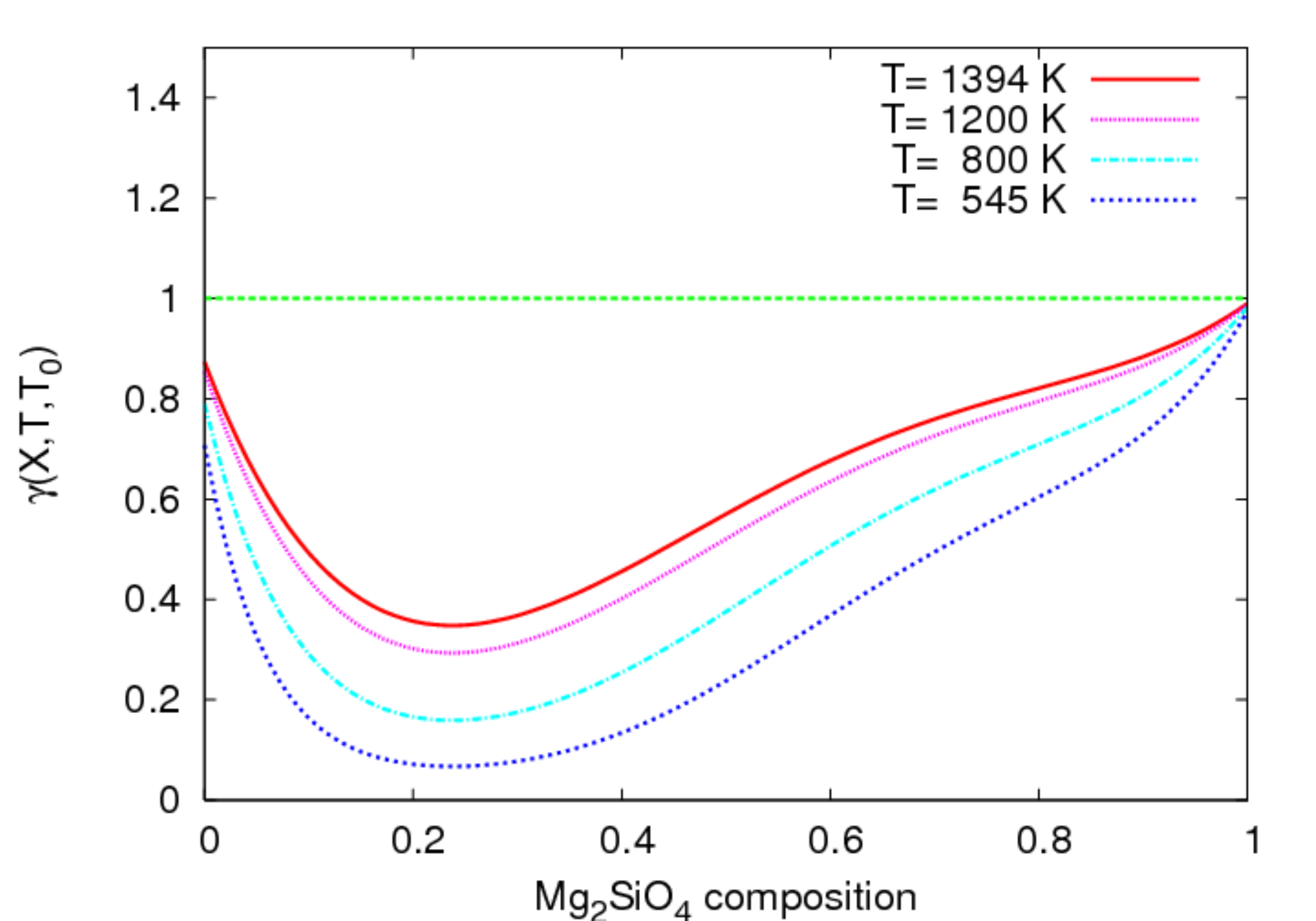}\\
\caption{Activity coefficient of forsterite (\ce{Mg2SiO4}) across the entire range of composition at different temperatures derived from equation~(\ref{acoefforst1}). The $\gamma$=1 line represents the ideal.}
\label{gamforst}
\end{center}
\end{figure}

\begin{figure}
\begin{center}
\includegraphics[width=8.5cm]{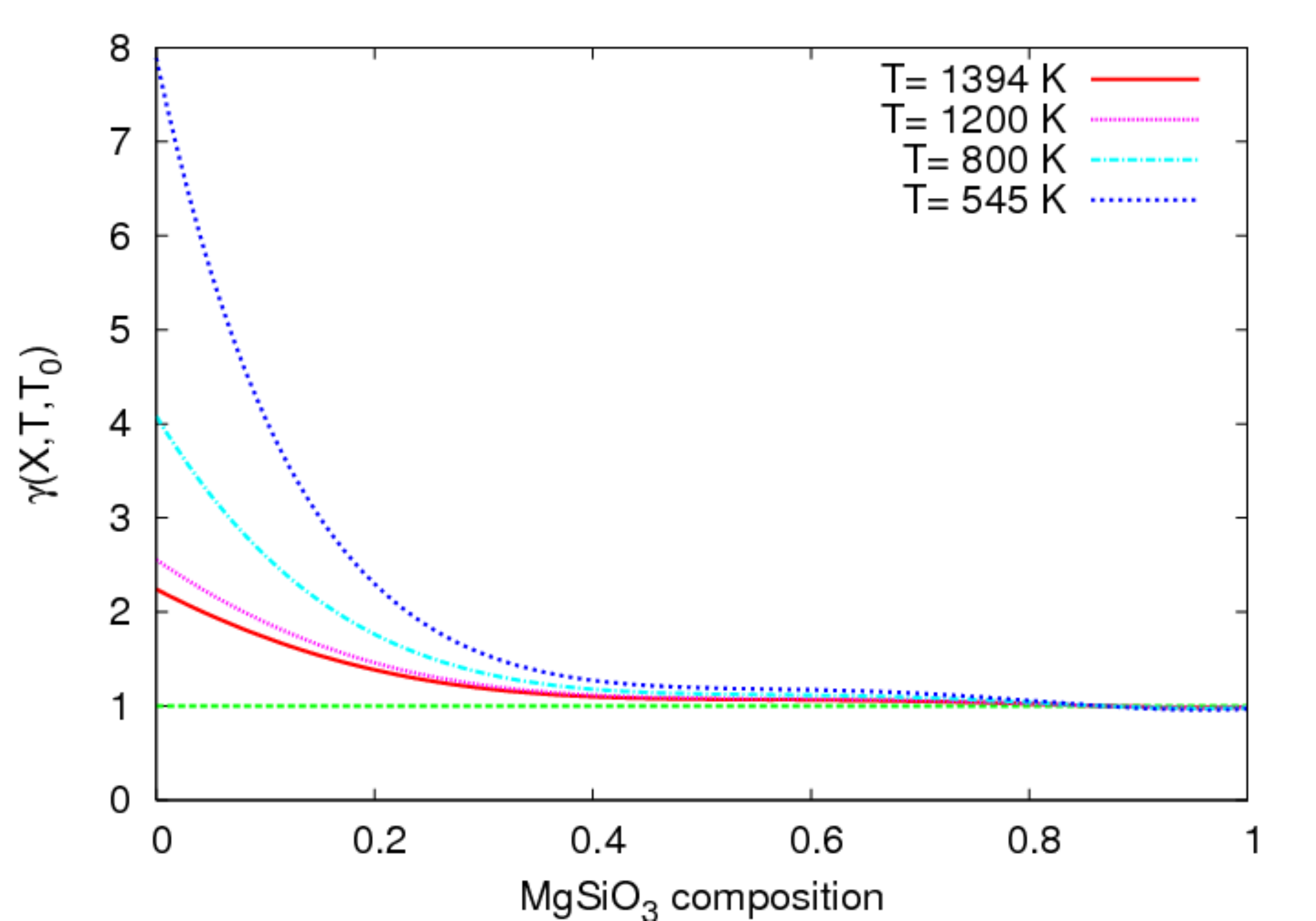}\\
\caption{Activity coefficient of enstatite (\ce{MgSiO3}) across the entire range of composition at different temperatures derived from equation~(\ref{genstatite1}). The $\gamma$=1 line represents the ideal.}
\label{gamenstat}
\end{center}
\end{figure}

In order to understand the differences between the regular and the ideal solution models, we need to return to the definition of activity, $a$, and activity coefficient, $\gamma$.
Activity--composition diagrams derived from experimental data provide information about the departure from the ideal behaviour of compounds according to their composition at the reference temperature (see Fig.~\ref{fig-ActCompRelations}).
Activity coefficient--composition diagrams, derived from equations in section~\ref{ssb}, also provide information about the departure from the ideal with varying temperature across the entire composition range of a selected compound.
Figures~\ref{gamforst} and~\ref{gamenstat} show activity coefficient polynomials for forsterite and enstatite at different values of temperature. 
The $\gamma$ polynomials curves, from equations~(\ref{acoefforst1})  and~(\ref{genstatite1}), move closer the ideal ($\gamma =1$) with increasing temperature. This behaviour is seen for every compound in the system.
Furthermore, every $\gamma$ curve for different temperatures return values close to one when $X\rightarrow 1$ for all temperatures.

\begin{figure*}
\centering 
{\includegraphics[width=.6\columnwidth]{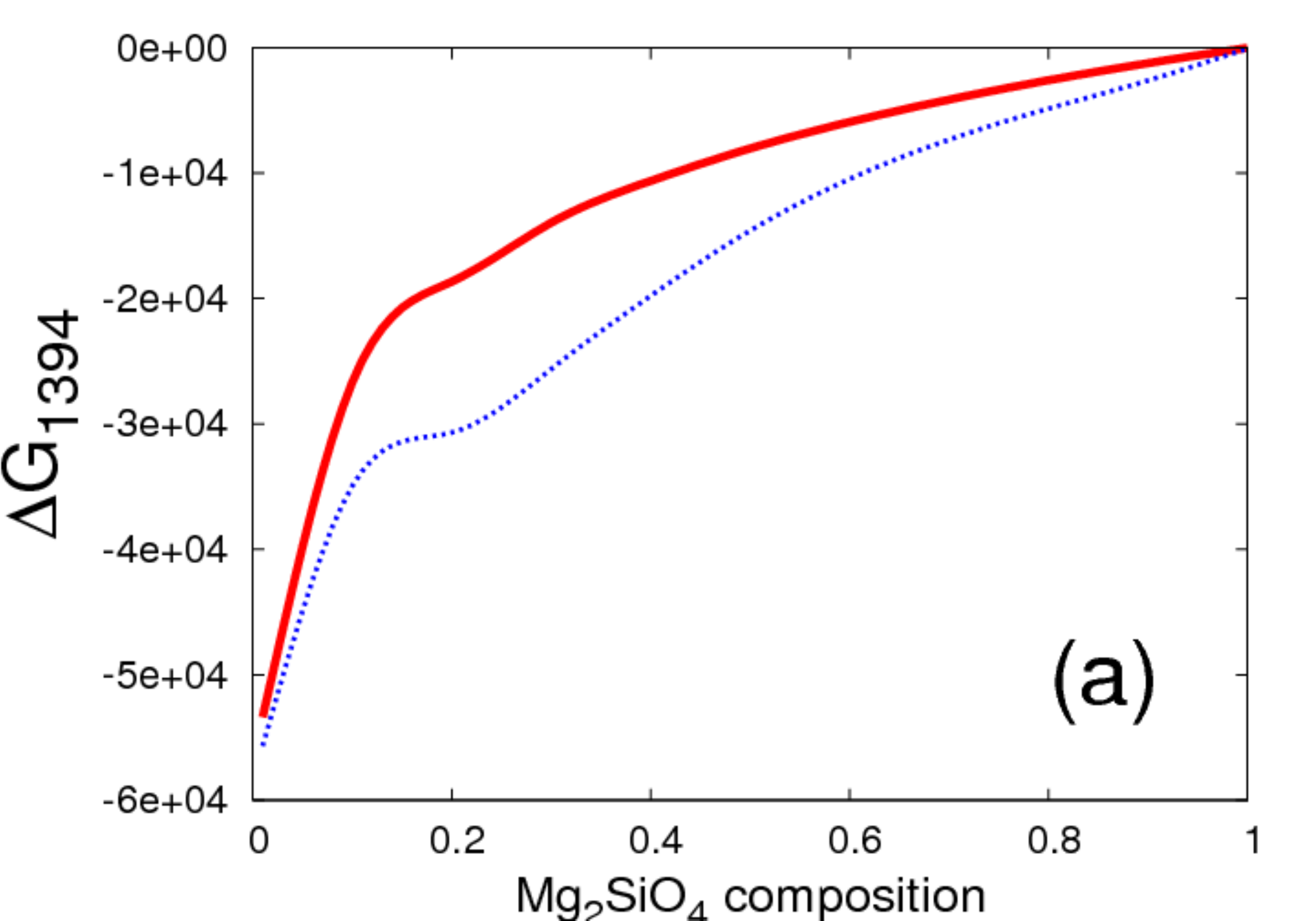}} 
{\includegraphics[width=.6\columnwidth]{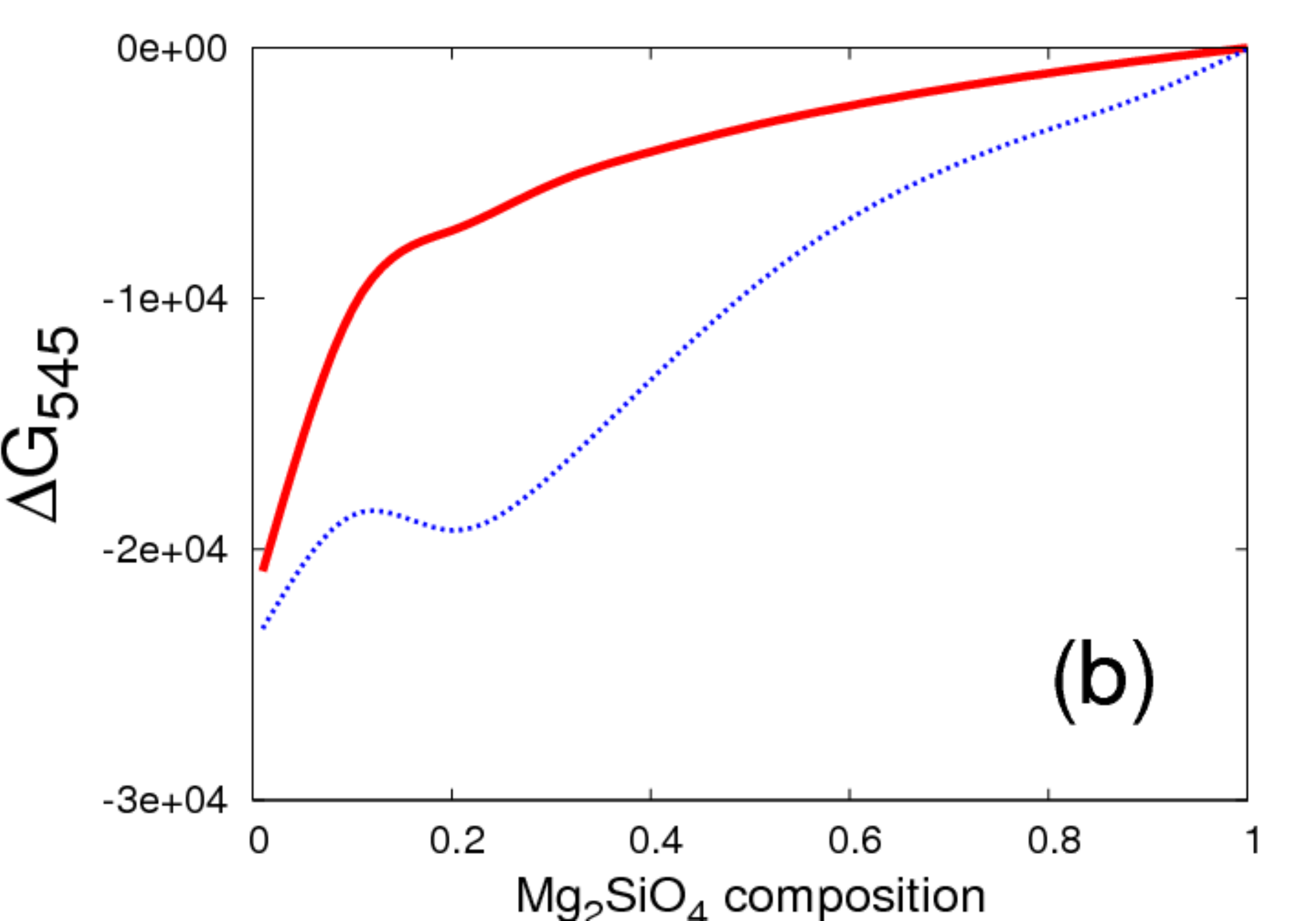}} \\
{\includegraphics[width=.6\columnwidth]{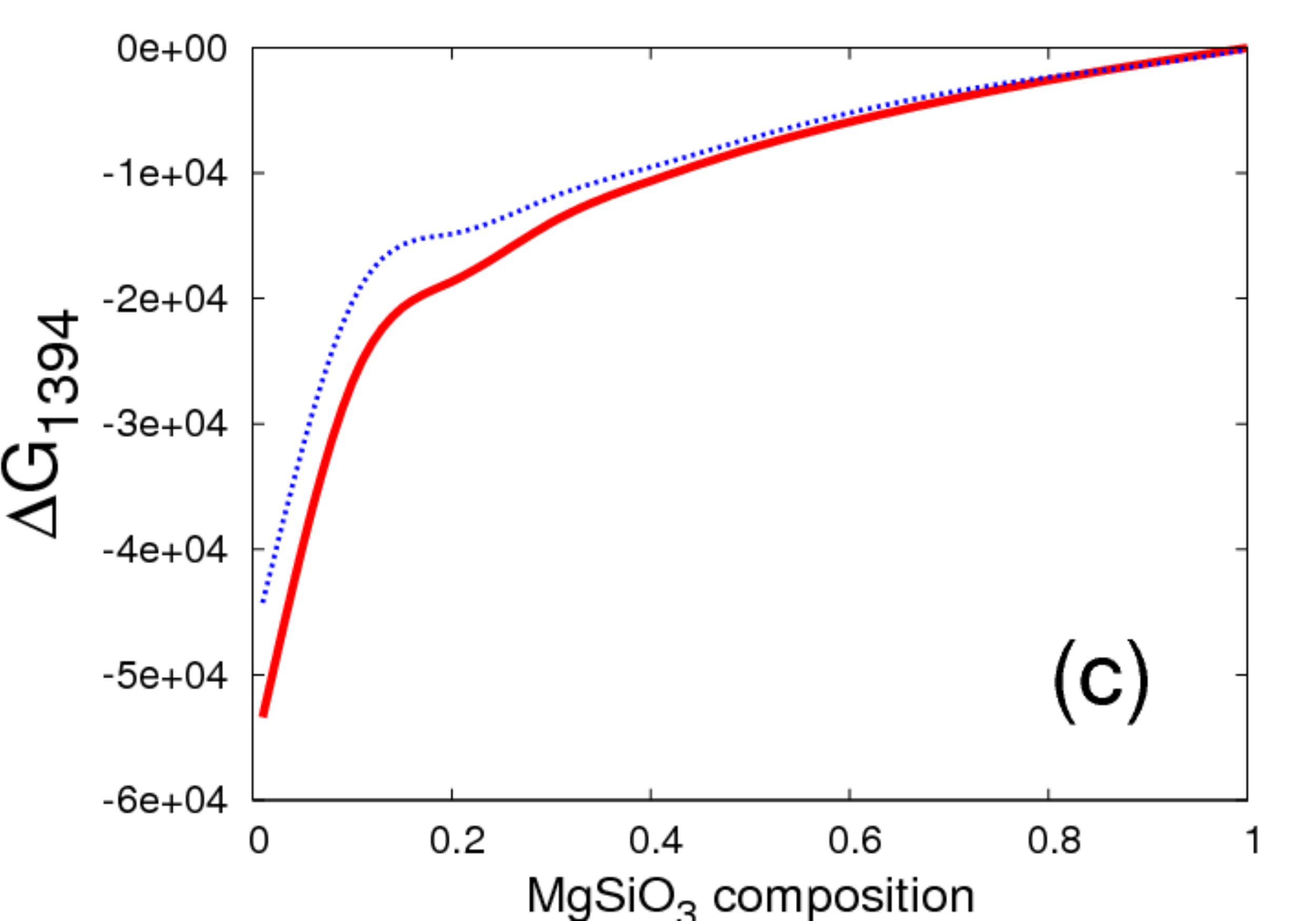}} 
{\includegraphics[width=.6\columnwidth]{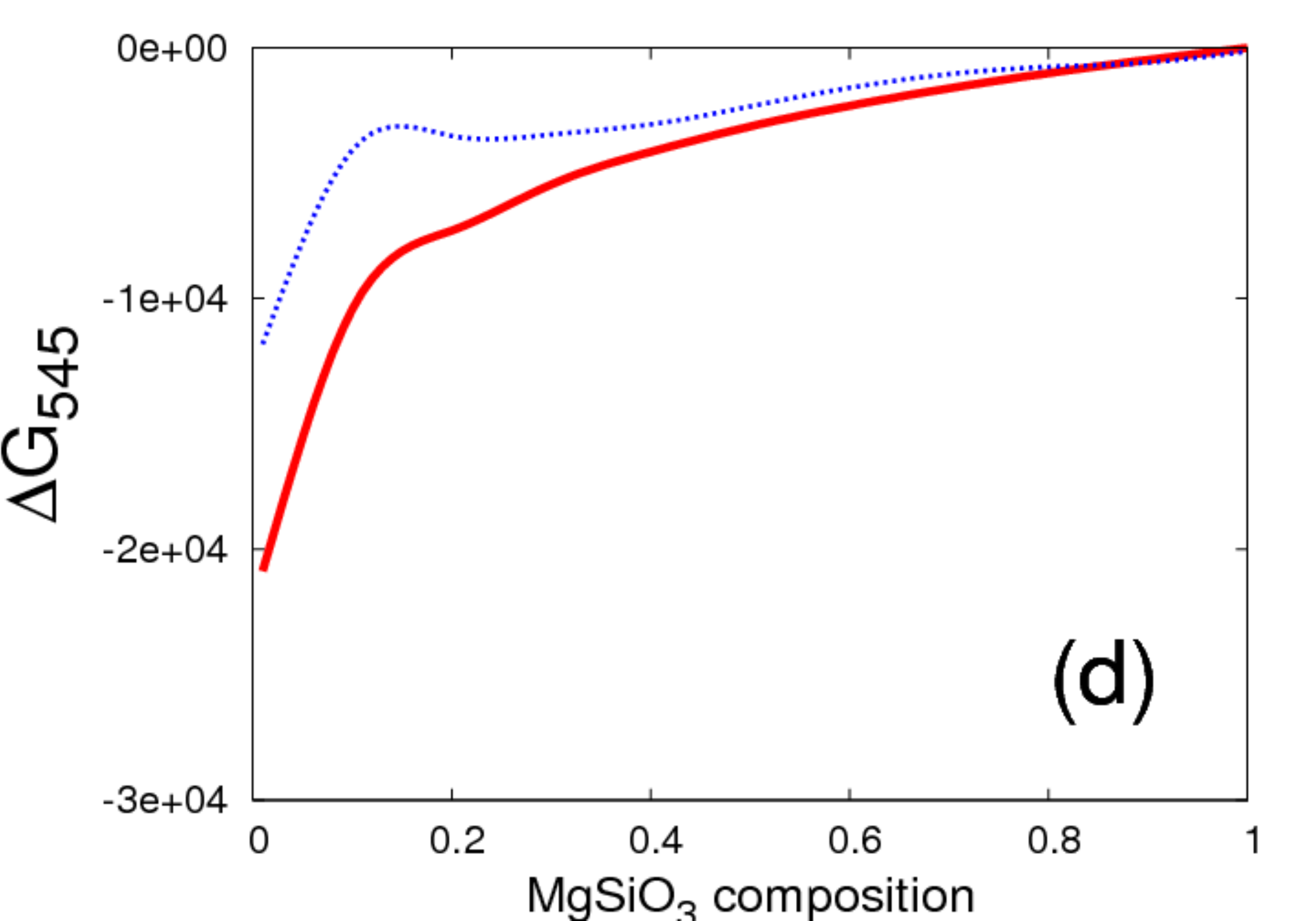}}
\caption{ $\Delta G_{i}^{id}$ and $\Delta G_{i}^{ex}+\Delta G_{i}^{id}$ from equation~(\ref{biboba}) in the whole range of composition for forsterite, (a) T=1394 K, (b) T=545 K, and enstatite, (c) T=1394 K, (d) T=545 K. Red thick line is $\Delta G_{i}^{id}$, blue thin line is $\Delta G_{i}^{ex}+\Delta G_{i}^{id}$. } 
\label{fig:subfig} 
\end{figure*}

Equation~(\ref{eqn-gibbs})  for the Gibbs free energy of compounds at a given temperature could also be written as
\begin{equation}
\label{biboba}
\Delta G_{i}=G_i-G_{i}^{0}=RT\ln{\gamma}_{i}+RT\ln{X_{i}}=\Delta G_{i}^{ex}+\Delta G_{i}^{id} ,
\end{equation}
where $RT\ln{X_{i}}$ is the partial molal Gibbs free energy of mixing for an ideal solution, and $RT\ln{{\gamma}_{i}}$ is the departure from ideal behaviour of component {\it i} in the mixture. We define the former as $\Delta G_{i}^{id}$ and the latter $\Delta G_{i}^{ex}$, where the superscripts {\it id} means ideal and  {\it ex} is excess. 
We can study the non-ideal contribution in the computation of $\Delta G_{i}$ by plotting $\Delta G_{i}^{ex}+\Delta G_{i}^{id}$ from equation (\ref{biboba}) compared to the ideal $\Delta G_{i}^{id}$ at fixed values of temperature. This is shown in Fig.~\ref{fig:subfig} for forsterite and enstatite, and we find that $\Delta G_{i}^{ex}$ could be high enough to change the value of the $\Delta G_{i}$.

\begin{figure*}
\centering 
{\includegraphics[width=.65\columnwidth]{figure6a.pdf}} 
{\includegraphics[width=.65\columnwidth]{figure5a.pdf}}
{\includegraphics[width=.65\columnwidth]{figure4a.pdf}} \\
{\includegraphics[width=.65\columnwidth]{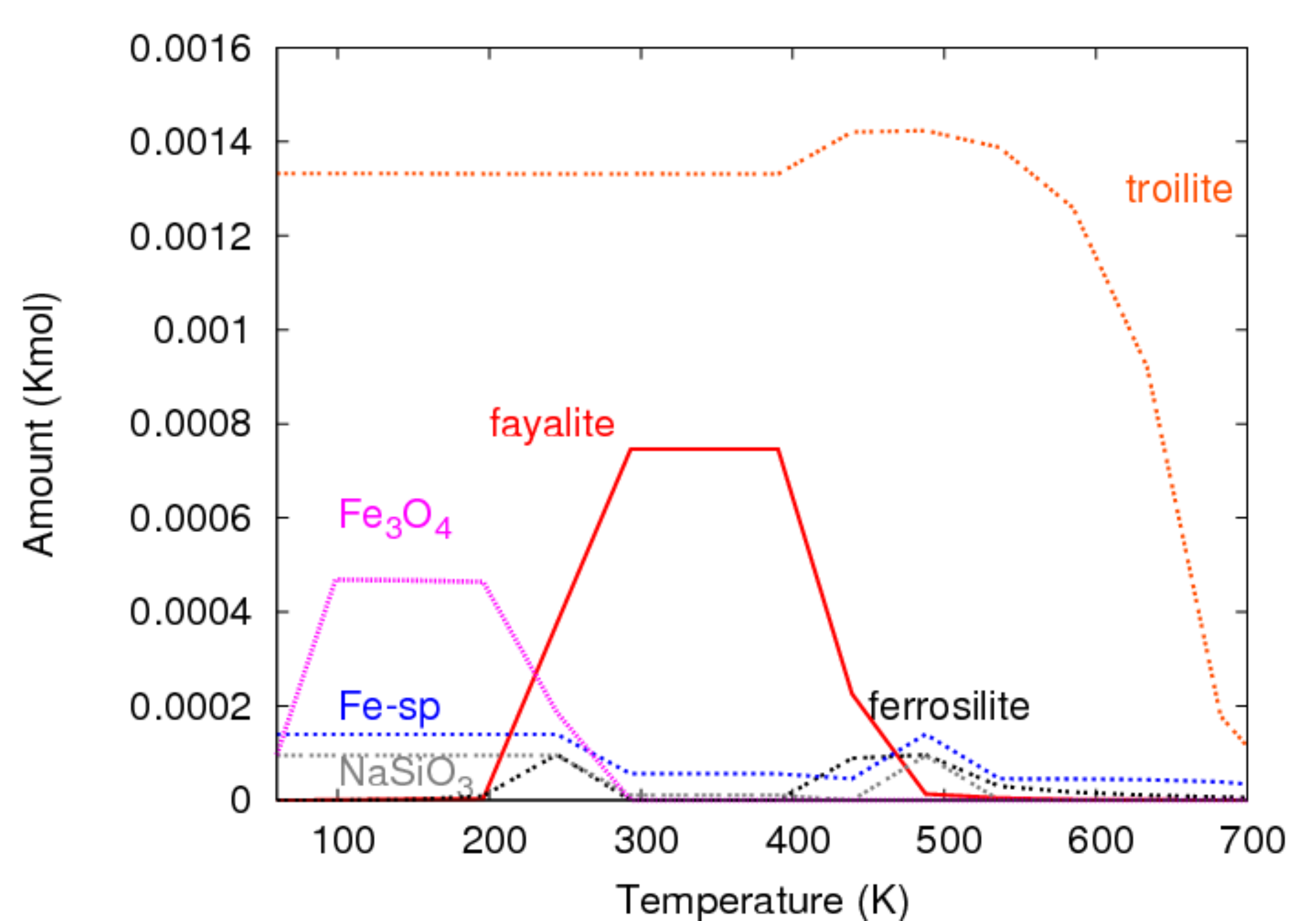}}
{\includegraphics[width=.65\columnwidth]{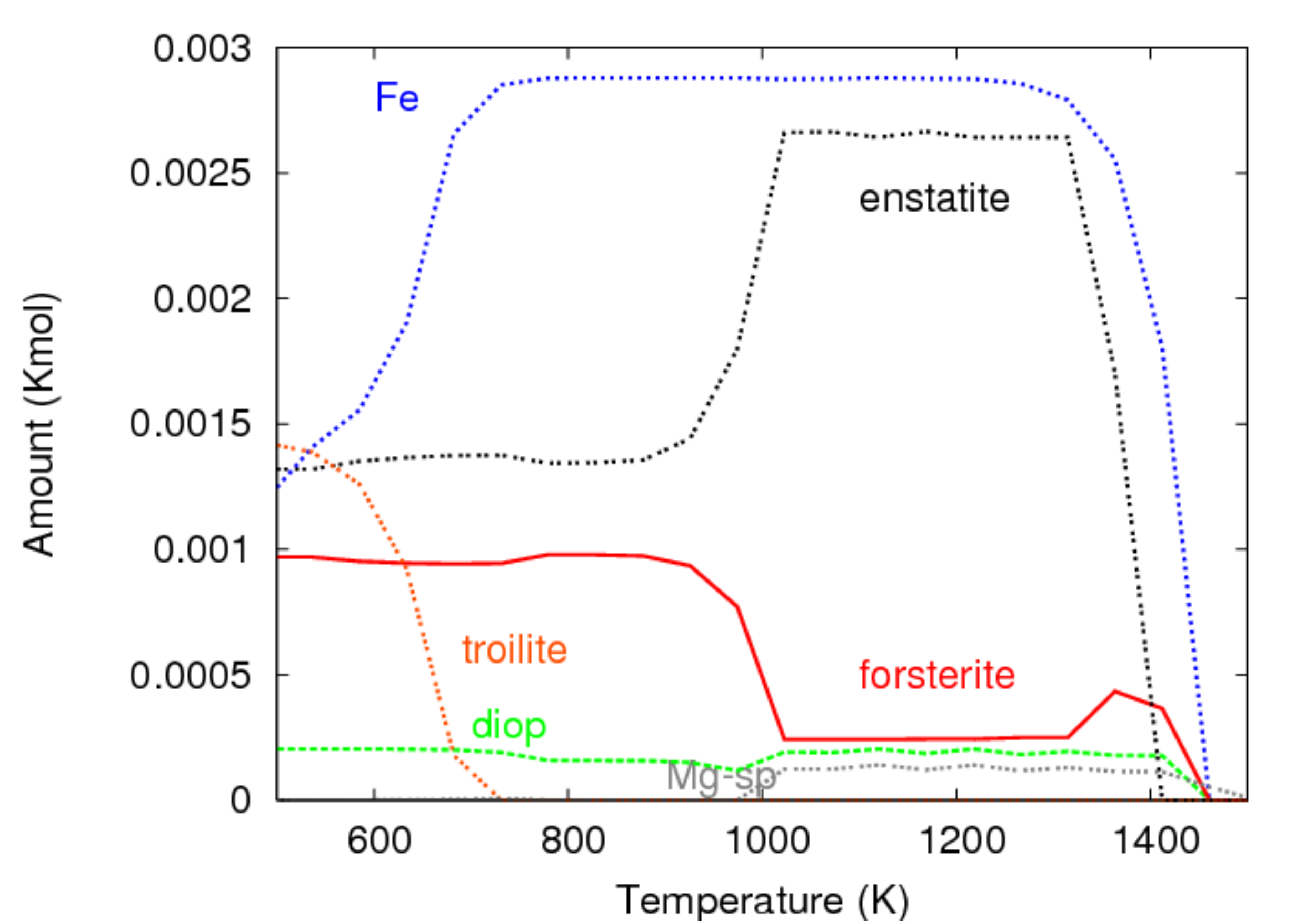}} 
{\includegraphics[width=.65\columnwidth]{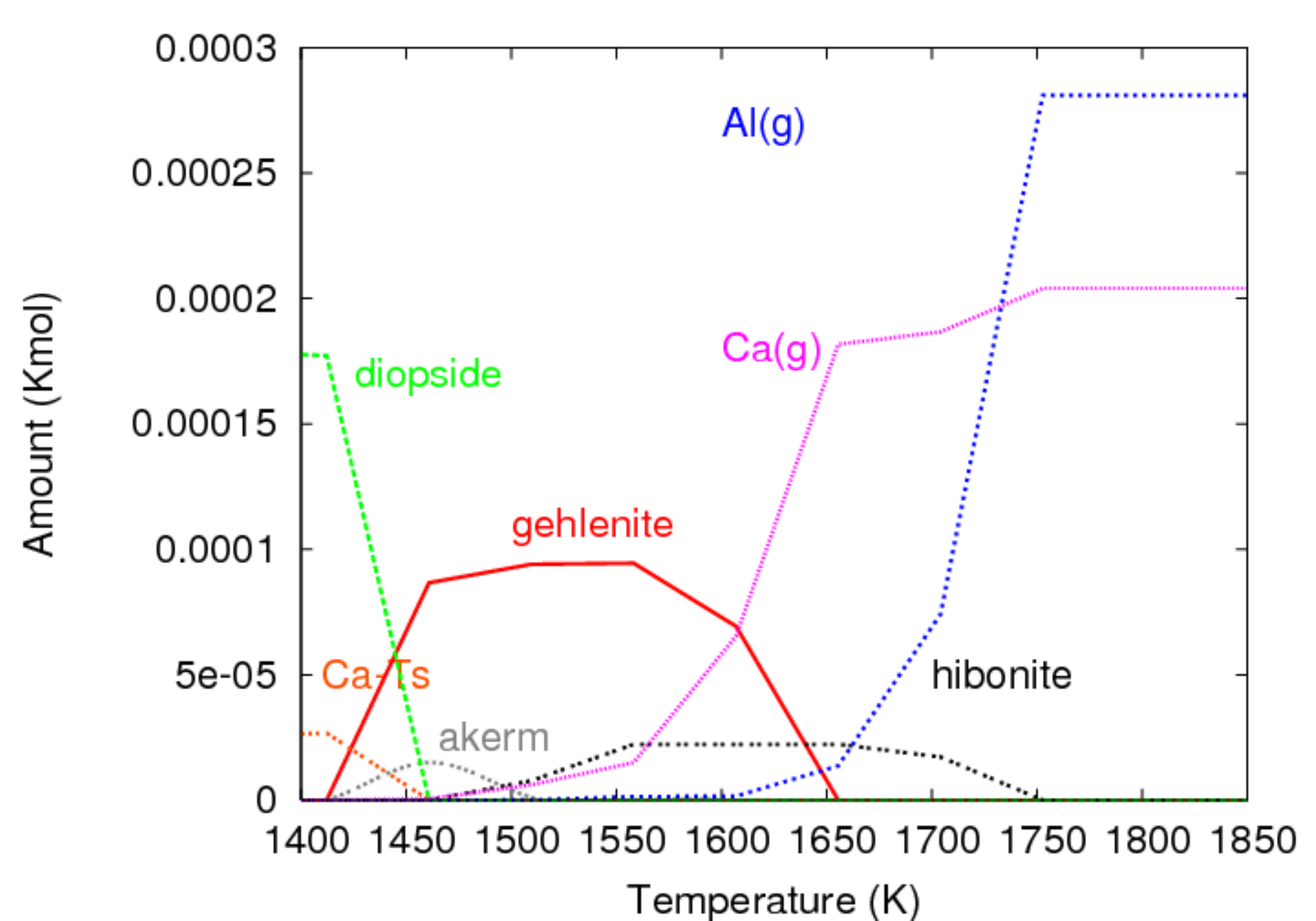}}
\caption{Condensation sequences in the low temperature region (left), the middle temperature region (middle) and the high temperature region (right) for the regular (top) and the ideal (bottom) solution models. Pressure is $10^{-3}$ bar.} 
\label{idealregular} 
\end{figure*}

We compare the condensation sequences calculated using the ideal solution and the regular solution to show how the activity coefficients can affect the results. Fig.~\ref{idealregular} shows the condensation sequences with both the ideal solution and regular solution models at a pressure of $10^{-3}$ bar.  Similar plots (not shown) were obtained for $P=10^{-6}$. In tables~\ref{3regularideal} and~\ref{tab-tot} we also report the temperature of appearance and disappearance for the ideal and regular solution models.

We see some differences in the values of the temperatures of appearance and disappearance.
The first peak of forsterite at 1394 K is more abundant in the regular solution than for the ideal case at $P=10^{-3}$ bar. In the low temperature region, the regular solution model returns a larger range of temperatures in which fayalite (\ce{Fe2SiO4}) is stable. In the ideal solution, this range is smaller and there is more Fe available for the the formation of FeS, Fe metal, ferrosilite (\ce{FeSiO3}) and Fe-spinel for $T\ge 400$~K, and at very low temperature magnetite (\ce{Fe3O4}) replaces fayalite (this also happen at the pressure of $10^{-6}$ bar) -- see Fig.~\ref{idealregular}. Furthermore, there is no evidence of formation of magnetite (\ce{Fe3O4}) in the regular solution at the temperature and pressure range used in these calculations (see Fig.~\ref{fig-lowt}).

 The ideal solution model also shifts the turning point of enstatite and forsterite to lower temperatures (see Fig.~\ref{idealregular}), which also explains the lack of fayalite (see section~\ref{disenfor}) in these temperature regions.
 
 Note that the errors in temperatures are 5~K due to the step size used in the HSC calculations.

\section{Discussion}
\label{sec-results}

In this section we compare the results our regular solution model with previous thermodynamic models and observational data.

\subsection{Comparison with models}

In order to check the results of our regular solution model, we compare our simulation results to those of earlier studies by \citet{Pasek2005},  \citet{Yoneda1995} and \citet{Gail1998}. The main difference between our model and theirs is the regular solution model we described in section~\ref{sec-solidsoln}.

\citet{Pasek2005} carried out their calculation using the HSC software package, using 150 gas species and 100 solid species from the 20 most abundant solar elements, but they assumed their solid solutions as ideal. Their calculations assumed a temperature range of $400\le T(\rm{K}) \le2000$.

\citet{Yoneda1995} created a larger system with solid, liquid, and gas states. They used 120 gas species, 5 liquid species, and 110 solid species from the 20 most abundant solar elements. They studied the behaviour of the liquid solution and each solid phase using data from the literature to define the list of possible compounds and the activity--composition model from \citet{Berman1983} for non-ideal liquid solution. They calculated equilibrium for temperatures from 900~K to 1800~K, and with a larger range of pressures (from $10^{-6}$ to $\sim 1$~bar). Their results show that the liquid phase appears only at high pressures ($P \geq 0.1$~bar), confirming the solid--gas assumption in the pressure range used in our model.

\citet{Gail1998} studied the chemical equilibrium between the gas phase and the solid species, and determined relationships between the partial pressure of the gas involved in the creation of the solid species and the equilibrium constant of the reaction. 

Our calculations are in good general agreement with \citet{Pasek2005}, \citet{Yoneda1995} and  \citet{Gail1998} in terms of the condensation sequence and the temperature of appearance and disappearance. However, there are some important differences. Table~\ref{tab-resultT} shows our condensation temperatures  compared with these three models.
Small differences between the temperature of appearance and disappearance can be seen, except for the plagioclase phase for which large discrepancies are seen, and the order of appearance of forsterite and enstatite in the low pressure regime. However, our regular solution model and our ideal solution model (see tables~\ref{3regularideal} and~\ref{tab-tot}) are consistent.  
We will discuss forsterite and enstatite and the discrepancy in the plagioclase in more details in the next section.

Overall,  the condensation sequence is the same for the 12 most abundant species in our model and those in \citet{Pasek2005} and  \citet{Yoneda1995} except for the corundum (\ce{Al2O3}). Indeed, the previous models show that the corundum is created just before the hibonite species, whereas in our model, hibonite is the most refractory species of the system. This difference between the models comes from the closeness of the melting point between Ca and Al. In the model of \citet{Yoneda1995}, Ca begins to condense in hibonite at a lower temperature than the Al melting point, and thus \ce{Al2O3} can be created between these two condensation temperatures, whereas our system does not show these difference of melting point temperatures until the pressure drops to $P=10^{-6}$ bar. \citet{Gail1998} reports corundum as the first solid formed, followed by melilite. The absence of hibonite (\ce{CaAl12O19}) in his model does not allow us to compare the difference of our corundum and hibonite condensation sequence, however the condensation temperature of corundum at $P=10^{-6}$ bar show similar values.

\subsubsection{Enstatite and forsterite}
\label{disenfor}

The behaviour of forsterite and enstatite seen by \citet{Yoneda1995}, \citet{Gail1998}, \citet{2004A&A...413..571G}, \citet{Pasek2005} and this work all show differences from one another.  

\citet{Yoneda1995} report in their Fig.~2 an initial peak of forsterite at high temperatures. The abundance of forsterite then drops and enstatite replaces it as the most stable compound at $T=1366$~K. There is then a plateau of constant enstatite/forsterite ratio until $T=1188$~K where the amount of forsterite increase while enstatite remains constant.  \citet{Yoneda1995}  have $T= 900$~K as the lower limit of their calculations and so it is not possible to compare results in the lower temperature regions.

\citet{Gail1998} shows the condensation sequence of enstatite and forsterite in terms of radial variation of the fraction of silicon condensed into these compound.  The first peak of forsterite is followed at $T=1319$~K by an increase of enstatite that becomes the most stable compound -- see their Fig.~21. Then there is a plateau of constant enstatite/forsterite ratio and there is no further change in either enstatite or forsterite (out to $R=2$~AU and $T\le 300$~K).

In \citet{2004A&A...413..571G} an initial peak of forsterite precedes the formation of enstatite. Then enstatite overtakes forsterite at 0.55 AU ($T\sim1350$~K) -- see their Figs.~10 and 11 -- and then rather than a plateau of constant enstatite/forsterite ratio, both the enstatite and forsterite abundances drop with decreasing temperature. 

\citet{Pasek2005} show the equilibrium condensation sequence in terms of atom fraction of cationic elements. Differences from previous work can been seen in their Fig.~2, in which the amount of enstatite continues to increase once fayalite forms at  $T\sim450$~K.
Our condensation sequences shows a complex behaviour, demonstrating that forsterite and enstatite follow a condensation sequence of forsterite--enstatite--forsterite, going from the warmer region to the cooler outer region of the disc (see Fig.~\ref{fig-middlet}).  Forsterite first condenses at $T=1415$~K and is then  replaced by enstatite (with has a temperature of appearance of $T=1384$~K ). As the temperature decreases ($1000 \le T(\rm{K}) \le1300$), a plateau of constant  enstatite/forsterite ratio can be seen. At $T\sim1000$~K there is a sudden drop of enstatite and an increase in the abundance of forsterite. A second plateau of constant enstatite/forsterite ratio is seen through $600 \le T(\rm{K}) \le 900$. At $T\sim 600$~K forsterite increases its abundance, overtaking enstatite at $T\sim550$~K,  while the enstatite abundance continues to drop with decreasing temperature. The temperature values reported here refer to a pressure of $P$=$10^{-3}$ bars. 

The differences in the condensation sequence reported by these four references and our work could be due to several factors, such as different initial compositions, different lists of possible compounds considered in the final system, different solution model and different thermodynamical databases used in the calculations. To resolve this issue, which is of great importance to astrophysics and in understanding observational evidence from both protoplanetary disks and meteorite data, a controlled laboratory experiment  using a complex system over a range of pressures and temperatures is required.

While we cannot explain the differences in the forsterite and enstatite behaviour seen in these four references, we can explain the forsterite and enstatite  behaviour seen in our results.  Here we will discuss (i) the formation of enstatite after the first peak of forsterite, (ii) the drop of enstatite and the increase of forsterite at $T\sim 1000$~K, and (iii) the turning point between enstatite and forsterite around $T\sim 600$~K.

\citet{2006A&A...456..535S} showed that the reaction from forsterite to enstatite is possible if quartz (\ce{SiO2}) is present  according to the following reaction given by \citet{1986Icar...66..211R}:
\begin{equation}
\ce{Mg2SiO4}+\ce{SiO2} \rightarrow \ce{2MgSiO3}. 
\end{equation}
In our simulation, there is no \ce{SiO2} present as an isolated oxide. This is because simulations made using the Gibbs free energy minimization do not provide information about the rate of the reaction and the steps between the initial and the final conditions. In our equilibrium conditions, silica is always involved in the formation of more complex compounds. The amount of \ce{SiO}(g) in our simulation drops by four orders of magnitude in the range of temperature in which the two silicates form ($1152 \le T(\rm{K}) \le1450$). 

The formation of enstatite from forsterite at higher temperature could be also chemically explained with reactions involving \ce{Mg}(g) with the condensation of \ce{SiO}(g):
\begin{equation}
\ce{Mg2SiO4}+\ce{O(g)}+\ce{SiO(g)}\rightarrow2\ce{MgSiO3} ,
\end{equation}
but this reaction is related to the availability of oxygen and to the kinetic processes that drive its rate \citep[see][and reference therein]{1985A&A...142..451P}.
Given the low free oxygen abundance assumed to exist in protoplanetary disks, a more likely reaction is the following \citep{Imae1993}:
\begin{equation}
\ce{Mg2SiO4}+\ce{H2O(g)}+\ce{SiO(g)}\rightarrow2\ce{MgSiO3}+\ce{H2(g)} .
\end{equation}
This reaction is also supported by the decrease of \ce{H2O}(g) and \ce{SiO}(g) we see in Fig.~\ref{fig-majorgases} in the high temperature region.

\begin{figure}
\centering
\includegraphics[width=8.5cm]{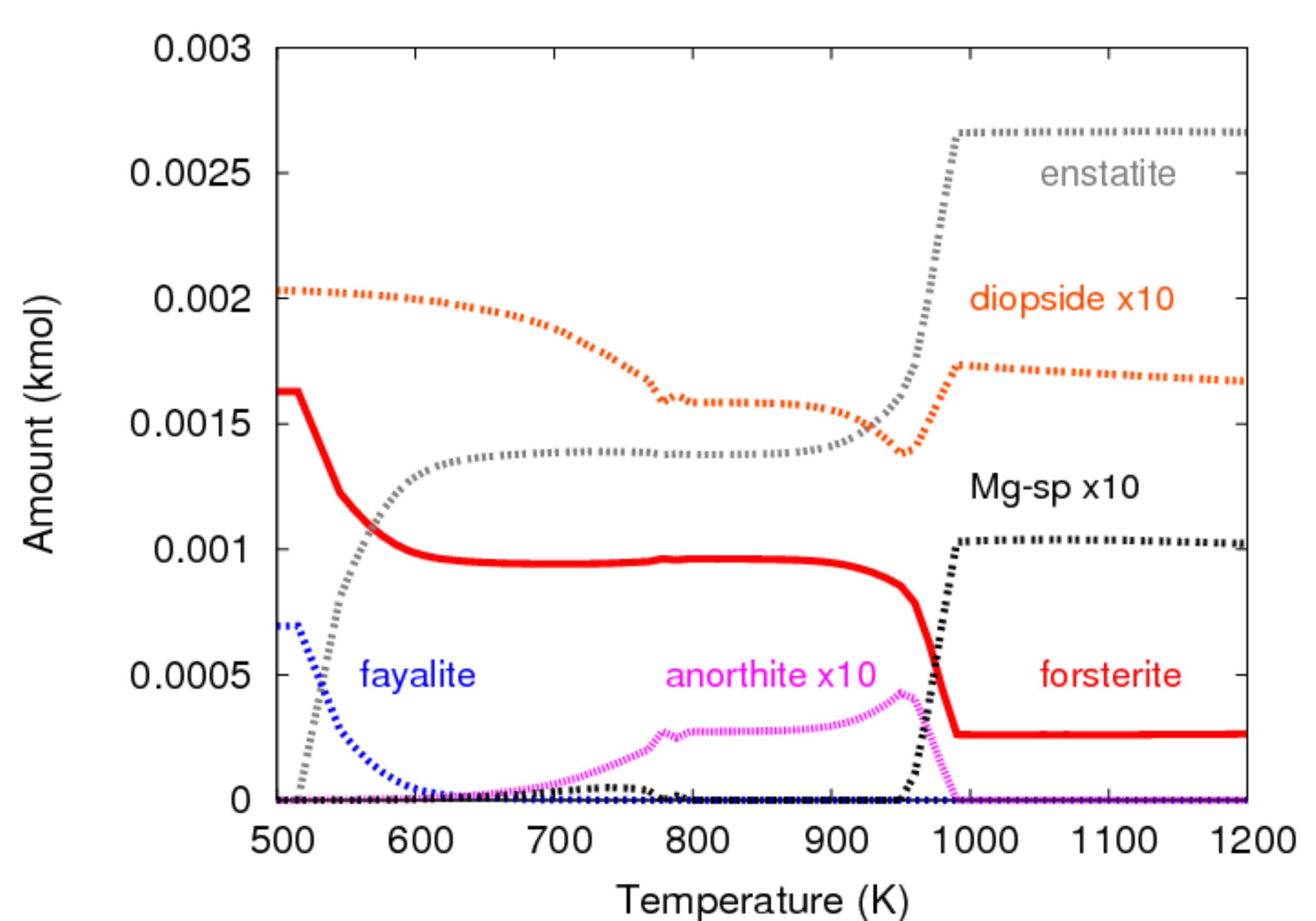}
\includegraphics[width=8.5cm]{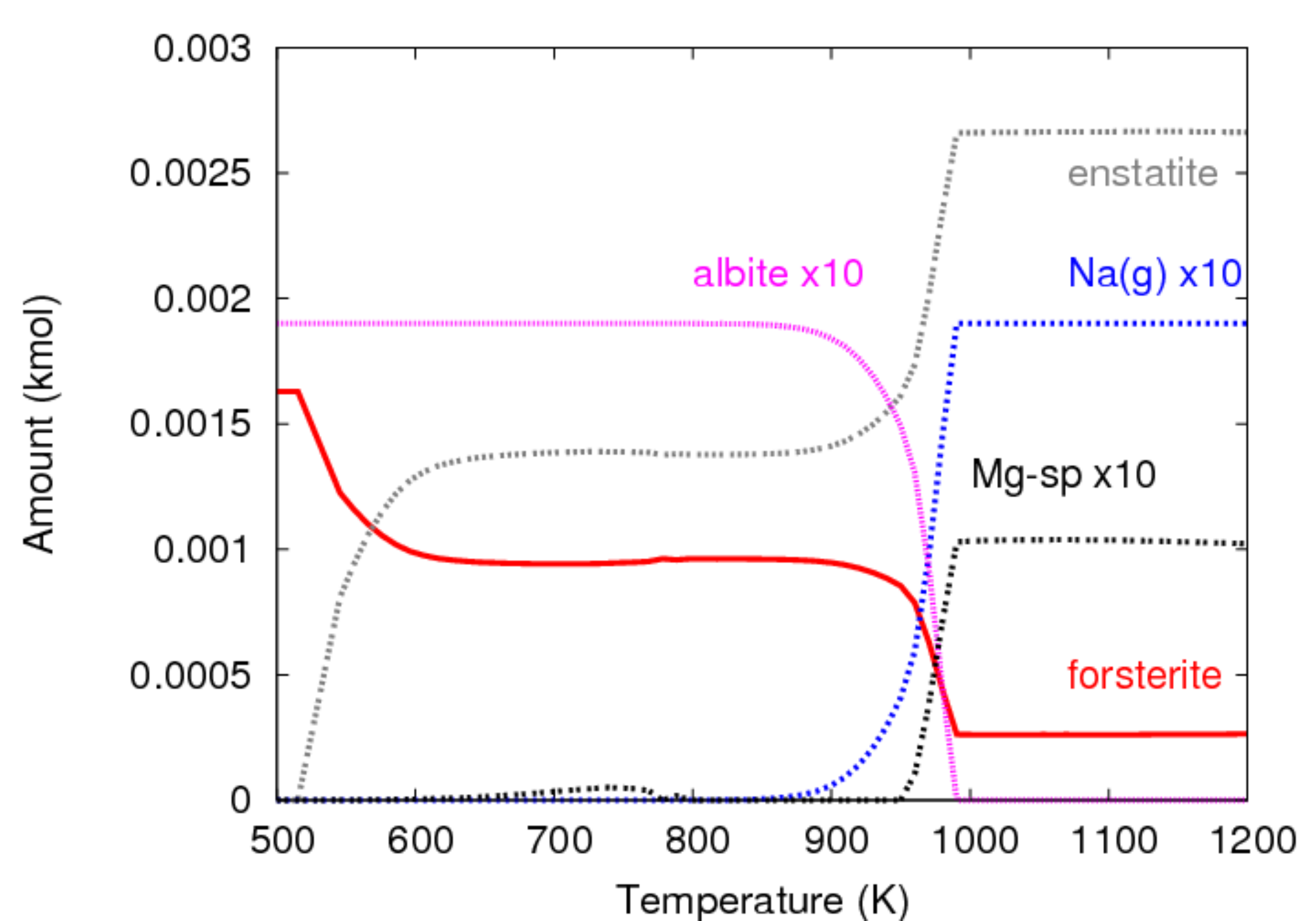}
\includegraphics[width=8.5cm]{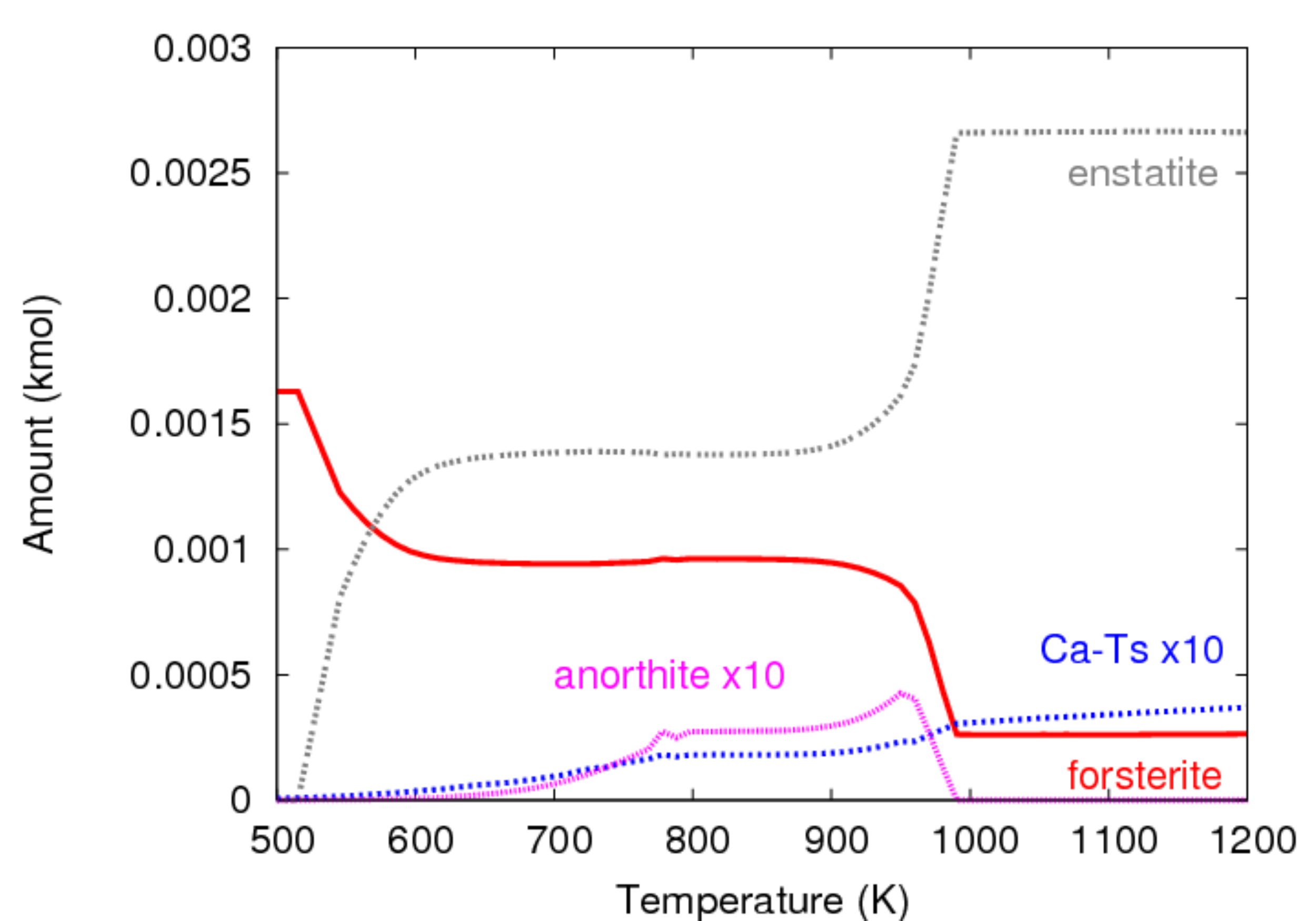}
\caption{Condensation sequence from $T=500$~K to $T=1200$~K for $P=10^{-3}$~bar for compounds involved in the reactions~(\ref{drop}) [top], (\ref{drop2}) [middle] and (\ref{drop3}) [bottom] using the regular solution model. The amount of anorthite, Mg-spinel, diopside, albite, Ca-Ts and Na(g) are multiplied by a factor of 10 for clarity in this plot.}
\label{1000}
\end{figure}

\begin{figure}
\centering
\includegraphics[width=8.5cm]{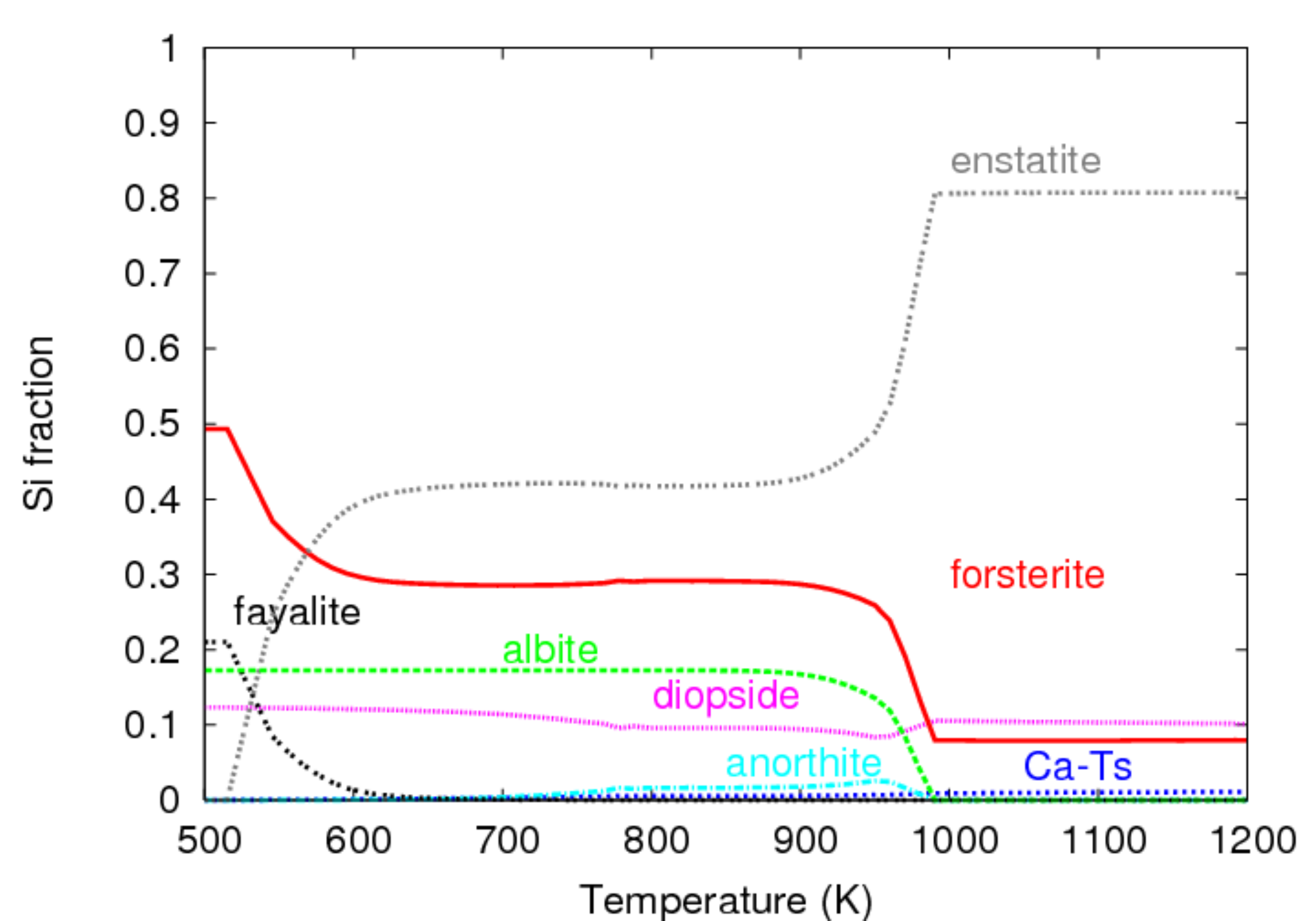}
\caption{{\bf Fraction of Si-bearing compounds in the 500--1200~K temperature range, where reactions~(\ref{drop}), (\ref{drop2}), (\ref{drop3}) and (\ref{fayalite}) take place.}}
\label{1000b}
\end{figure}

%
%

In our model we find that when $T\sim 1000$~K, the amount of enstatite drops dramatically and forsterite starts to form again. 

This sudden drop of enstatite can be explained with the following three reactions that occur simultaneously in this temperature region:
\begin{eqnarray}
\label{drop}
\lefteqn{2 \ce{MgSiO3}  + \ce{CaMgSi2O6}  + \ce{MgAl2O4}  \rightarrow} \nonumber \\ 
\lefteqn{\ce{CaAl2Si2O8} + 2\ce{Mg2SiO4},}  \
\end{eqnarray}
from \citet{Marfunin1998},
\begin{eqnarray}
\label{drop2}
\lefteqn{13\ce{MgSiO3}+\ce{MgAl2O4}+2\ce{Na}(g)+\ce{H2O}(g) \rightarrow }  \nonumber\\
\lefteqn{7\ce{Mg2SiO4}+2\ce{NaAlSi3O8}+\ce{H2}(g),} \
\end{eqnarray}
from \citet{Lewis2004}, and
\begin{eqnarray}
\label{drop3}
\lefteqn{2\ce{MgSiO3}+\ce{CaAl2SiO6} \rightarrow \ce{Mg2SiO4} +\ce{CaAl2Si2O8}}
\end{eqnarray}
from \citet{Marfunin1998} (see Fig.~\ref{1000}).
The forsterite:enstatite mole ratio produced by these three reactions is $10/17=0.59$.  We can compare this forsterite:enstatite ratio with the data returned by our calculation. We define  $\Delta X_{en}$ and $\Delta X_{f\!o}$  by
\begin{eqnarray}
\Delta X_{en} &=& |X_{en}^{1000}-X_{en}^{950}|  \\
\Delta X_{f\!o} &=& |X_{f\!o}^{1000}-X_{f\!o}^{950}| \,
\end{eqnarray}
which represent the difference between the amount of compound at $T=1000$~K and  $T=950$~K for enstatite, $\Delta X_{en}$, and forsterite, $\Delta X_{f\!o}$ respectively. Our calculation returns $\Delta X_{en}=$0.0010477 and $\Delta X_{fo}=$0.000593. Thus $\Delta X_{fo}$/$\Delta X_{en}=0.56$, in good agreement with the mole ratio involved in these three reactions above.

According our simulations, reactions~(\ref{drop}), (\ref{drop2}) and (\ref{drop3}) become most favourable when the temperature reaches $T\sim1000$~K. 
These reactions also explain the difference between our temperature of appearance of plagioclase ($T=980$~K) and those derived by previous works.

In the mid-temperature range $650\le T(\rm{K}) \le900$ the enstatite--forsterite ratio is constant, with $T\sim550$~K being the turning point where forsterite increases its amount over enstatite (see Figs.~\ref{fig-middlet} and \ref{1000}).
Thermodynamic predictions by  \citet{2000SSRv...92..177F} involving Fe show that, with decreasing temperature, the following reaction becomes more favourable:
\begin{eqnarray}
\label{fayalite}
\lefteqn{2 \ce{Fe}  + 2 \ce{MgSiO3}  + 2 \ce{H2O}(g) \rightarrow} \nonumber \\ 
\lefteqn{\ce{Mg2SiO4}  + \ce{Fe2SiO4}  + 2 \ce{H2}(g).}  \
\end{eqnarray}
This reaction explains the conversion of enstatite to forsterite at $T\sim 600$~K (the turning point) and also the formation of fayalite (\ce{Fe2SiO4}) at lower temperature (see Figs.~\ref{fig-lowt} and \ref{1000}). 

Fig.~\ref{1000b} show the Si-bearing in the 500--1200~K temperature region in which reactions~(\ref{drop}), (\ref{drop2}), (\ref{drop3}) and (\ref{fayalite}) take place.

In terms of maximum abundance, we find that enstatite is the dominant compound among the crystalline silicates, as is demonstrated in previous works.  However, our model suggests this is only the case in a well determined range of temperatures going from $T\sim 600$~K to $T\sim 1400$~K.

As briefly introduced in Section~\ref{solidsequence}, according to our calculations made using HSC with the regular solution model, the first peak of forsterite is pressure sensitive (see Fig.~\ref{fig-middlet}). We also see that the difference between the temperatures of appearance of forsterite and enstatite becomes smaller as the pressure decreases.  When the pressure drops to $P=10^{-6}$ bars, HSC returns $T_a=1212$~K as the temperature of appearance for forsterite and $T_a=1223$~K for enstatite. This in contrast with previous results reported by \citet{Yoneda1995}, \citet{Gail1998},  \citet{2004A&A...413..571G}  and \citet{Pasek2005}.  Table~\ref{tab-enstfor} reports the temperatures of appearance of forsterite and enstatite at different values of pressure (from $P=10^{-3}$ bars  to $P=10^{-6}$ bars) in our calculations. The temperature difference, $\Delta T = T_{a}^{for} - T_{a}^{ens}$, is also shown.

\begin{figure}
\centering
\includegraphics[width=8.5cm]{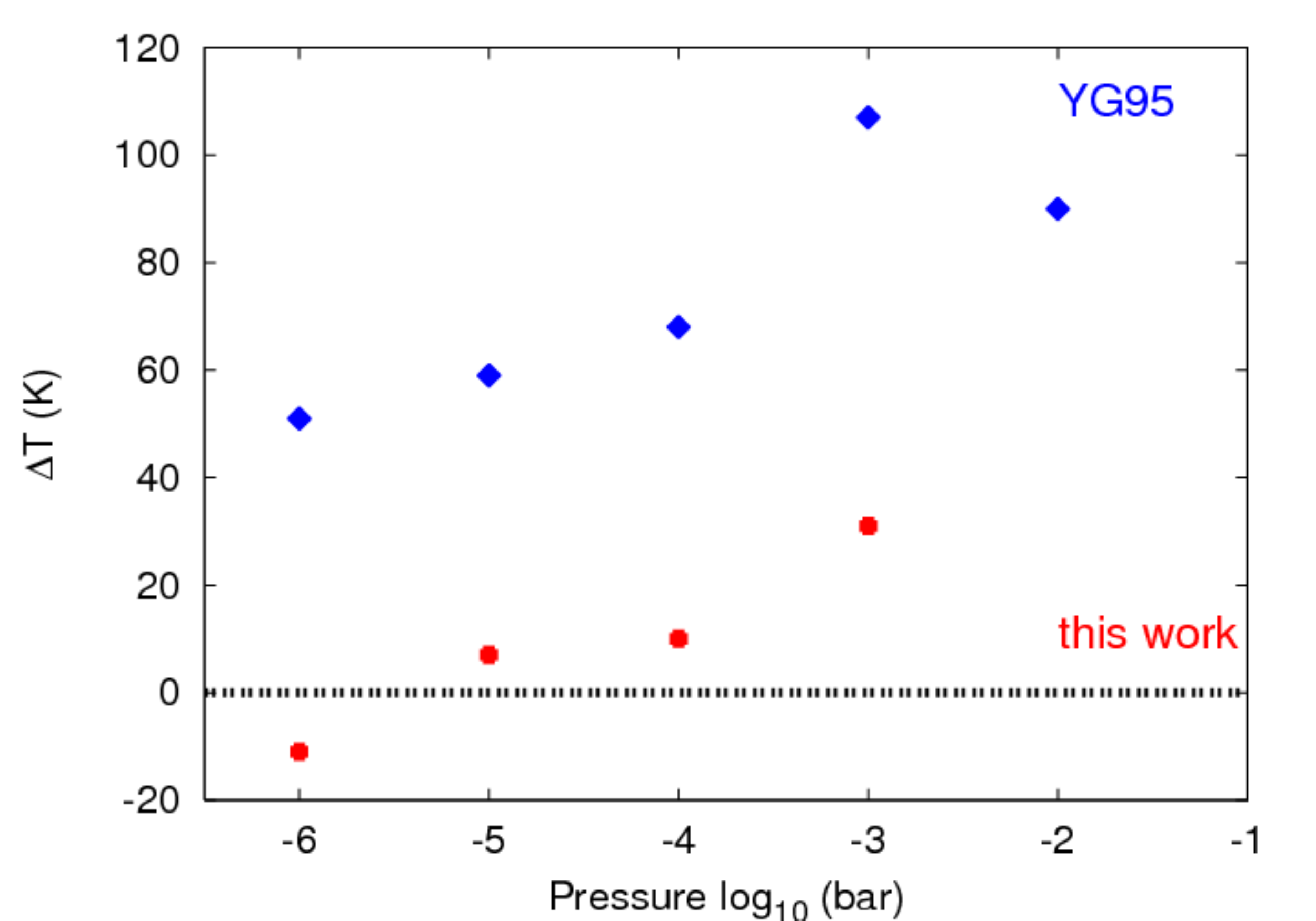}
\caption{Difference between the temperatures of appearance of forsterite and enstatite ($\Delta T$) at different pressure values as reported by  \citet{Yoneda1995} (blue diamonds) and this work (red dots). (See Table~\ref{tab-enstfor} for data values from this work). The $\Delta T=0$ line represents the point in which $T_a$(forsterite) = $T_a$(enstatite). }
\label{delta}
\end{figure}

The decreasing difference between the temperatures of appearance of forsterite and enstatite as the pressure decreases was also noticed by \citet{Grossman1972}. Indeed, the data reported by \citet{Yoneda1995} shows a similar behaviour when the pressure decreases from $P=10^{-3}$~atm to $P=10^{-6}$~atm.  In Fig.~\ref{delta} we plot this temperature difference, $\Delta T$, with pressure as found by \citet{Yoneda1995} and in our work. Similar results were also seen by \citet{Gail1998} and \citet{2004A&A...413..571G}.
\citet{Yoneda1995}, \citet{Gail1998}, \citet{Gail1998, 2004A&A...413..571G} and \citet{Pasek2005}  do not show the swap in the order of of appearance of forsterite and enstatite at $P=10^{-6}$ bar which we find with our regular solution model.  We can only assume that it is due to either the solution models used, or more likely, the thermodynamics data used Ð or indeed a combination of both. 

\begin{table}
\centering
\caption{The temperature of appearance, $T_a$, of forsterite and enstatite at different values of pressure, $P$ , and the difference in their temperatures of appearance, $\Delta T$. }
\begin{tabular}{l c c c c c}
\hline

$P$ (bar) & $T_a$ (K) & $T_a$ (K) & $\Delta T$ (K)  \\
  & Forsterite &  Enstatite &   \\
\hline									
$10^{-3}$	&	1415  &	1384	 &	31		\\
$10^{-4}$	&	1344  &	1334	 &	10		\\
$10^{-5}$	&	1285  &	1278 &	7		\\
$10^{-6}$	&	1212	 &	1223	 &	-11		\\
\hline
\end{tabular}									
\label{tab-enstfor}									
\end{table}

\subsection{Comparison with observations}

\subsubsection{Discs}

\citet{Pollack1994} studied the mass fraction of the eight most abundant elements (H, C, O, N, Mg, Si, Fe, and S) in molecular clouds, the interstellar medium, and solar system bodies. They also deduced a model for the composition of each {\bf compound} for the outer parts of protoplanetary discs where $T\le700$~K. Table~\ref{tab-pollack} shows the distribution of elements  in a typical disc, considering three kinds of matter: ices, refractory grains (which condense at high temperature), and gas. The main refractory elements observed by \citet{Pollack1994} are the same as in our model: olivine, orthopyroxene, troilite, and Fe species. Indeed these species are present for a large scale of temperature and for a high mol amount.

\begin{table*}
\centering
\caption{Fractional abundance of elements between gas, ice, and refractories in protoplanetary discs \citep{Pollack1994}.}
\begin{tabular}{l c c c}
\hline
Elements & Gas & Refractory & Ice \\ 
\hline
Si & 0 & 1 & 0 \\
 & & \begin{footnotesize}(\ce{Fe2SiO4}, \ce{Mg2SiO4}, \ce{FeSiO3}, \ce{MgSiO3})\end{footnotesize} & \\
Mg & 0 & 1 & 0 \\
 & & \begin{footnotesize}(\ce{Mg2SiO4}, \ce{MgSiO3})\end{footnotesize} & \\
Fe & 0 & 1 & 0 \\
 & & \begin{footnotesize}(\ce{Fe2SiO4}, \ce{FeSiO3}, FeS, Fe)\end{footnotesize} & \\
S & 0.20 & 0.75 & 0.05 \\
 & \begin{footnotesize}(\ce{H2S})\end{footnotesize} & \begin{footnotesize}(FeS)\end{footnotesize} & \begin{footnotesize}(\ce{H2S}, SO, \ce{SO2})\end{footnotesize} \\
O & 0.13 & 0.28 & 0.59 \\
 & \begin{footnotesize}(CO, O, \ce{H2O})\end{footnotesize} & \begin{footnotesize}(Silicates, CHON)\end{footnotesize} & \begin{footnotesize}(\ce{H2O}, CO, Volatile Organics)\end{footnotesize} \\
C & 0.28 & 0.55 & 0.17 \\
 & \begin{footnotesize}(CO, C)\end{footnotesize} & \begin{footnotesize}(CHON\end{footnotesize}) & \begin{footnotesize}(CO, CO2, \ce{CH4})\end{footnotesize} \\
N & 0.79 & 0.20 & 0 \\
 & \begin{footnotesize}(\ce{N2}, N, \ce{NH3})\end{footnotesize} & \begin{footnotesize}CHON\end{footnotesize} & \\
 \hline
\end{tabular}
\label{tab-pollack}
\end{table*}

The results derived by \citet{Pollack1994}, based on astronomical observations and modeling, show that most of the silicates are present in two main phases, olivine and orthopyroxene, with the former dominating. In the middle temperature region of our condensation sequence (see Fig.~\ref{fig-middlet}), moving toward lower temperatures we see that forsterite, the main olivine species in our system, replaces enstatite as the most stable compound. \citet{Pollack1994} indicate troilite (\ce{FeS}) as the most abundant Fe-compound ($T\le680$~K), also stating that its temperature of condensation is almost insensitive to the pressure. Furthermore, as the temperature increases,  troilite reacts with molecular hydrogen forming \ce{H2S}(g) and Fe(s).
Our results are in good agreement with \citet{Pollack1994}.  The condensation temperature for troilite derived with our regular solution model is $T=687$~K (see Tab.~\ref{tab-tot}) over a wide pressure range and \ce{H2S}(g) is the most abundant sulfur-gas when the temperature raises above $T=650$~K -- see Fig.~\ref{fig-majorgases}.

Silicates are observed almost everywhere that dust can survive: in the interstellar medium (ISM), in protoplanetary discs, in the Earth's mantle, in comets and meteorites in different structures and shapes. Infrared spectral analysis shows that the ISM is characterized by amorphous silicates, while in protoplanetary discs there is evidence of crystalline silicates, clear proof of processing and growth of grains \citep[]{2009A&A...507..327O,2009A&A...497..379M}. Among them, forsterite (\ce{Mg2SiO4}) and enstatite (\ce{MgSiO3}) are the best tracers of the condensation processes involved in the protoplanetary dics because of their related chemistry (section~\ref{disenfor}). \citet{2008ApJ...683..479B} used infrared (5--35~$\umu$m) spectra to study silicates in seven protoplanetary discs around pre-main sequence stars with Spitzer. This wavelength range probes both warm ($\lambda \sim10$~$\umu$m, $T\sim500$~K) and cool ($\lambda \sim30$~$\umu$m, $T\sim120$~K) region of the disc. They found variations in the abundance of the two main silicates with enstatite dominating the warmer region of the discs and forsterite the colder region. \citet{2009A&A...497..379M} also studied infrared spectra (7.5--35~$\umu$m) of 12 T Tauri systems and report the forsterite--enstatite ratio ($R_{(f/e)}$) for seven objects. The warmer region (7.5--17~$\umu$m) is characterized by $R_{(f/e)} \sim0.3$ while the cooler region (17--37~$\umu$m) has $0.9\le R_{(f/e)}\le 1.2$. the observation and modeling by \citet{2008ApJ...683..479B} and \citet{2009A&A...497..379M} are in good agreement and support our condensation results. Our equilibrium simulation shows that forsterite becomes more stable than enstatite in the low temperature region ($150\le T(\rm{K}) \le500$), with enstatite more abundant when $T\ge500$~K.

\subsubsection{Meteorites}

Additional information can be inferred from the study of chondrites. Chondrites are thought to be the oldest known components of the solar system and are assumed to have been created during the protoplanetary disc phase. Thus, they provide important evidence of the conditions present in the early stages of the solar system's formation. 
Chondrites are composed of three major parts: Calcium and Aluminium Inclusions (CAIs), chondrules and matrix \citep{2007AREPS..35..577S}. CAIs are the oldest objects in the chondrites and they are created at highest temperatures. CAIs are composed of different phases, including spinel (\ce{MgAl2O4},\ce{FeAl2O4}), melilite (\ce{Ca2Al2SiO7},\ce{Ca2MgSi2O7}) and compounds like perovskite (\ce{CaTiO3}), hibonite (\ce{CaAl12O19}), calcic pyroxene (\ce{CaMgSi2O6},\ce{CaAl2SiO6},\ce{CaTiAl2O6},\ce{CaTiAlSiO6}), anorthite (\ce{CaAl2Si2O8}), and forsterite (\ce{Mg2SiO4}), and also include less common compounds like grossite (\ce{CaAl4O7}), Ni-Fe metal, corundum (\ce{Al2O3}), rhonite (\ce{Ca4}(Mg,Al,Ti)$_{12}$--(Si,Al)$_{12}$\ce{O40}) and \ce{CaAl2O4} \citep{MacPherson2006}.

All the most important elements are present in our  system at high temperatures ($T\geq1300$~K at $P=10^{-3}$~bar). This first comparison provides a validation of the condensation process.  


Chondrules are composed of less refractory species and their structure is very diverse. However most chondrules contain the same main species of olivine (\ce{Mg2SiO4}, \ce{Fe2SiO4}), pyroxene (\ce{MgSiO3}, \ce{FeSiO3}), metallic Fe-Ni and Fe-Ni sulfides (FeS, \ce{Ni3S2}, NiS).  The results of our regular solution model support these observations, as these species are the main Fe, Mg, Ni compounds in our model creating in the middle and low temperatures regions ($T\leq1400$~K with a pressure of $10^{-3}$ bar).

The high temperature region between 1850~K and 1400~K  provides information about the first solids that condensed in protoplanetary discs. It is the region in which Ca--Al compounds start to condense into a variety of species. According our model, the condensation temperatures for Al(g) and Ca(g) are sensitive to the pressure (as shown in Fig.~\ref{fig-hight}). At higher values of pressure, when Ca(g) condenses at the same temperature as Al(g), no corundum (\ce{Al2O3}) forms as an isolated compound. This is because, at equilibrium the corundum is directly involved in the formation of Ca--Al compounds like hibonite according to the reaction
\begin{equation}
\ce{CaO}(g)+6\ce{Al2O3} \rightarrow \ce{CaAl12O19} .
\end{equation}
Lowering the pressure to $P=10^{-6}$ provides a small temperature window in which there is no \ce{CaO}(g) available for reaction and hence \ce{Al2O3} is the first solid compound. As soon as Ca(g) condenses, all the corundum is then involved in the formation of hibonite.
Observational evidence of isolated corundum has been reported by \citet{Nakamura2007}. They found forty-three corundum grains (1--11 ${\rm \umu}$m in size) in the matrix of the carbonaceous chondrite Acfer094 by using cathodoluminescence imaging. Their petrographic observations indicate that, in the framework of the homogeneous condensation theory, the pressure had to be lower than $3\times10^{-5}$~bar. They calculate the homogeneous condensation temperature of corundum as a function of total pressure and cooling time, showing that at pressures higher than few times $10^{-5}$~bar, the homogeneous condensation temperature of corundum is lower than the equilibrium condensation temperature of hibonite. This means that at these conditions hibonite is the first compound formed. The evidence of corundum grains in Acfer94 constrains the total pressure in the environment in which they formed to be lower than $10^{-5}$~bar. 
Thus corundum, as an isolated compound, could be tracer of low pressure environment.

\section{Conclusions}
\label{sec-conc}

Determining the condensation sequence and the physical chemistry that drives it is an important step in understanding the first stages of the planet formation process. Previous work made use of the ideal solution model ($a=X$) to describe the thermodynamic systems. However, laboratory experiments show that the behaviour of most species is non-ideal. In this work we investigated the thermodynamic properties and the activities of several solid compounds in a protoplanetary discs environment specifically forming on the regular solution model. Many compounds in our system deviate from the ideal across the entire range of composition, while others may be relatively close to ideal for most of the composition range. 
Decreasing the temperature causes further deviations from the ideal and, at fixed values of temperature, the composition of compounds can also vary.
As a consequence, the activity coefficients affect the results of the computation of the Gibbs free energy minimization.
From the experimental data describing the behaviour of the activity of several compounds in solution, and the regular solution model we derived, it is clear that a non-ideal solution must be taken in account outside the very high temperature regions (where most compounds are close the ideal) and when the compounds' composition in the phase is not dominant (i.e. when $X$ is low). We introduced activity coefficient relations for nine phases and twenty-one compounds. These relations can easily be used in Gibbs free energy minimization tools requiring polynomials describing $\gamma_{k}(X_{k},T)$.

Our regular solution model shows that the solids which condense in protoplanetary discs are mainly composed of silicates for a large range of temperatures and of water ice for the cooler regions of the disc. Calcium--aluminum compounds and iron compounds characterise the high temperature region and the low temperature region respectively.

Our results also demonstrate that pressure plays an important role determining the temperature of appearance and disappearance of compounds, their amounts  (e.g. the peak of forsterite in the middle temperature region) and formation (e.g. corundum in the high temperature region). 
  
 This work provides a starting point for the study of the non-ideal behaviour and the resulting composition of grains in protoplanetary discs.  
In future work, we will introduce 1D and 2D disc models to study in more detail the condensation processes and the resulting condensation sequence in well-defined locations in the disc.

\section*{Acknowledgments}

We thank the anonymous referee for making us investigate the behaviour of forsterite and enstatite in greater detail, which  has improved the manuscript.  This work was partly supported by a Swinburne Faculty of ICT Deans Collaborative Grant Scheme (VT), Swinburne Special Studies Program (STM) and  the CSIRO astrophysics group.

\bibliographystyle{mn2e}
\bibliography{biblio}

\newpage

\appendix

\onecolumn
\section{Full List of Species}
\begin{table*}
\caption{Full list of species used in the models.}

\begin{tabular}{l l l l l} \\ \hline
\multicolumn{2}{c}{\textbf{Gas species}} & & \multicolumn{2}{c}{\textbf{Solid species}} \\ \hline							
Al        	& 	Mg	& & 	\ce{(NH4)2SO4}	&	\ce{Na2SO4}	 \\	
AlH       	& 	MgH	& & 	Al	&	\ce{NaAlSi3O8}	 \\	
AlO	&	MgS	& & 	\ce{Al2O3}	&	Ni	 \\	
\ce{AlO2}	&	 N  	& &	\ce{Al2S3}	&	\ce{Ni3S2}	 \\	
AlOH	&	\ce{N2}	& &	\ce{Al2SiO5}	&	\ce{Ni3S4}	 \\	
\ce{Al2O}	& 	NCO 	& &	\ce{Al6Si2O13}	&	NiS	 \\	
\ce{Al2S}	& 	NH  	& & 	AlN	&	\ce{NiS2}	 \\	
\ce{AlS2}	&	\ce{NH2} 	& & 	C	&	Si	 \\	 
AlS	&	\ce{NH3}    	& &	\ce{Ca2Al2SiO7}	&	\ce{Si3N4}	 \\	
Ar	&	\ce{N2H2}	& &	\ce{Ca2MgSi2O7}	&	SiC	 \\	
C	& 	NO 	& &	\ce{CaAl12O19}	&	SiS	 \\	
CH	& 	\ce{NO2}	& & 	\ce{CaAl2O4}	&	\ce{SiS2}	 \\	
\ce{CH2}	&	\ce{N2O}	& & 	\ce{CaAl2Si2O8}	&	\ce{Si2N2O}	 \\	
\ce{CH3}	&	Na  	& &	\ce{CaAl2SiO6}	&		 \\	
\ce{CH4}	&	 \ce{Na2O2H2}	& &	\ce{CaFe(SiO3)2}	&		 \\
\ce{C2H2}	& 	\ce{Na2SO4}	& &	\ce{CaMgSi2O6}	&		 \\
\ce{C2H4}	& 	NaCN    	& & 	CaO	&		 \\
\ce{C2H6}	&	NaH 	& & 	CaS	&		 \\
\ce{CH2O}	&	NaO   	& &	\ce{CaSiO3}	&		 \\
CN	&	NaOH	& &	\ce{CaSO4}	&		 \\
\ce{C4N2}	& 	Ne	& &	Fe	&		 \\
CO	& 	 Ni  	& & 	{Fe}$_{0.947}$O	&		 \\
\ce{CO2}	&	\ce{Ni(CO)4}	& & 	\ce{Fe2(SO4)3}	&		 \\
COS	&	NiH	& &	\ce{Fe2O3}	&		 \\
CS  	&	NiO	& &	\ce{Fe2SiO4}	&		 \\
\ce{CS2}	& 	NiS  	& &	\ce{Fe3C}	&		 \\
Ca 	& 	O	& & 	\ce{Fe3O4}	&		 \\
CaH 	&	\ce{O2}	& & 	\ce{Fe3Si}	&		 \\
CaOH	&	 OH 	& &	\ce{FeAl2O4}	&		 \\
CaS	&	S	& &	FeO	&		 \\
Fe 	& 	\ce{S2}	& &	FeS	&		 \\
FeH 	& 	\ce{S2O}	& & 	\ce{FeS2}	&		 \\
FeO	&	SN	& & 	FeSi	&		 \\
FeS	&	SO	& &	\ce{FeSi2}	&		 \\
H 	&	\ce{SO2}	& &	\ce{FeSiO3}	&		 \\
\ce{H2}	& 	 Si 	& &	\ce{FeSO4}	&		 \\
HCN	& 	 SiC 	& & 	\ce{H2O}	&		 \\
HCO 	&	SiH   	& & 	\ce{Mg2SiO4}	&		 \\
HNO	&	\ce{SiH2}	& &	MgO	&		 \\
HNCO                	&	\ce{SiH4}	& &	\ce{MgAl2O4}	&		 \\
\ce{H2O}	&	SiN 	& &	MgS	&		 \\
HS	&	SiO   	& &	\ce{MgSiO3}	&		 \\
\ce{H2S}	&	\ce{SiO2}	& &	\ce{Na2S}	&		 \\
\ce{H2SO4}	&	 SiS 	& &	\ce{Na2SiO3}	&		 \\
He                  	&		& &	\ce{Na2SO4}	&		 \\

\hline							 
\end{tabular}
\label{table-fullspecieslist}
\end{table*}


\newpage

\section[]{Appearance and disappearance}
 
 \begin{table*}
\centering
\caption{Temperature of appearance and disappearance for solid compounds in the system at different pressures for the ideal solution model. The temperature of appearance (disappearance) is defined as the temperature in which the amount of compound raises above (falls below) $10^{-7}$ kmol and we indicate with {\it trace}, compounds that are present in the results with an amount of $10^{-9}\le$ kmol $\le10^{-7}$ and with {\it $<$50}, compounds present with an amount not less than $10^{-7}$ kmol at the lower limit of our range of temperature (50~K).  Compounds that do not appear in our calculations are indicated with --.}
\begin{tabular}{l l c c c c c c}
\hline
\multicolumn{2}{c}{Ideal Solution} & \multicolumn{2}{c}{$P=10^{-3}$ bar} & \multicolumn{2}{c}{$P=10^{-6}$ bar} \\	
Phase &  Species &  $T_{a}$~(K) & $T_{d}$~(K) & $T_{a}$~(K) & $T_{d}$~(K) \\
\hline
Melilite & Gehlenite  &  1606 & 1412  & 1359 & 1250 \\
Melilite & Akermanite & 1509 & 1412  & 1413 & 1250 \\
Fassaite & Ca--Ts &   1460  & 633  & 1413 & 704 \\
Fassaite & Diopside &   1460  & $<$50  & 1250 & 104 \\
Plagioclase & Anorthite & 974 & 536  & 813 & 540 \\
Plagioclase & Albite &  974& 244  & 813 & 213 \\
Olivine & Fayalite &  682 & 98  & 540 & 50 \\
Olivine & Forsterite &  1412 & $<$50  & 1195 & $<$50 \\
Spinel & Fe--Sp &  1606 & 974  & 1140  & $<$50 \\
Spinel & Mg--Sp &  1460 & 974  & 1250 & 540 \\
Sulfide & Troilite & 682 & $<$50  & 650 & $<$50 \\
Sulfide & \ce{Ni3S2} &  633 & $<$50  & 540 & $<$50 \\
Sulfide & \ce{NiS} & 682 & 147  & 540 & 159 \\
Orthopyr. & Ferrosilite & 1071 & 390  & 977 & 431 \\
Orthopyr. & Enstatite &  1363 & 390  & 1195 & 431 \\
Metal & Fe &  1412 & 390  & 1195 & 431 \\
Metal & Si &  1455 & 1182  & trace & trace \\
Metal &  Ni&  1455 & 424  & 1195 & 486 \\
Metal & Al &  trace & trace  & trace & trace \\
Magnesio-Wus & FeO &  -- & --  & -- & -- \\
Magnesio-Wus &  MgO & -- & --  & -- & -- \\
Pure & Corundum &  -- & --  & 1522 & 1468 \\
Pure & Ice &  147 & $<$50  & 104 & $<$50 \\
Pure & Hibonite &  1704 & 1460  & 1468 & 1304 \\
Pure & \ce{Na2SiO3} &  487 & $<$50  & 431 & $<$50 \\
Pure & CaO &  -- & --  & -- & -- \\
Pure & \ce{Fe2O3} &  -- & --  & -- &--  \\
Pure & \ce{Fe3O4} &  244 & 50  & 213 & 50 \\
Pure & \ce{SiO2} &  trace & trace  & -- & -- \\
Pure & C &  -- & --  & -- & -- \\
\hline
\end{tabular}
\label{3regularideal}
\end{table*}

\begin{table*}															
\centering		
\caption{Temperature of appearance and disappearance for solid compounds in the system at different pressures for the regular solution model. The temperature of appearance (disappearance) is defined as the temperature in which the amount of compound raises above (falls below) $10^{-7}$ kmol and we indicate with {\it trace}, compounds that are present in the results with an amount of $10^{-9}\le$ kmol $\le10^{-7}$ and with {\it $<$50}, compounds present with an amount not less than $10^{-7}$ kmol at the lower limit of our range of temperature (50~K). Compounds that do not appear in our calculations are indicated with --.}																												\begin{tabular}{l l c c c c c c}															
\hline															
\multicolumn{2}{c}{Regular Solution}  &\multicolumn{2}{c}{$P=10^{-3}$ bar} & \multicolumn{2}{c}{$P=10^{-6}$ bar} & \multicolumn{2}{c}{$P=10^{-8}$ bar} \\															
Phase	&	Species	&	$T_a$~(K)	&	$T_d$~(K)	& 	$T_a$~(K)	&	$T_d$~(K)	&	$T_a$~(K)	&	$T_d$~(K)	\\
\hline															
Melilite	&	Gehlenite	&	1556	&	1445	&	1364	&	1253	&	1202	&	1152	\\
Melilite	&	Akermanite	&	1546	&	1445	&	1364	&	1273	&	1202	&	1152	\\
Fassaite	&	Ca-Ts	&	1445	&	515	&	1253	&	505	&	1253	&	484	\\
Fassaite	&	Diopside	&	1445	&	$<$50	&	1253	&	60	&	1182	&	484	\\
Plagioclase	&	Anorthite	&	980	&	606	&	828	&	575	&	757	&	565	\\
Plagioclase	&	Albite	&	980	&	151	&	828	&	151	&	757	&	151	\\
Olivine	&	Fayalite	&	717	&	70	&	656	&	70	&	656	&	60	\\
Olivine	&	Forsterite	&	1415	&	$<$50	&	 1212 	&	$<$50	&	1081	&	$<$50	\\
Spinel	&	Fe-Sp	&	1233&	$<$50	&	1212	&	$<$50	&	1111	&	$<$50	\\
Spinel	&	Mg-Sp	&	1455	&	950	&	1253	&	586	&	1152	&	575	\\
Sulfide	&	FeS	&	687	&	$<$50	&	687	&	$<$50	&	687	&	$<$50	\\
Sulfide	&	\ce{Ni3S2}	&	687	&	60	&	687	&	$<$50	&	676	&	50	\\
Sulfide	&	NiS	&	687	&	505	&	trace	&	trace	&	trace	&	trace	\\
Orthopyro	&	Ferrosilite	&	757	&	545	&	687	&	505	&	676	&	484	\\
Orthopyro	&	Enstatite	&	1384	&	545	&	1223	&	505	&	1132	&	464	\\
Metal	&	Fe	&	1455	&	464	&	1202	&	464	&	1081	&	444	\\
Metal	&	Si	&	1455	&	1182	&	trace	&	trace	&	trace	&	trace	\\
Metal	&	Ni	&	1455	&	424	&	1202	&	464	&	1081	&	454	\\
Metal	&	Al	&	trace	&	trace	 &	trace	&	trace	&	trace	&	trace	\\
Magnesio-Wus	&	FeO	&	--	&	--	&	--	&	--	&	--	&	--	\\
Magnesio-Wus	&	MgO	&	--	&	--	&	--	&	--	&	--	&	--	\\
Pure	&	Corundum	&	--	&	--	&	1536	&	1506	&	1425	&	1415	\\
Pure	&	Ice	&	191	&	$<$50	&	151	&	$<$50	&	121	&	$<$50	\\
Pure	&	Hibonite	&	1738	&	1455	&	1506	&	1253	&	1405	&	1253	\\
Pure	&	\ce{Na2SiO3}	&	464	&	$<$50	&	464	&	$<$50	&	444	&	$<$50	\\
Pure	&	CaO	&	--	&	--	&	--	&	--	&	--	&	--	\\
Pure	&	\ce{Fe2O3}	&	--	&	--	&	--	&	--	&	--	&	--	\\
Pure	&	\ce{Fe3O4}	&	--	&	--	&	--	&	--	&	--	&	--	\\
Pure	&	\ce{SiO2}	&	trace	&	trace	&	trace	&	trace	&	trace	&	trace	\\
Pure &	C &	--	&	--	&	--	&	--	&	--	&	--	\\
\hline
\end{tabular}													
\label{tab-tot}															
\end{table*}

\newpage

\section[]{Comparison with other works}
\begin{landscape}
\begin{table}
\centering
\caption{{\bf Temperature of appearance ($T_a$) and disappearance ($T_d$) for pressures of  $10^{-3}$ (left columns) and $10^{-6}$~bars (right columns) for our model, compared to those of YG95 \citep{Yoneda1995}, G98 \citep{Gail1998} and P05 \citep{Pasek2005}.}}
\begin{tabular}{l c c c  c c c c c c c c c c c c}
\hline																					
 & \multicolumn{6}{c}{$P=10^{-3}$ bar} & & \multicolumn{8}{c}{$P=10^{-6}$ bar} \\ 
 & \multicolumn{2}{c}{This model} & \multicolumn{2}{c}{YG95} &  \multicolumn{2}{c}{G98}& & \multicolumn{2}{c}{This model} & \multicolumn{2}{c}{YG95} & \multicolumn{2}{c}{P05} & \multicolumn{2}{c}{G98}\\  
 & $T_a$ (K) & $T_d$ (K) & $T_a$ (K) & $T_d$ (K) &   $T_a$ (K)   &     $T_d$ (K)       &  & $T_a$ (K) & $T_d$ (K) & $T_a$ (K) & $T_d$ (K) & $T_a$ (K) & $T_d$ (K) &  $T_a$ (K) & $T_d$ (K) \\ 
 \hline																					
Corundum	&	--	&	--	&	1770	&	1740	&           1770 & --&                                           &	1546	&	1506	&	1571	&	1481	&	1535	&	1501	& 1550	& --\\
Hibonite	&	1738	&	1455	&	1743	&	1500	&		    -- & --&   					     &	1536	&	1253	&	1485	&	1292	&	1501	&	1292	&	--&-- \\
Melilite	&	1556	&	1445	&	1628	&	1444	&		    1560 & --&   					     &	1364	&	1253	&	1405	&	1264	&	1430	&	1251	&	1360 & --\\
Plagioclase	&	980	&	606	&	--	&	--	&	    1030 & --&   				     &	828	&	575	&	1268	&	1250	&	1261	&	--&	-- &	--\\
Fassaite	&	1445	&	515	&	1449	&	--	&		    -- & --&   				     &	1253	&	505	&	1257	&	--	&	1201	&	--&	--&--	\\
Spinel (Mg)	&	1455	&	950	&	1501	&	1409	&	    1460 & --&   				     &	1253	&	586	&	1252	&	1227	&	1228	&	--&	1270& --	\\
Forsterite	&	1415	&	$<$50	&	1443	&	--	&	    1440 & --&   				     &	{\bf 1212} 	&	$<$50	&	1246	&	--	&	1228	&	--& 1250	& --\\
Plagioclase	&	980	&	151	&	1416	&	--	&	    -- & --&   				     &	828	&	151	&	1234	&	--	&	1197	&	--	&	-- &-- \\
Metal (Fe)	&	1455	&		&	1464	&	--	&	             1450 & --&   					     &	1202	&	--	&	1214	&	--	&	1201	&	--&	1200 &	--\\
Enstatite	&	1384	&	545	&	1366	&	--	&	             1380 & --&   					     &	1223	&	505	&	1195	&	--	&	1208	&	--&	1205& --	\\
Spinel (Fe)	&	1445	&	$<$50	&	--	&	--&	    -- & --&   			     &	1223	&	$<$50	&	--	&	--	&	--	&	&	--	&-- \\
Fayalite	&	717	&	70	&	--	&	--	&		    -- & --&   				     &	656	&	70	&	--	&	--	&	--	&	--&	--	&-- \\
Ferrosilite	&	757	&	545	&	--	&	--	&		    -- & --&   				     &	687	&	505	&	--	&	--	&	--	&	--&	--	&-- \\
Troilite	&	687	&	$<$50	&	--	&	--	&	    720 & --&   				     &	687	&	$<$50	&	--	&	--	&	--	&	--	& 720	&-- \\
Magnetite	&	--	&	--	&	--	&	--	&		    -- & --&   				     &	--	&	--	&	--	&	--	&	--	&	--	&--	&-- \\
\hline
\end{tabular}																					
\label{tab-resultT}																					
\end{table}
\end{landscape}

\label{lastpage}

\end{document}